\documentclass[floats,floatfix,amssymb,prd,twocolumn,superscriptaddress,nofootinbib,nolongbibliography,reprint]{revtex4-1}

\usepackage{amsmath}
\usepackage{amsfonts}
\usepackage{amssymb}	
\usepackage{graphicx}
\usepackage{bm}
\usepackage{color}
\usepackage{commath}
\allowdisplaybreaks
\usepackage{makecell}

\usepackage{color}
\definecolor{linkcolor}{rgb}{0.1,0.3,0.6}
\usepackage[unicode, colorlinks=true, linkcolor=linkcolor, citecolor=linkcolor, filecolor=linkcolor, urlcolor=linkcolor, linktocpage, breaklinks]{hyperref}

\def\F{{\cal F}}
\def\de{\partial}
\def\l{{\ell}}
\def\lm{{\ell m}}
\def\ha{{\hat{a}}}
\def\TEOBResumS{\texttt{TEOBResumS}}
\def\RWZ{{\texttt{RWZhyp}}}

\def\oe{{\rm (o/e)}}
\def\tA22{{t_{A_{22}}^{\rm peak}}}
\def\tAlm{{t_{A_\lm }^{\rm peak}}}
\def\Amx{{A_{\rm peak}}}

\def\d2Amx{{\ddot{A}_{\rm peak}}}
\def\omgmx{{\omega_{\rm peak}}}
\def\Deltaomgmx{{\Delta \omega_{\rm peak}}}

\def\Amxq{{A_\tA22}}
\def\dAmxq{{\dot{A}_\tA22}}
\def\ddAmxq{{\ddot{A}_\tA22}}

\def\tOmgOrb{{t_{\Omega_{\rm orb}}^{\rm peak}}}
\def\tLR{{t_{\rm LR}}}
\def\tNQC{{t_\lm^{\rm NQC}}}

\def\t{{\tau}}
\def\tQNM{{t_0^{\rm QNM} }}
\def\tsep{{t_{\rm sep}}}
\def\esep{{e_{\rm sep}}}
\def\tplunge{{t_{\rm plunge}}}
\def\Omgpk{{\Omega_{\rm pk}}}
\def\bmrg{{\hat{b}_{\Omgpk}}}
\def\tmatch{{t_\lm^{\rm match}}}

\newcommand\be{\begin{equation}}
\newcommand\ee{\end{equation}}

\usepackage[nolist,nohyperlinks]{acronym}

\allowdisplaybreaks

\begin{document}

\title{Faithful effective-one-body waveform of small-mass-ratio coalescing black hole binaries: the eccentric, nonspinning, case}

\author{Simone Albanesi}
\affiliation{INFN sezione di Torino, Torino, 10125, Italy}
\affiliation{Dipartimento di Fisica, Universit\`a di Torino, Torino, 10125, Italy}
\author{Sebastiano Bernuzzi}
\affiliation{Theoretisch-Physikalisches Institut, Friedrich-Schiller-Universit{\"a}t Jena, 07743, Jena, Germany}
\author{Thibault Damour}
\affiliation{Institut des Hautes Etudes Scientifiques, 35 Route de Chartres, Bures-sur-Yvette, 91440, France}
\author{Alessandro Nagar}
\affiliation{INFN sezione di Torino, Torino, 10125, Italy}
\affiliation{Institut des Hautes Etudes Scientifiques, 35 Route de Chartres, Bures-sur-Yvette, 91440, France}
\author{Andrea Placidi}
\affiliation{Dipartimento di Fisica e Geologia, Universit\`a di Perugia, Via A. Pascoli, 06123 Perugia, Italia}
\affiliation{Niels Bohr Institute, Copenhagen University,  Blegdamsvej 17, DK-2100 Copenhagen \O{}, Denmark}
\affiliation{Galileo Galilei Institute for Theoretical Physics, Largo Enrico Fermi, 2, 50125 Firenze, Italy}

\begin{abstract}
We present a new effective-one-body (EOB) waveform for eccentric, nonspinning, binaries
in the extreme mass ratio limit, with initial eccentricities up to $0.95$. The EOB analytical waveform, 
that includes noncircular corrections up to second post-Newtonian order, is completed by a 
phenomenological ringdown model that is informed by Regge-Wheeler-Zerilli (RWZ) 
type waveforms generated by a point-particle source. 
This model notably includes the beating between positive and negative frequency quasi-normal-modes (QNMs).
We analyze various prescriptions to faithfully complete the analytical EOB waveform
in the transition from plunge to merger. In particular, we systematically explore the effect 
of: (i) the generic Newtonian prefactor; (ii) next-to-quasi-circular (NQC) corrections to amplitude
and phase; (iii) the point where NQC corrections are determined; (iv) the ringdown attachment point.
This yields EOB/RWZ quadrupolar phase differences through merger and ringdown $\lesssim 0.01$~rad 
for the quasi-circular case and $\lesssim 0.05$~rad for the eccentric case. Higher modes are
also modeled up to the $\ell=m=5$ multipole. We finally discuss the excitation of the QNMs and 
present a heuristic model to motivate it in correlation with the presence of a point-particle source.
\end{abstract}

\date{\today}

\maketitle

\section{Introduction}
\label{sec:introduction}
The gravitational waves (GWs) that have been detected by the interferometers of the 
LIGO-Virgo-KAGRA collaboration~\cite{LIGOScientific:2016aoc,LIGOScientific:2018mvr,LIGOScientific:2020ibl,LIGOScientific:2021djp} were generated by the coalescence of comparable mass
binaries, mainly black hole binaries. Since accurate waveform models are required to analyze 
these signals without biases,
many efforts have been devoted to build accurate and fast semi-analytical models using different 
approaches. However, there are no analytical solutions of the Einstein field equations 
for the merger of the two objects, even if one considers only vacuum-solutions. 
Therefore, all the currently available semi-analytical models that are able to provide waveforms 
for the inspiral, plunge merger and ringdown of the binaries are informed/calibrated using numerical results.
 
One of the semi-analytical approaches that has been proved to be accurate and fast enough to perform parameter 
estimations~\cite{Gamba:2021gap,Bonino:2022hkj,LIGOScientific:2018mvr,LIGOScientific:2020ibl,LIGOScientific:2021djp} 
is the Effective-One-Body 
(EOB) model~\cite{Buonanno:1998gg,Buonanno:2000ef,Damour:2000we,Damour:2001tu,Damour:2015isa}. 
While the pure analytical EOB approach can be used to
faithfully describe only the inspiral of compact binaries, it is possible to use numerical data to 
extend the model and describe also the merger and ringdown of the system~\cite{Buonanno:2006ui,Damour:2007xr}. 
Notably, EOB models can be improved and completed using both numerical results from comparable mass binaries 
and from the extreme-mass-ratio regime. This is linked to the fact that, given a system of two compact objects 
with masses $m_1$ and $m_2$, the EOB metric is a $\nu$-deformation of a black hole solution, 
being $\nu=\mu/M$ the symmetric mass ratio and $\mu=m_1 m_2/(m_1+m_2)$
and $M=m_1+m_2$ the reduced and total mass of the binary, respectively. Consequently, the test-mass limit
is smoothly connected to the comparable mass case. 
The dawn of what is nowadays the established gravitational waveform modeling via the EOB 
approach informed by numerical simulations can be traced back to two seminal
papers, one by Buonanno, Cook and Pretorius~\cite{Buonanno:2006ui}, and the other 
by Damour and Nagar~\cite{Damour:2007xr}. The former used the pioneering Numerical Relativity (NR) 
simulation of equal-mass, nonspinning black hole binary~\cite{Pretorius:2005gq}, while the latter considered the 
quasi-circular inspiral and plunge of a non-spinning test-particle in a Schwarzschild black hole~\cite{Nagar:2006xv}.
Since the test-mass limit is a controlled theoretical laboratory to test 
prescriptions to use also in the comparable mass case, it has been explored
in many EOB works, see e.g. Refs.~\cite{Nagar:2006xv,Damour:2007xr,Bernuzzi:2010ty,Bernuzzi:2010xj,Yunes:2010zj,Barausse:2011kb,Albanesi:2021rby}
Moreover, this limit is also interesting by itself, since the space-based mission 
LISA~\cite{LISA:2017pwj} will be able
to detect Extreme-Mass-Ratio-Inspirals (EMRIs)~\cite{Babak:2017tow,Berry:2019wgg}. 
However, to model these astrophysical systems, an accurate description of the dynamics is needed. The 
Gravitational Self-Force (GSF) community has devoted many efforts in tackling this problem~\cite{Pound:2015tma,Barack:2018yvs,Warburton:2021kwk,VanDeMeent:2018cgn}, 
with techniques and results that are beyond the scope of this work. We just mention that the EOB approach was recently
shown to accurately describe the dynamics of GSF-evolved EMRIs in the quasi-circular case~\cite{Albertini:2022rfe,Albertini:2022dmc}
once informed by GSF-results~\cite{Damour:2009sm,Barack:2010ny,LeTiec:2011dp,Barausse:2011dq,Antonelli:2019fmq,Nagar:2022fep}. 
In this work we focus instead on the EOB prescription to compute the waveform at infinity for 
a non-spinning test-particle plunging in a Schwarzschild black hole after an eccentric inspiral.
The numerical data used to build the ringdown model, and thus complete the EOB 
waveform, are obtained solving numerically the Regge-Wheeler and Zerilli (RWZ) 
equations~\cite{Regge:1957td,Zerilli:1970se,Nagar:2005ea,Martel:2005ir} 
using the time-domain code \RWZ{}~\cite{Bernuzzi:2010ty,Bernuzzi:2011aj,Bernuzzi:2012ku}. 
We argue that a description based exclusively
on Quasi-Normal-Modes (QNMs) cannot be used to describe the ringdown waveform starting from
the peak of the amplitude. We rather use a modified version of the phenomenologically agnostic 
ringdown model presented in Ref.~\cite{Damour:2014yha}. 
We model the (2,2) multipole and also all the $m>0$ higher modes up to $\l=4$, plus the (5,5) mode. 
The complete waveform obtained is, to our knowledge, the most accurate EOB waveform for a 
nonspinning  test particle on quasi-circular inspirals in Schwarzschild spacetime, and also 
generalize well to highly eccentric dynamics. 
To conclude, we also discuss how the QNMs build-up during the last stages of the evolution of 
the binary. Among other phenomenological results, we find that the overtones are excited before the fundamental QNM, 
confirming the qualitative arguments presented in Ref.~\cite{Damour:2007xr}. 

The paper is structured as follows.
In Sec.~\ref{sec:configuration} we discuss how we compute the approximate dynamics of the particle, the numerical
RWZ waveform and the modeling of the ringdown. In Sec.~\ref{sec:modeling} we discuss the inspiral waveform
and how we match it to the ringdown model using the NQC corrections. 
The complete EOB waveform obtained is then compared to the original numerical results in Sec.~\ref{sec:model_test}.
In Sec.~\ref{sec:qnm_phenom} we further analyze the QNMs contributions in the postpeak waveform and we
revisit the matching of the ringdown model using an extended time interval.
The build up of the QNMs excitation is discussed in Sec.~\ref{sec:toymodel}.
The rescaled phase-space variables that we use in this work are related to the physical ones by $t=T/(GM)$, 
$r=R/(GM)$, $p_{r}=P_{R}/\mu$ and $p_\varphi=P_\varphi/(\mu GM)$. We
will also use geometric units $G=c=1$.

\section{From eccentric inspiral to plunge, merger and ringdown}
\label{sec:configuration}
%
\begin{table}[t]
	\caption{\label{tab:ID}Configurations considered and relevant quantities at merger time, defined as the peak of the orbital frequency.	
	From left to right: initial semilatus rectum, initial eccentricity, eccentricity at the separatrix-crossing time,
	time difference between the peak of the orbital frequency and the peak of the quadrupolar amplitude,
	energy, angular momentum and quadrupolar amplitude at the peak of the orbital frequency (that
	corresponds to the light-ring crossing). }
	\begin{center}
		\begin{ruledtabular}
			\begin{tabular}{c c c c c c c c} 
				$\#$ & $p_0$  & $e_0$ & $e_{\rm sep}$ & $t^{\rm peak}_{\Omega_{\rm orb}} - \tA22$ &
				$\hat{E}^{\rm LR}$ & $p_{\varphi}^{\rm LR}$ & $A_{22}^{\rm LR}$  \\	
				\hline
				\hline
 1 & 7.0 & 0.00 & 0.000 & 2.559 & 0.9422 & 3.4574 & 0.2928 \\ 
 2 & 7.3 & 0.05 & 0.061 & 2.585 & 0.9424 & 3.4598 & 0.2932 \\ 
 3 & 7.0 & 0.10 & 0.113 & 2.657 & 0.9429 & 3.4668 & 0.2942 \\ 
 4 & 8.0 & 0.15 & 0.141 & 2.708 & 0.9433 & 3.4717 & 0.2950 \\ 
 5 & 7.5 & 0.20 & 0.201 & 2.856 & 0.9444 & 3.4855 & 0.2970 \\ 
 6 & 8.0 & 0.25 & 0.229 & 2.925 & 0.9450 & 3.4932 & 0.2981 \\ 
 7 & 8.0 & 0.30 & 0.276 & 3.080 & 0.9462 & 3.5081 & 0.3003 \\ 
 8 & 8.0 & 0.35 & 0.321 & 3.240 & 0.9476 & 3.5243 & 0.3027 \\ 
 9 & 7.5 & 0.40 & 0.393 & 3.538 & 0.9503 & 3.5544 & 0.3070 \\ 
10 & 8.0 & 0.45 & 0.415 & 3.651 & 0.9513 & 3.5644 & 0.3085 \\ 
11 & 7.7 & 0.50 & 0.482 & 4.004 & 0.9545 & 3.5984 & 0.3133 \\ 
12 & 8.0 & 0.55 & 0.514 & 4.189 & 0.9563 & 3.6159 & 0.3158 \\ 
13 & 8.0 & 0.60 & 0.563 & 4.479 & 0.9592 & 3.6447 & 0.3199 \\ 
14 & 8.0 & 0.65 & 0.615 & 4.839 & 0.9625 & 3.6768 & 0.3245 \\ 
15 & 8.0 & 0.70 & 0.670 & 5.298 & 0.9665 & 3.7142 & 0.3298 \\ 
16 & 8.0 & 0.75 & 0.728 & 5.877 & 0.9710 & 3.7554 & 0.3359 \\ 
17 & 8.0 & 0.80 & 0.778 & 6.284 & 0.9751 & 3.7912 & 0.3408 \\ 
18 & 8.2 & 0.85 & 0.818 & 6.628 & 0.9784 & 3.8202 & 0.3451 \\ 
19 & 8.3 & 0.90 & 0.869 & 7.497 & 0.9839 & 3.8684 & 0.3523 \\ 
20 & 8.5 & 0.95 & 0.904 & 8.160 & 0.9878 & 3.9004 & 0.3566 
			\end{tabular}
		\end{ruledtabular}
	\end{center}
\end{table}

\begin{figure*}[t]
	\begin{center}
	\includegraphics[width=0.45\textwidth]{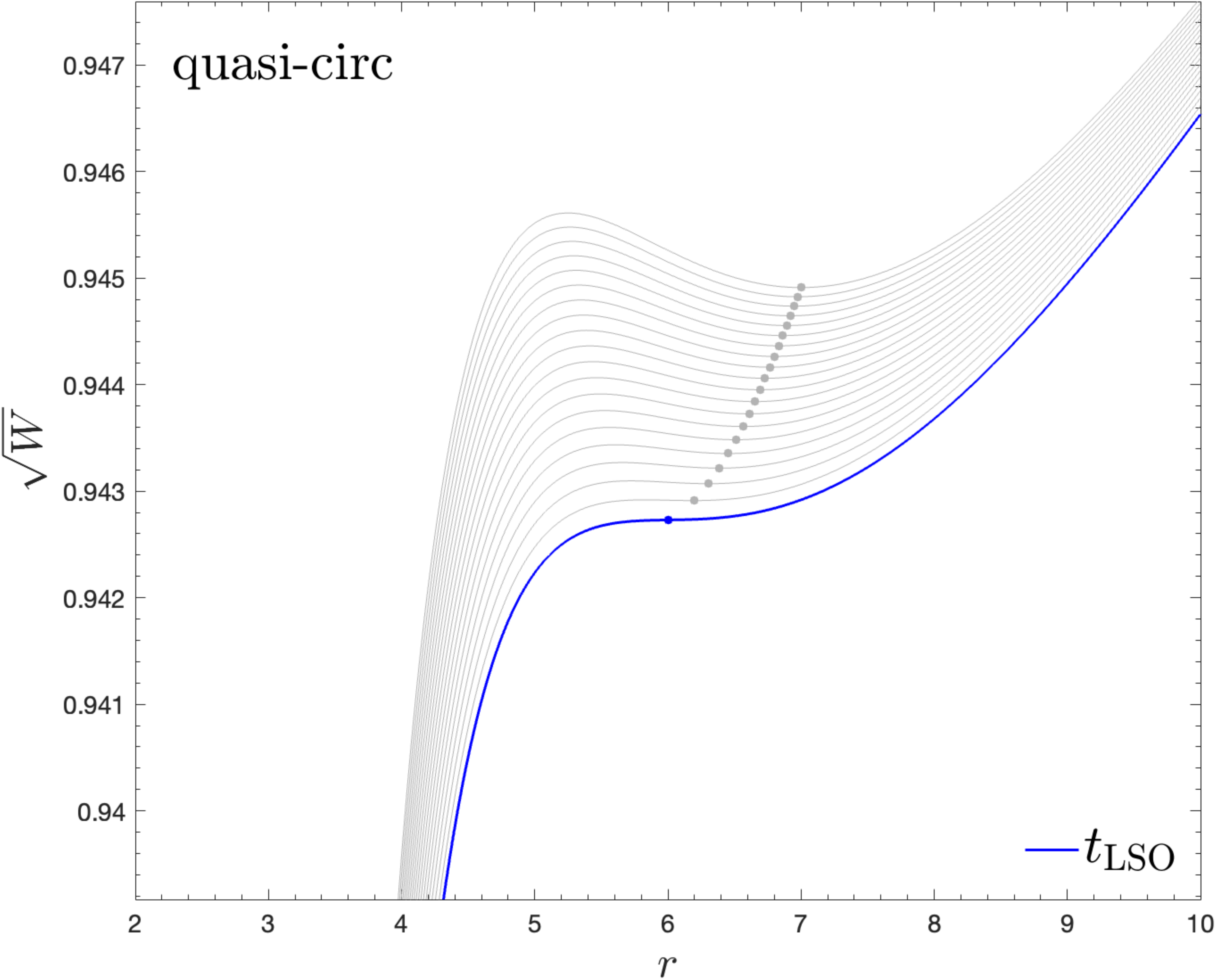}
	\hspace{5mm}
	\includegraphics[width=0.45\textwidth]{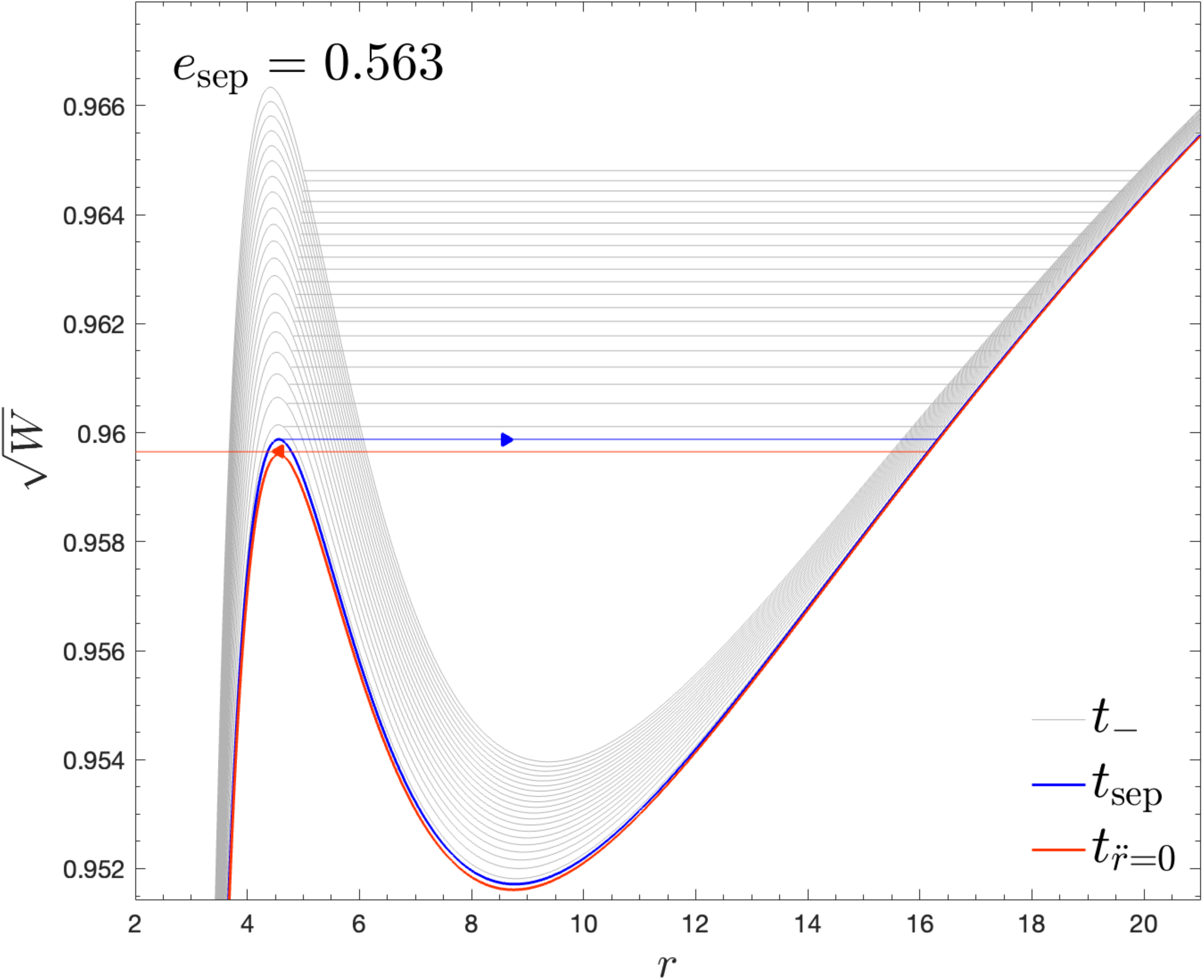}	
	\caption{\label{fig:potential}
	 Left panel: the evolution of radial effective potential $W$, Eq.~\eqref{eq:W}, 
	along the transition
	from quasi-circular inspiral up to the Last Stable Orbit. The LSO potential is highlighted, while 
	the minima (identified by visible markers) indicate the radius and energy of the particle.
	Right panel: eccentric case with initial eccentricity $e_0=0.6$ ($\esep=0.563$).
	We show a sequence of potentials $W$ (that change due to radiation reaction) and corresponding
	energies of the particle at the periastron (horizontal line, gray) during the bound motion of the inspiral.
	The quantities at the separatrix crossing are shown in blue, while the red lines correspond to the
	beginning of the plunge. Note that in this case one has $V_{\rm max}\lesssim E$.
	The horizontal arrows mark the radial location (and energy) of the particle along the orbit 
	 at $t=\tsep$ and at $t=t_{\ddot{r}=0}$. Note that the arrows point toward the corrisponding 
	 direction of the radial motion, outgoing for $t=\tsep$ and ingoing  for $t=t_{\ddot{r}=0}$. This latter
	 point can be considered a {\it missed} periastron and practically marks the beginning of the plunge.}
	\end{center} 
\end{figure*}
The radiation-reaction-driven transition from eccentric inspiral to plunge, merger and ringdown 
in the large mass ratio limit and the emitted gravitational waveform (computed using black-hole 
perturbation theory) were first discussed in Ref.~\cite{Chiaramello:2020ehz} and then more in extenso 
in Sec.~VB of Ref.~\cite{Albanesi:2021rby} (notably also allowing the central black hole to spin). 
In the same Sec.~VB of Ref.~\cite{Albanesi:2021rby} we also presented a (preliminary) complete 
effective-one-body-based waveform model including merger and ringdown for the $\ell=m=2$ mode.
In this section we build upon Ref.~\cite{Albanesi:2021rby} and complement the description of the
dynamics and waveform phenomenology described there. In particular, we: (i) present a precise
description of the transition from inspiral to plunge and its dependence on the eccentricity; (ii) explicitly
present an analytical description of the postpeak waveform, that improves the one presented in
Ref.~\cite{Albanesi:2021rby} and that is crucial (as we will see) to construct a complete EOB waveform,
that is the main goal of Sec.~\ref{sec:modeling} below.
Here we focus on the nonspinning case only, while the spinning case will be discussed elsewhere.
The radiation-reaction-driven relative dynamics is obtained solving Hamilton's equations in the presence
of driving forces, that read 
\begin{subequations}
\label{eq:Hamilton}
\begin{align}
\dot{r} =& \frac{A}{\hat{H}} p_{r_{*}}  , \\ 
\dot{\varphi} =&\frac{A}{\hat{H}}\frac{p_\varphi}{r^2} \equiv \Omega  , \label{eq:freq} \\ 
\dot{p}_{r_*} =& A \hat{\F}_r - \frac{A}{r^2\hat{H}}\left[ p_\varphi^2\left( \frac{3}{r^2} 
                - \frac{1}{r} \right) +1 \right]  , \\
\dot{p}_\varphi =& \hat{\F}_{\varphi},
\end{align}
\end{subequations}
where $A=1-2/r$ is the metric Schwarzschild potential, $p_{r_*}$ is the conjugate momentum of the tortoise coordinate 
$r_* = r + 2 \log{(r/2-1)}$, and $\hat{H}$ is the $\mu$-normalized Hamiltonian of a test-particle
on Schwarzschild,
\be
\label{eq:H}
\hat{H} = \sqrt{A(r) \left(1+ \frac{p_\varphi^2}{r^2}\right) + p_{r_*}^2}.
\ee
The explicit form of $\hat{\F}_\varphi$ and $\hat{\F}_r$ can be found in Ref.~\cite{Chiaramello:2020ehz,Albanesi:2021rby}.

Let us also remind that, as in Ref.~\cite{Albanesi:2021rby}, we define the 
eccentricity $e$ and the semilatus rectum  $p$ in terms of the two radial turning points, 
the periastron $r_-$ and the apastron $r_+$,
\begin{align}
e = & \frac{r_+ - r_-}{r_+ + r_-} \\
p = & \frac{2r_+ r_-}{r_+ + r_-}.
\end{align}
Note that this definition yields $r_\pm = p/(1\mp e)$.
The link between $(e,p)$ and the energy and angular momentum, $(\hat{E}, p_\varphi)$,
is simply obtained by solving the energy condition $\hat{E} = \hat{H}|_{p_{r_*}=0}$ evaluated 
at the two radial turning points.
In order to have stable orbits, the semilatus rectum must satisfy the condition $p\geq p_s = 6 + 2 e$,
where $p_s$ is known as the separatrix and reduces to the Last-Stable-Orbit (LSO)
in the quasi-circular case. 
In this work we will consider configurations with initial eccentricities up to $e_0=0.95$ and semilatera
recta such that the particle undergoes at least a few radial orbits before plunging in the black hole. 
The simulations considered in this work are listed in Table~\ref{tab:ID}.
The dynamics is always started at the apastron, so that the initial radial momentum is zero. 
Note that we chose $(e_0,p_0)$ in order to have a clear geometrical intuition
of the orbit, but we immediately convert $(e_0,p_0)$ in energy and angular momentum
so that we have all the needed initial values to compute the dynamics from Hamilton's equations.
Note that since the dynamics is not conservative, $e$ and $p$ are not
constants of motion and are not defined through the whole evolution of the binary.
Indeed, after the separatrix-crossing time $\tsep$, i.e.~the time at which the 
condition $p(t)=p_s(t)$ is met, the periastron is no longer defined and thus neither 
the eccentricity and the semilatus rectum.
Since in the next sections we will focus on the last part of the dynamics, 
we will often use $\esep = e(\tsep)$ to refer to a certain simulation, rather than $e_0$.
Note however that this is only for labeling purposes, the eccentricity $e(t)$ is not actually 
used anywhere during the evolution.

In our conventions, the strain waveform is decomposed in multipoles as
\begin{equation}
\label{eq:strain}
h_+-{\rm i} h_\times = D_L^{-1} \sum_{\l=2}^{\infty} \sum_{m=-\l}^{\l} h_\lm \,{}_{-2}Y_\lm(\Theta, \Phi), 
\end{equation}
where $D_L$ is the luminosity distance and ${}_{-2}Y_\lm(\Theta, \Phi)$ are the spin-weighted spherical harmonics
with weight $s=-2$. The numerical waveform at linear order in $\nu$ is obtained solving the 
Regge-Wheeler and Zerilli~\cite{Regge:1957td,Zerilli:1970se} equations 
\be
\label{eq:RWZ}
\partial_t^2 \Psi_\lm^\oe - \partial_{r_*}^2 \Psi_\lm^\oe + V^\oe_\ell  \Psi_\lm^\oe = S_\lm^\oe,
\ee
where the superscripts (e) and (o) distinguish respectively even parity ($\ell+m$ even) and odd parity 
($\ell+m$ odd) solutions and the corresponding potentials $V^\oe_\ell$ and sources $S_\lm^\oe$.
The  $\Psi_\lm^\oe(t)$, are related to the waveform multipoles
of Eq.~\eqref{eq:strain} as  $\Psi_\lm^\oe(t) = h_\lm(t)/\sqrt{(\l+2)(\l+1)\l(\l-1)}$.
We solve the RWZ equations using the time-domain code 
\RWZ{}~\cite{Bernuzzi:2010ty,Bernuzzi:2010ty,Bernuzzi:2011aj,Bernuzzi:2012ku}.

\subsection{Transition from eccentric inspiral to plunge}
\begin{figure*}[t]
	\begin{center}
	\includegraphics[width=0.32\textwidth]{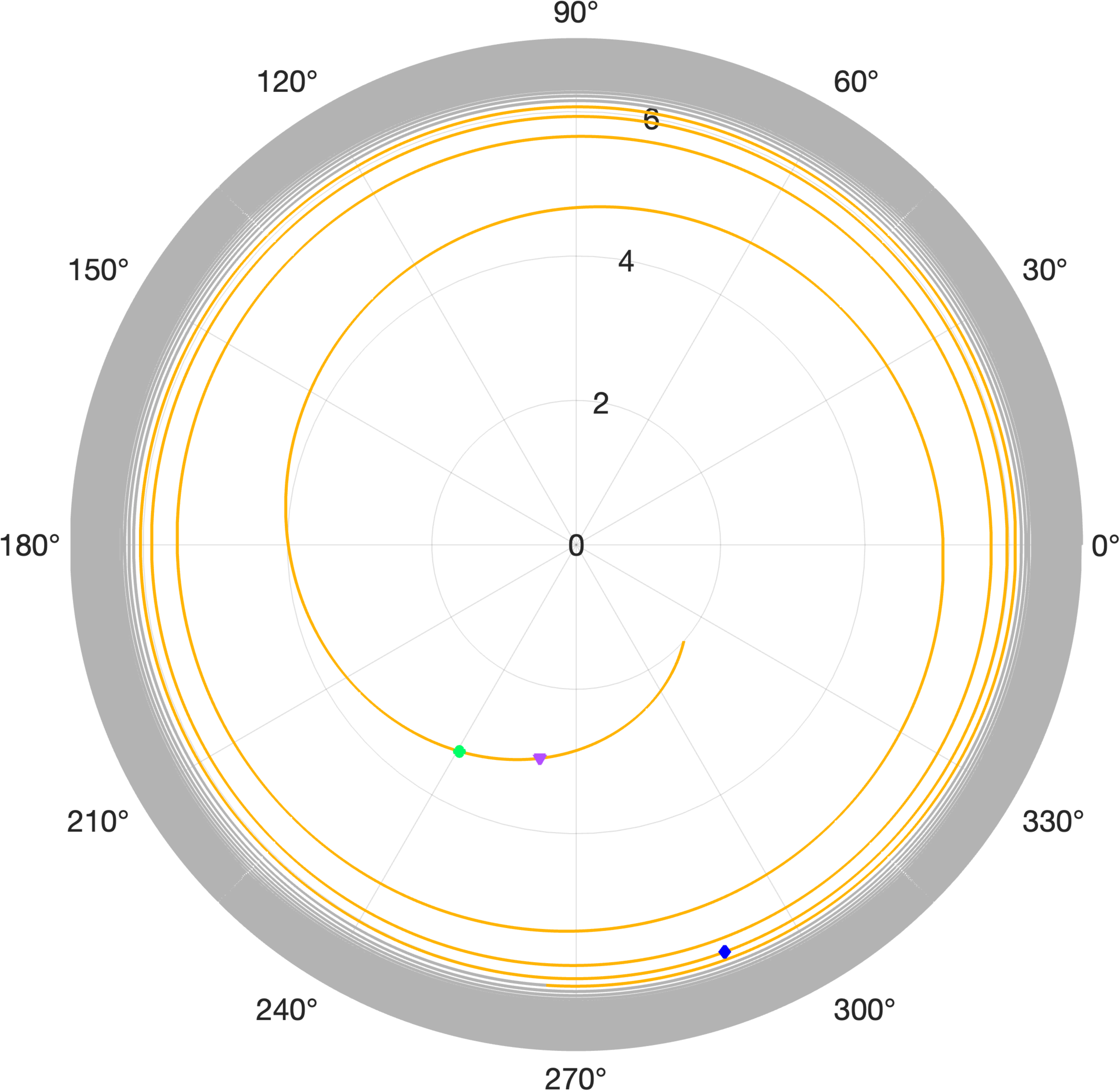}
	\includegraphics[width=0.32\textwidth]{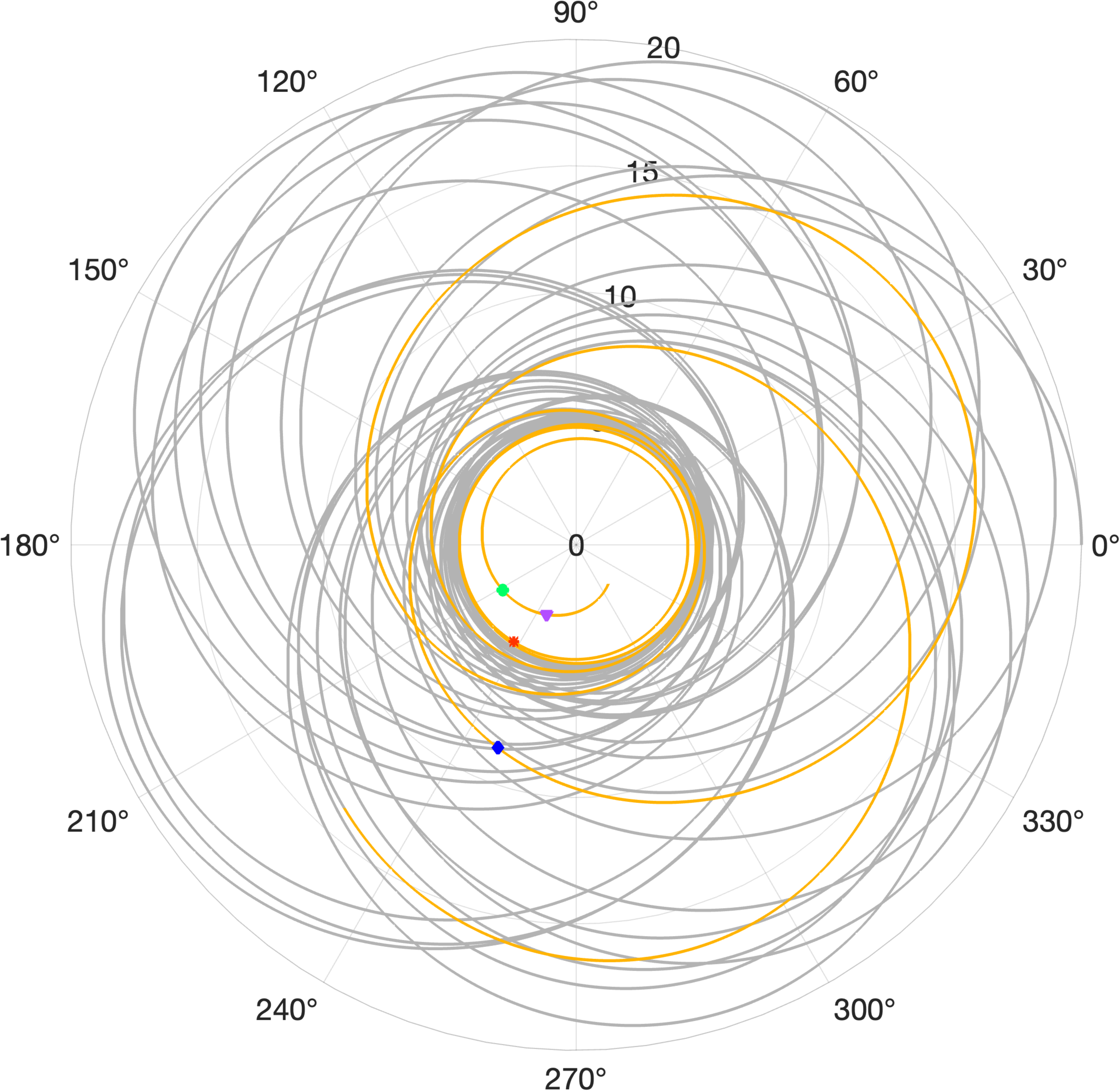}
	\includegraphics[width=0.32\textwidth]{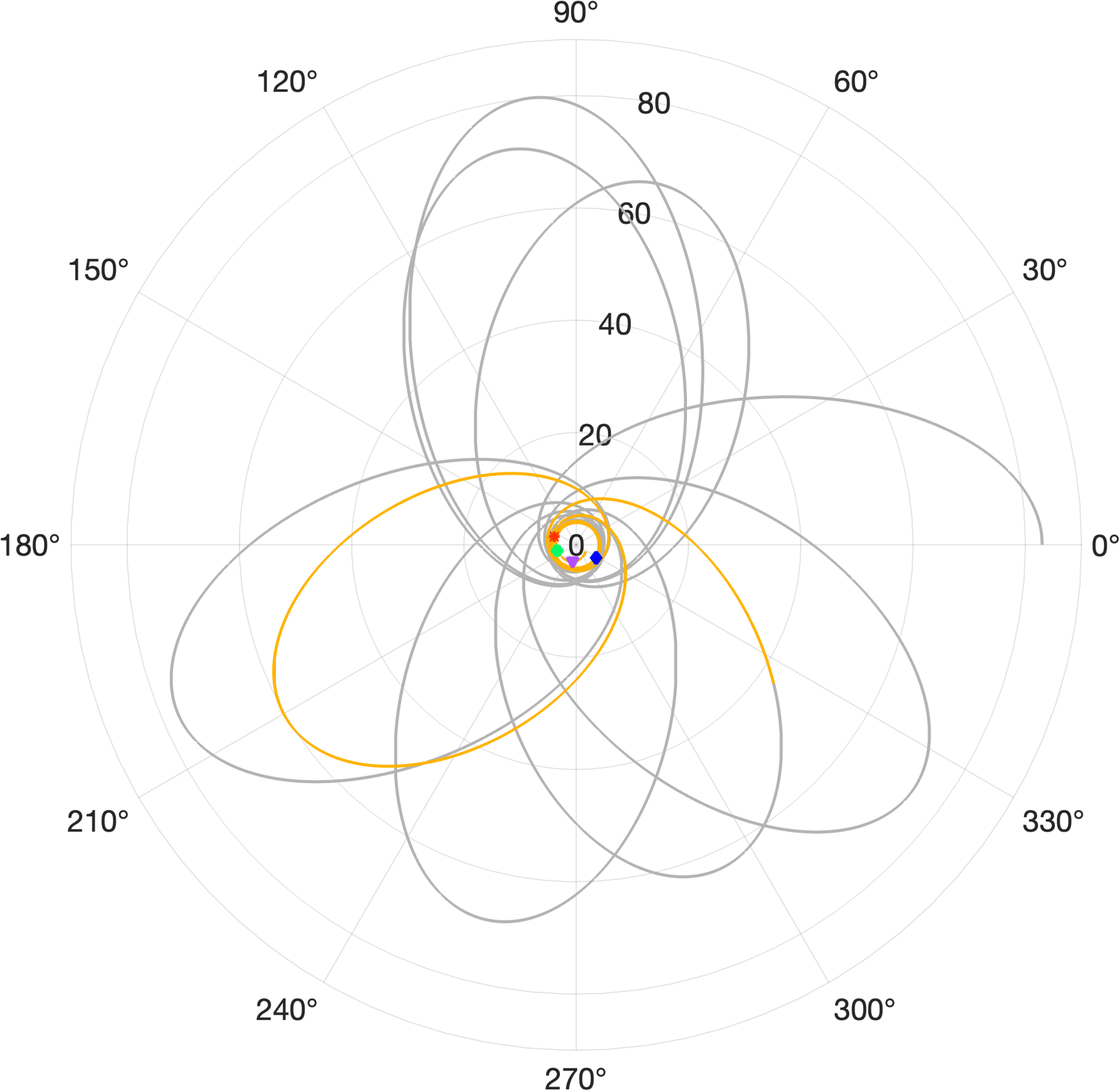}\\	
	\vspace{0.3cm}
	\includegraphics[width=0.32\textwidth]{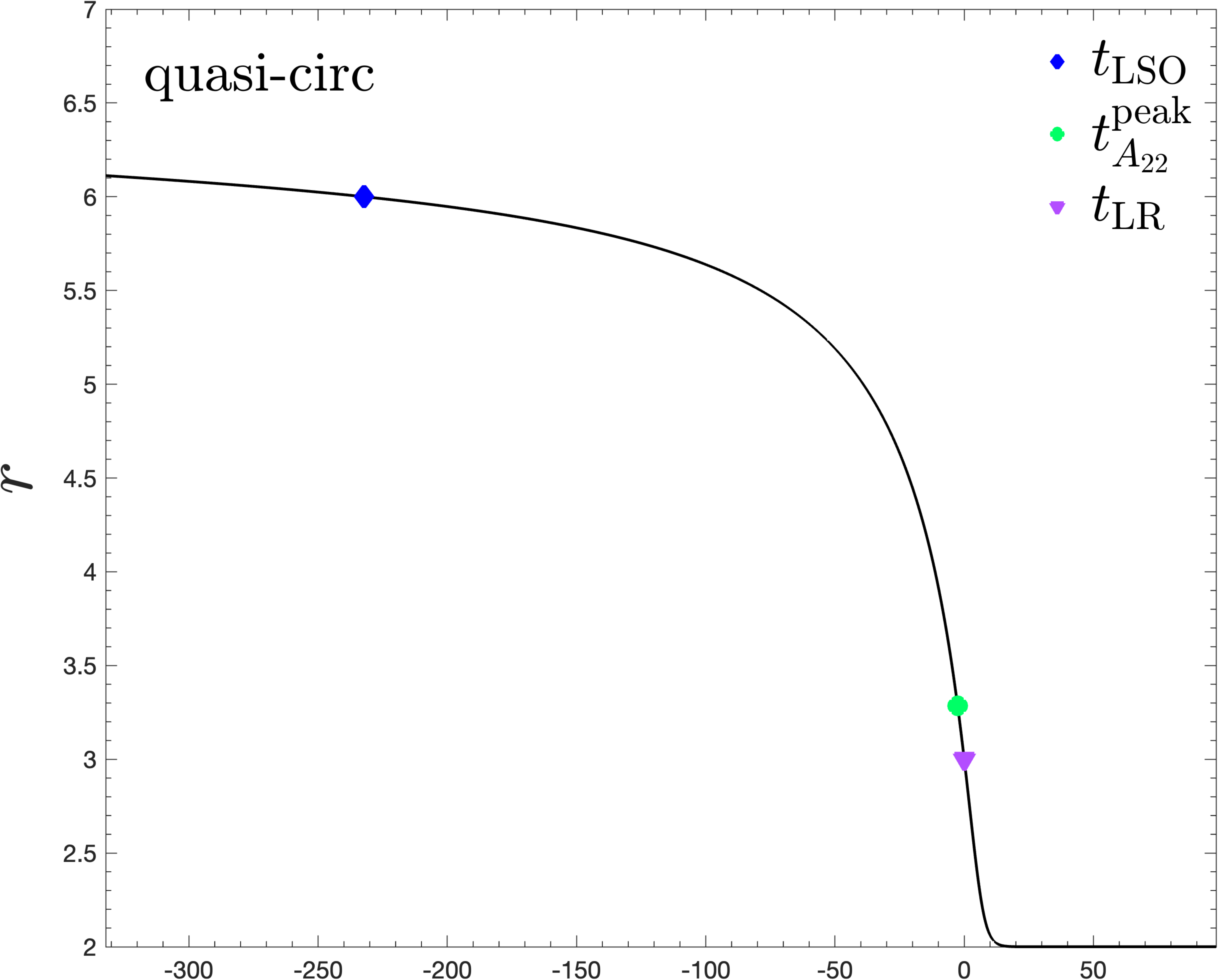}	
	\includegraphics[width=0.32\textwidth]{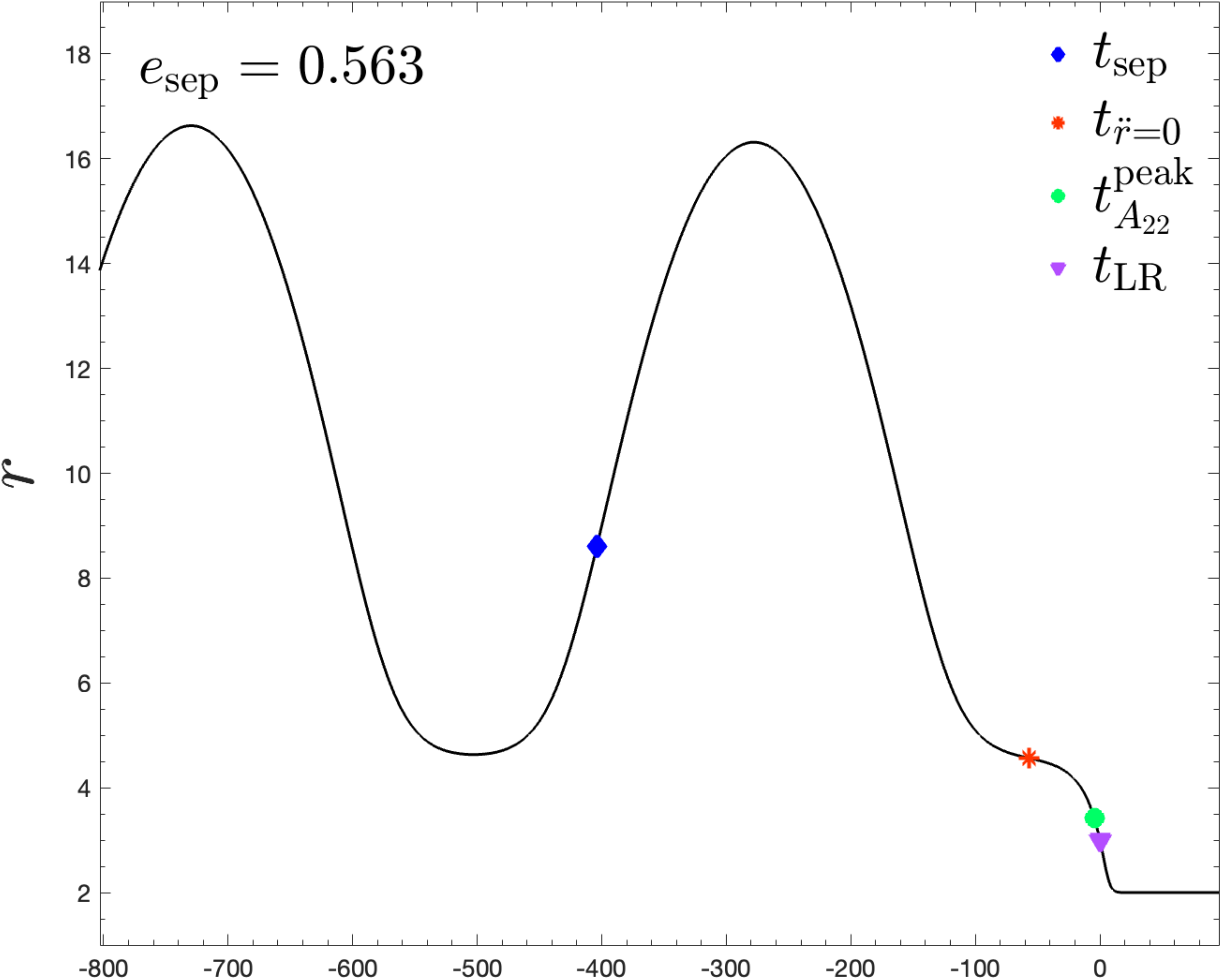}	
	\includegraphics[width=0.32\textwidth]{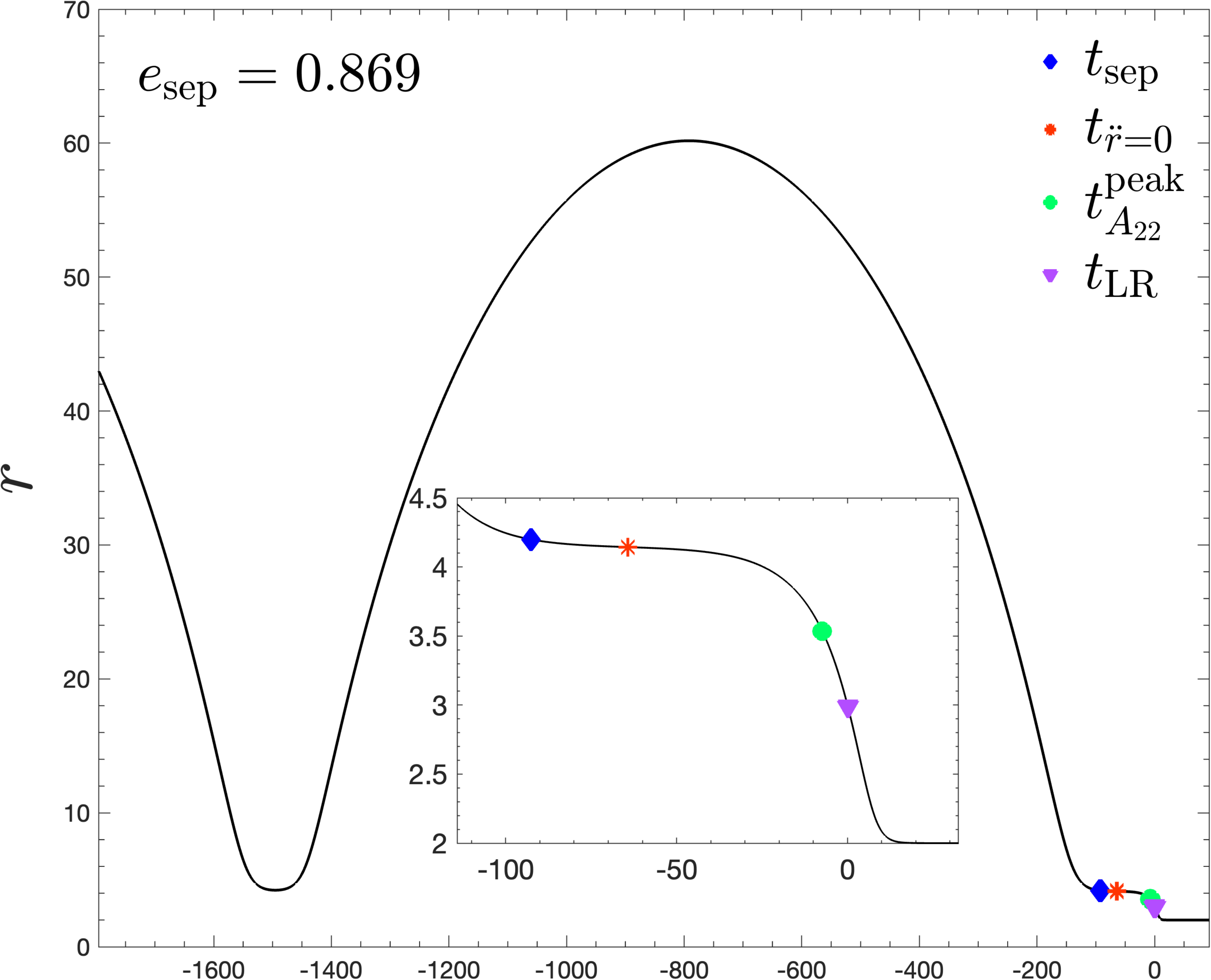}\\
	\vspace{0.3cm}
	\includegraphics[width=0.32\textwidth]{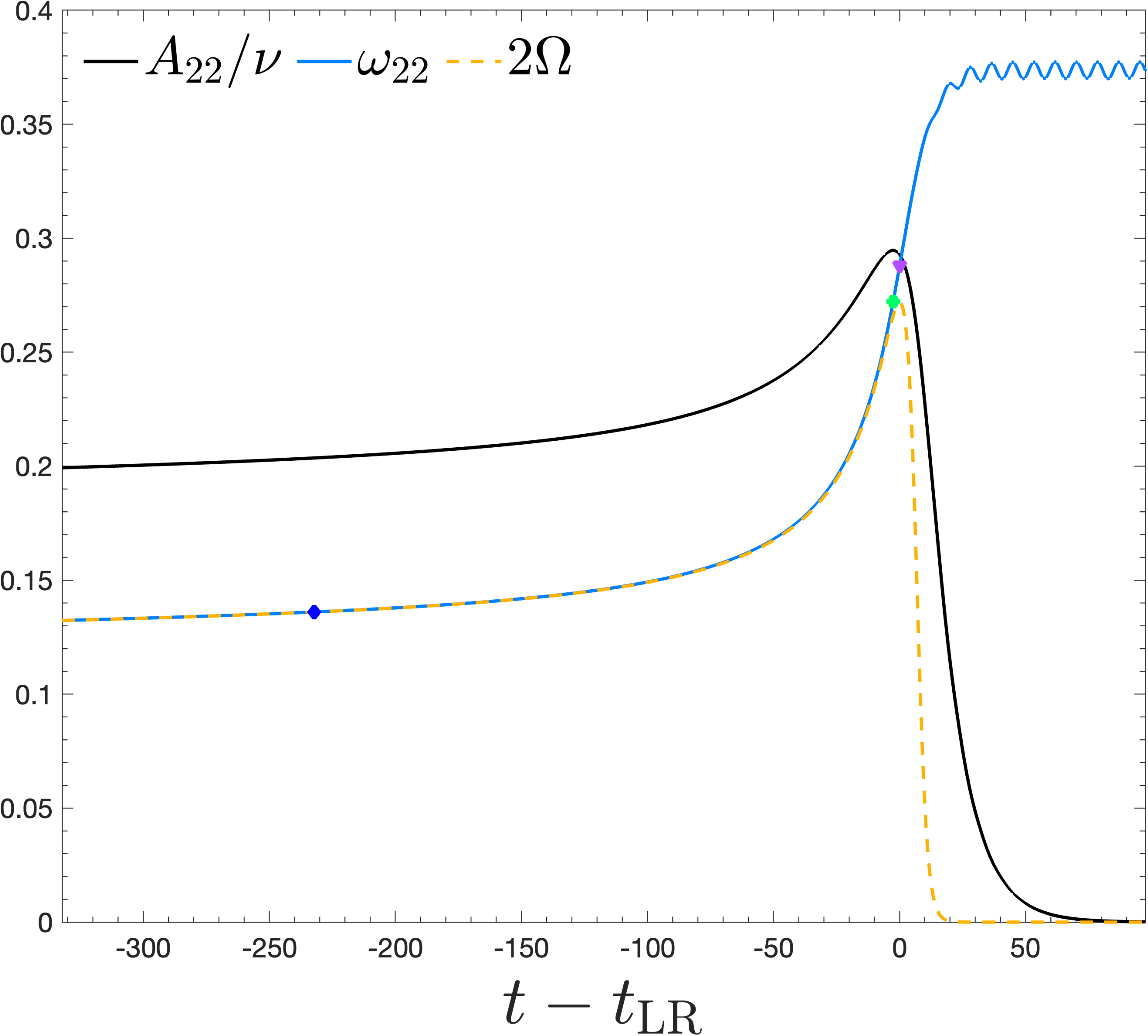}	
	\includegraphics[width=0.32\textwidth]{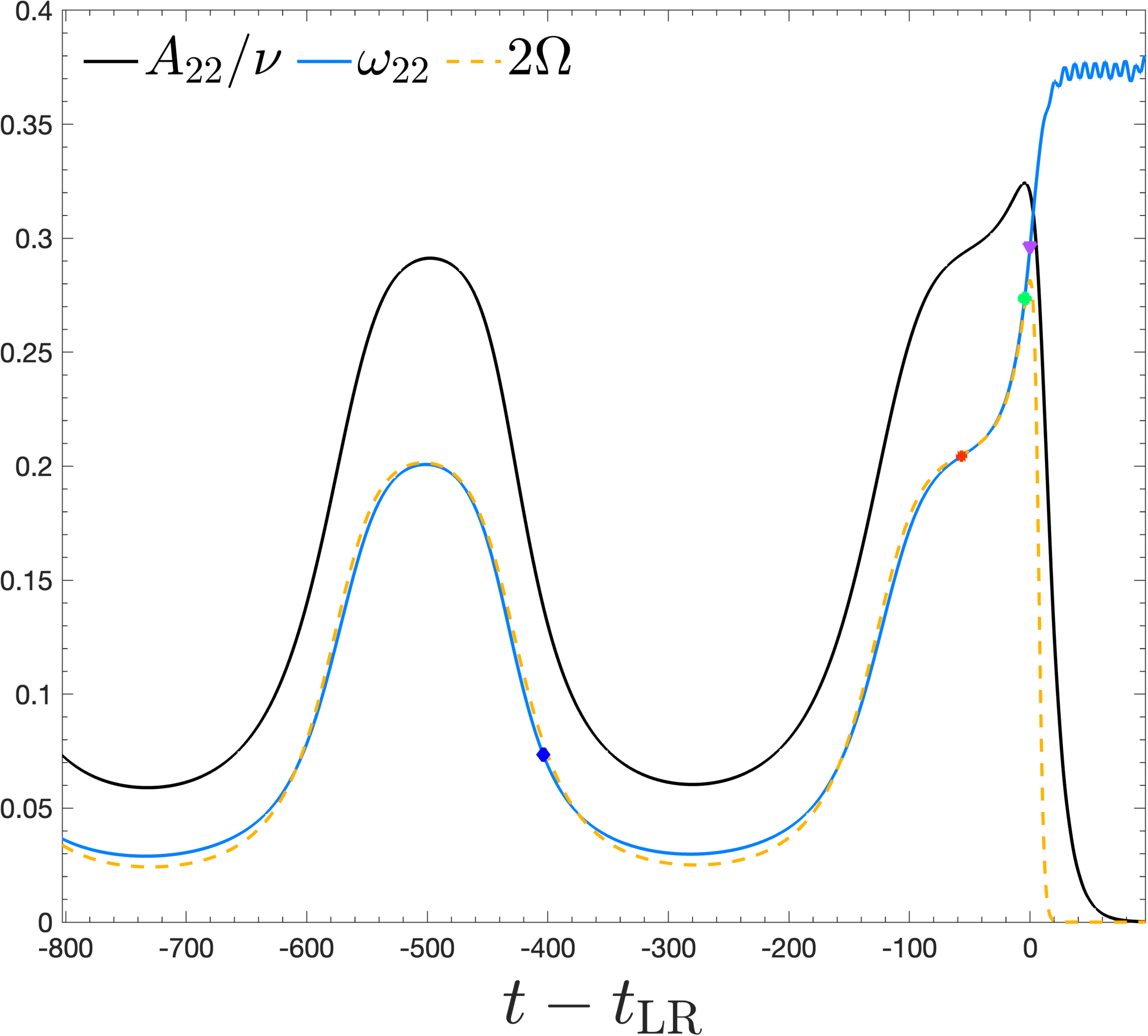}
	\includegraphics[width=0.32\textwidth]{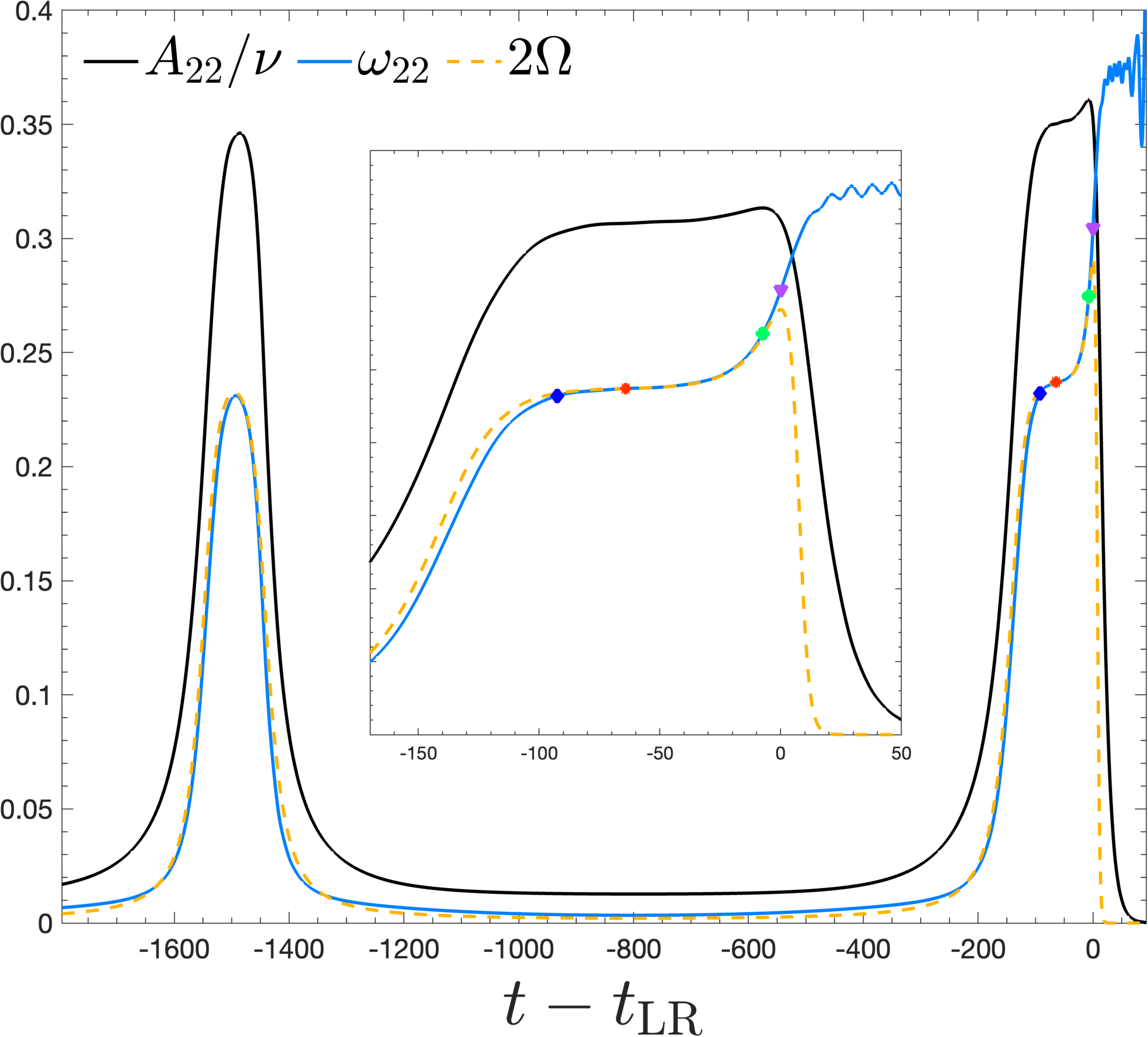}\\
	\vspace{0.3cm}	
	\caption{\label{fig:radii_waves}
	Top row: trajectories for the three configurations with initial eccentricity
	$e_0=\left\lbrace 0,0.6, 0.9 \right\rbrace$. We highlight in orange the portions that correspond
	to the parts shown in the lower panels.
	In all the cases we mark the separatrix crossing with a blue diamond
	(LSO in the quasi-circular case), the peak of the quadrupolar amplitude $\tA22$ with a green circle,
	and the light-ring crossing
	time $\tLR$ with a purple triangle. In the eccentric case we also highlight the inflection point of the radius 
	that marks the end of the last radial orbit, $t_{\ddot{r}=0}$, with a red star.
	Middle row: radius versus time, same markers as above. 
	Bottom row: corresponding amplitude (black) and frequency (blue) of the quadrupolar waveforms.
	We also show the orbital frequency (dashed orange).} 
	\end{center}
\end{figure*}
Let us discuss the main qualitative features of the transition from an eccentric inspiral to plunge
and merger. To do so pedagogically, let us first remind the reader how this transition occurs
in the quasi-circular case.
The (quasi)-circular inspiral is approximately representable as a sequence of circular orbits.
The radius $r_c$ of each circular orbit (with $r>r_{\rm LSO}$) corresponds to the local minimum
of the radial potential
\be
\label{eq:W}
W = \left( 1 - \frac{2}{r} \right) \left( 1 + \frac{p_\varphi^2}{r^2} \right) \ ,
\ee
i.e.~defined by the condition $\de_r(W)|_{r=r_c}=0$, and its energy is 
$\hat{E}=\sqrt{W_{\rm min}}$, where $W(r_c)=W_{\rm min}$
Since radiation reaction eliminates angular momentum from the system,
the potential $W$ is modified during the evolution, until the local maximum
and minimum fuse together in an inflection point at the Las Stable Orbit (LSO), 
$r=r_{\rm LSO}=6$, where $\de_r W=\de^2_rW=0$ and $p_\varphi^{\rm LSO} = 2\sqrt{3}$.
The evolution of the potential and the energy for the quasi-circular case up to $t_{\rm LSO}$ 
are shown in the first  panel of Fig.~\ref{fig:potential}. We highlight in blue the potential at
$t_{\rm LSO}$, after which the particle plunges into the black hole.

Eccentric orbits occur when $\sqrt{W_{\rm min}}<\hat{E}\leq\sqrt{W_{\rm max}}$,
and the radial motion is confined between the two turning points, apastron and periastron.
Since angular momentum is not conserved due to gravitational wave emission, the potential 
changes in time until a moment when $\hat{E} = \sqrt{W_{\rm max}}$. After this, the periastron 
no longer exists (and thus $e$ and $p$ are no longer defined). 
As the energy of the particle approaches the maximum of the potential, the radial velocity eventually reaches 
a local minimum $|\dot{r}|_{\rm min}\neq 0$ and the radial acceleration $\ddot{r}$ changes sign, 
forcing the particle to plunge into the black hole.
We identify this time, $t_{\ddot{r}=0}$, as the beginning of the plunge.
The evolution of the potential for a configuration with initial eccentricity $e_0=0.6$ 
is shown in the right panel of Fig.~\ref{fig:potential}.
In this example the particle undergoes many eccentric orbits, then it crosses 
the separatrix while moving away from the central black hole (blue marker). The particle then reaches the apastron,
inverts the motion and eventually crosses the potential barrier at approximately $t_{\ddot{r}=0}$ (red marker), 
where the plunge starts. 
We also show the corresponding trajectory, the radial evolution and the corresponding waveform in 
Fig.~\ref{fig:radii_waves}. Looking at $r(t)$, it is clear that $t_{\ddot{r}=0}$
can be though as a {\it missed} periastron.
In the same figure, we also show the quasi-circular configuration and another eccentric case with 
higher initial eccentricity, $e_0=0.9$. Note that in this highly eccentric case there is a long-lasting circular 
whirl around the plunge, while in the previous case the whirl 
around $t_{\ddot{r}=0}$ was much shorter. This phenomenology is linked to when $\tsep$ occurs.
In the $e=0.9$ case, the particle crosses the separatrix slightly before the plunge and thus the energy at 
$t_{\ddot{r}=0}$ is quite close to the maximum of the radial potential . 
As a consequence, the particle undergoes a 
long-lasting quasi-circular whirl\footnote{We recall that if the energy is close to the peak of the radial 
potential, the orbits show a zoom-whirl behavior, see e.g. Ref~\cite{Martel:2003jj}.}.
In the case with $e_0=0.6$, the separatrix crossing occurs slightly after the last periastron passage and thus 
the effect of the radiation reaction during the last radial orbit increases the
difference between the energy and the maximum of the radial potential at the beginning of the plunge
(in this case we have $\hat{E}-\sqrt{W_{\rm max}}\simeq 5 \times 10^{-5}$, while in the more eccentric one we had 
$2 \times 10^{-6}$).
Therefore, in our $e_0=0.6$ case the particle has a shorter quasi-circular whirl 
before the plunge with respect to our $e_0=0.9$ case.
For similar reasons, the configuration with $e_0=0.5$ has a longer whirl at $t_{\ddot{r}=0}$ 
than the configuration with $e_0=0.8$. We thus confirm that the length of the quasi-circular 
behavior occurring before the plunge does not simply depend on the value of eccentricity.

\begin{figure}
	\begin{center}
	\includegraphics[width=0.45\textwidth]{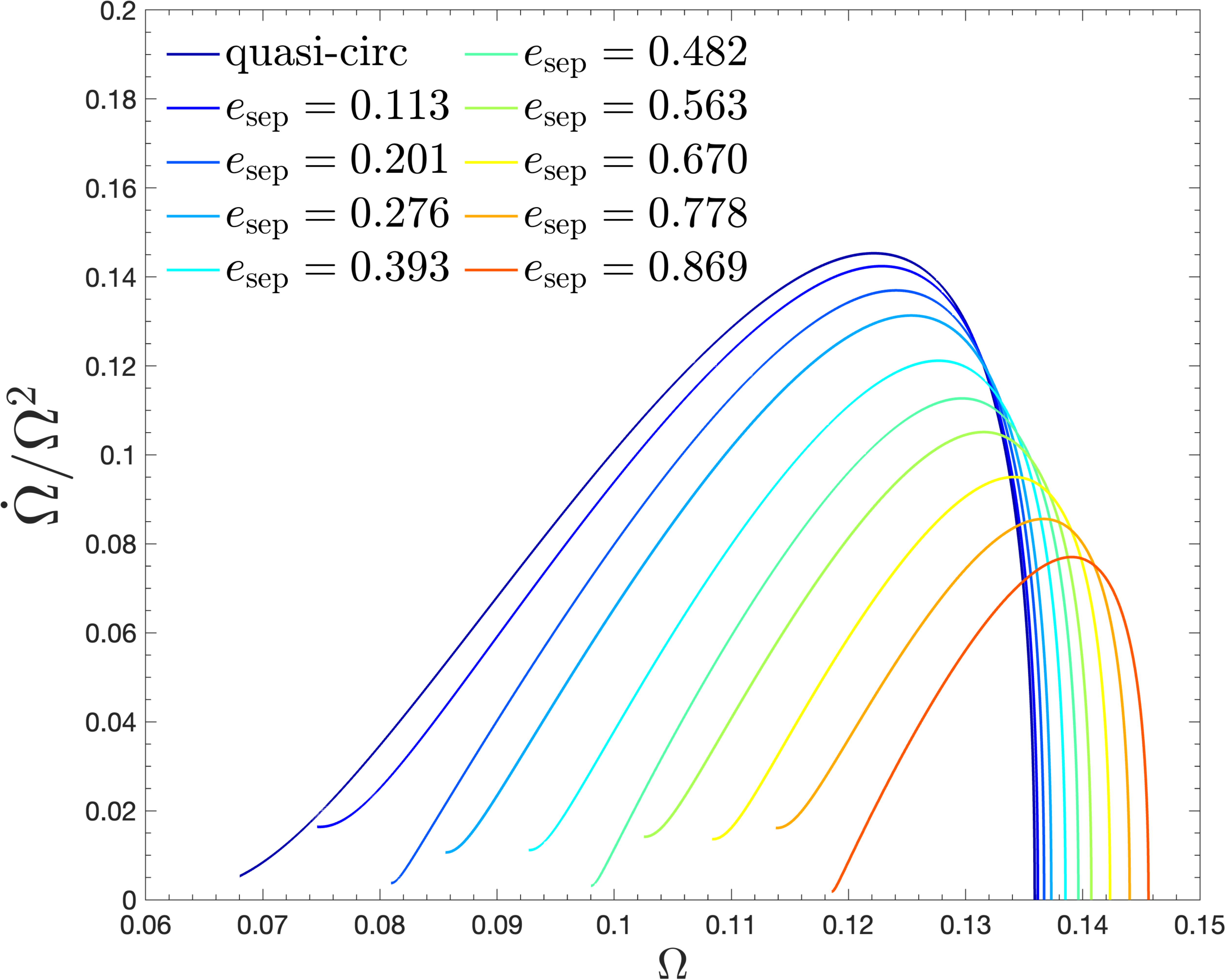}
	\caption{\label{fig:adiab} Measure of adiabaticity, $\dot{\Omega}/\Omega^2$, for
	the quasi-circular case and different eccentric configurations. The horizontal axis
	is restricted between $\Omega_{\ddot{r}=0}$ and $\Omega_{\rm pk}$.}
	\end{center}
\end{figure}

Since the beginning of the plunge is a {\it missed} periastron and the eccentricity is a slowly 
varying quantity, the radius that marks the beginning of the plunge can be approximated as
$r_{\rm plunge}\simeq(6+2\esep)/(1+\esep)$ and it is always smaller than $r=6$. The net 
result of this, together with the considerations above, is that the plunge is more adiabatic
in the presence of eccentricity than in the quasi-circular case.  This is made quantitative in 
Fig.~\ref{fig:adiab}, that depicts the adiabatic estimator $\dot{\Omega}/\Omega^2$ 
as function of $\Omega$ for some relevant configurations. For each dataset, the horizontal axis
is restricted between $\Omega_{\ddot{r}=0}$ and $\Omega_{\rm pk}$, that corresponds respectively
to the orbital frequency at the start of the plunge and at the light-ring crossing. 

%

\subsection{Waveform phenomenology}
\label{sec:RWZ}
\begin{figure*}[t]
	\begin{center}
	\includegraphics[width=0.48\textwidth,height=7.29cm]{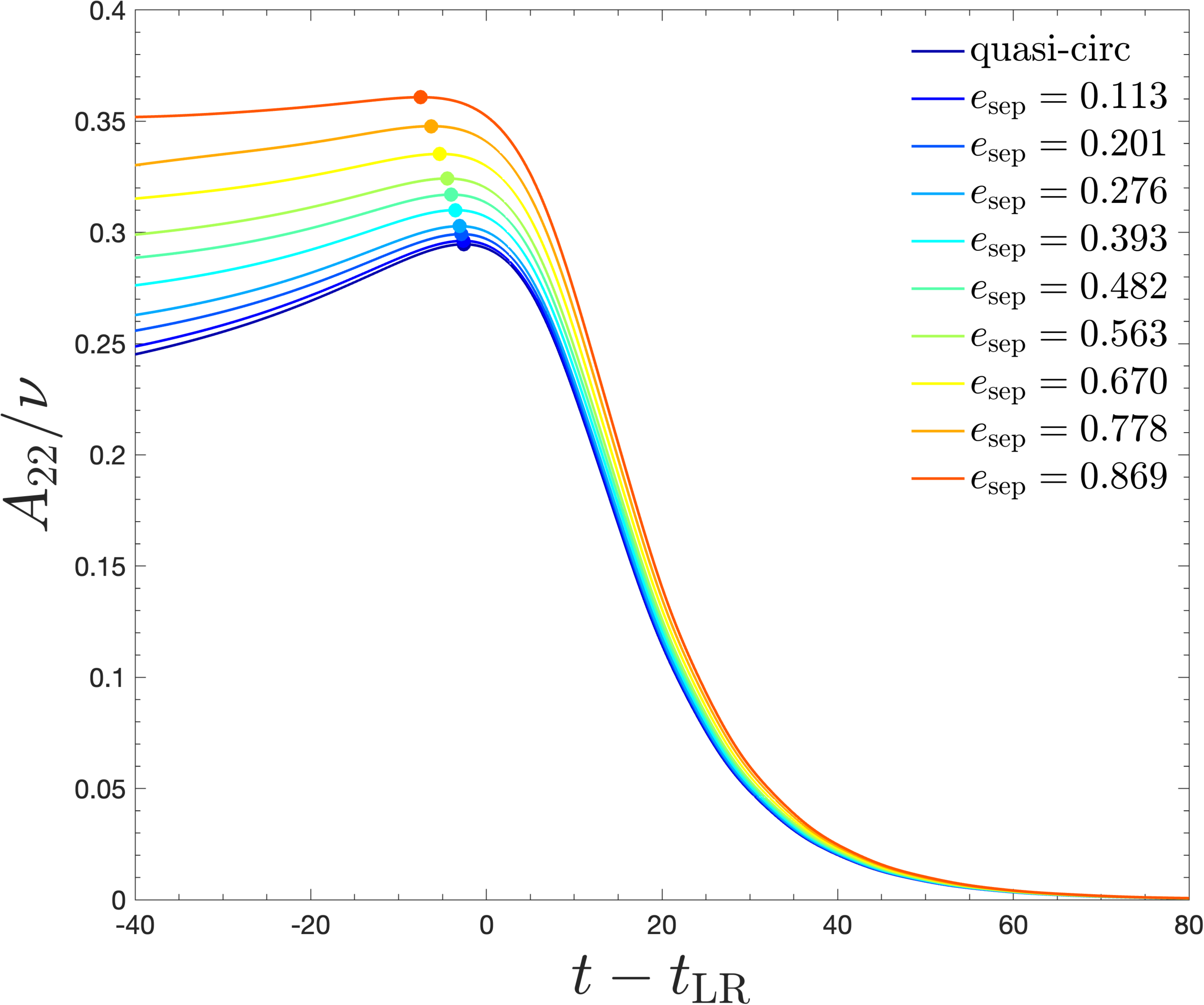}
	\includegraphics[width=0.48\textwidth]{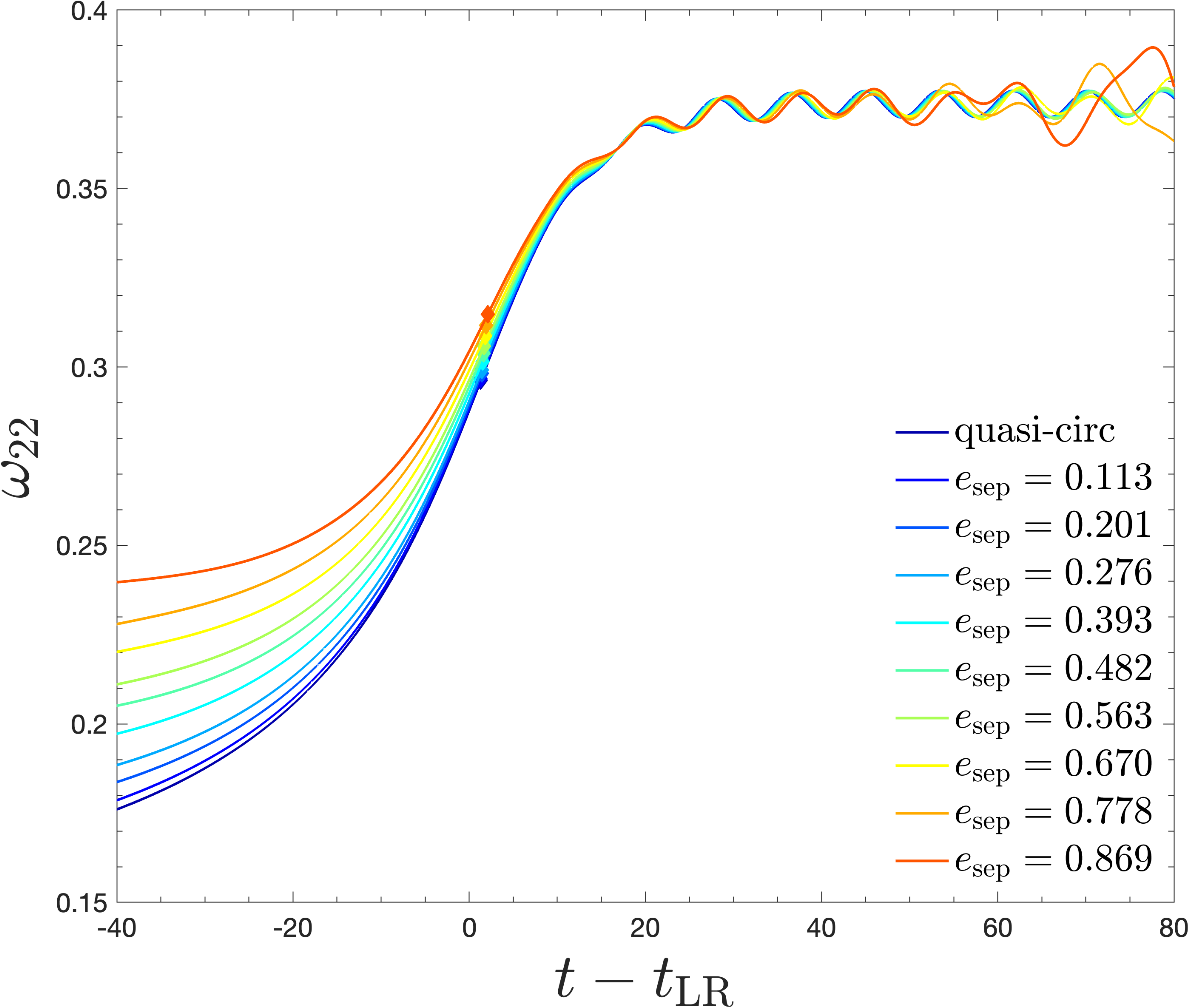}
	\caption{\label{fig:ecc_together} Left panel: quadrupolar Zerilli amplitudes for different eccentricities,
	aligned with respect to the light-ring crossing. The dots mark the maxima of the amplitudes.
	Right panel: Zerilli frequencies for different eccentricities, always aligned using the light-ring crossing. 
	The diamonds mark the inflection points of the frequencies.}
	\end{center}
\end{figure*}
The features of the dynamics that we have just discussed clearly reflect on the waveform
phenomenology, as shown by the Zerilli (2,2) waveforms reported in the bottom row of Fig.~\ref{fig:radii_waves};
the amplitude is shown in black, the frequency in blue. The latter is also compared with 
the orbital frequency, $\Omega$, shown in dashed orange. 
While in the circular case $2\Omega$ is a remarkably good approximation of the waveform frequency $\omega_{22}$,
in the two eccentric cases the noncircular effects increase the differences between these two quantities during the inspiral.
However, note that $\omega_{22}\simeq 2\Omega$ holds also during the plunge for the two eccentric cases,
up to the time of the quadrupolar amplitude peak, $\tA22$ (marked with a green circle), i.e.~shortly before the 
light-ring crossing $\tLR$ (marked with a purple triangle).
This can be easily understood considering that the eccentric plunge is rather adiabatic.

In order to better highlight the properties of the waveform for different eccentricities, 
in Fig.~\ref{fig:ecc_together} we plot the (2,2) mode of the waveforms for $e_0\in\left[0,0.9\right]$.
As a consequence of the fact that in highly eccentric configurations the plunge starts 
at smaller radii, the amplitude grows as the eccentricity increases 
and the peaks become wider. Moreover, the peaks
occur at earlier times with respect to the light-ring crossing, as shown by the markers in the left panel.
In the right panel of Fig.~\ref{fig:ecc_together} we show the corresponding 
frequencies. After the light-ring crossing all the frequencies 
reach the fundamental positive quasi-normal frequency of the Schwarzschild black hole. 
Notably, also the beating between the positive and negative 
fundamental quasi-normal frequencies is not influenced by the nature of the perturbation. However, at later
time the oscillations in the frequencies tend to grow for high eccentricity, but this is only an effect
of the power-law tail that begins to dominate on the quasi-normal-mode (QNM) contribution.
We postpone the discussion of the tail to Sec.~\ref{sec:tail}.
We also highlight the inflection point of the frequencies using diamond markers.
The location of this point is not strongly influenced by the eccentricity, but it is only slightly delayed with
respect to the light-ring crossing. On the contrary, the location of the amplitude peak is strongly
influenced by the eccentricity. This is a qualitative explanation of why the 
quasi-circular ringdown model used in \TEOBResumS{} is able to correctly reproduce the frequency
of highly eccentric comparable mass configurations, but not their amplitudes (see the supplemental
material of Ref.~\cite{Gamba:2021gap}).

\subsection{Ringdown (postpeak) modeling}
\label{sec:ringdown_primary}
\begin{figure*}
	\begin{center}
	 \includegraphics[width=0.32\textwidth]{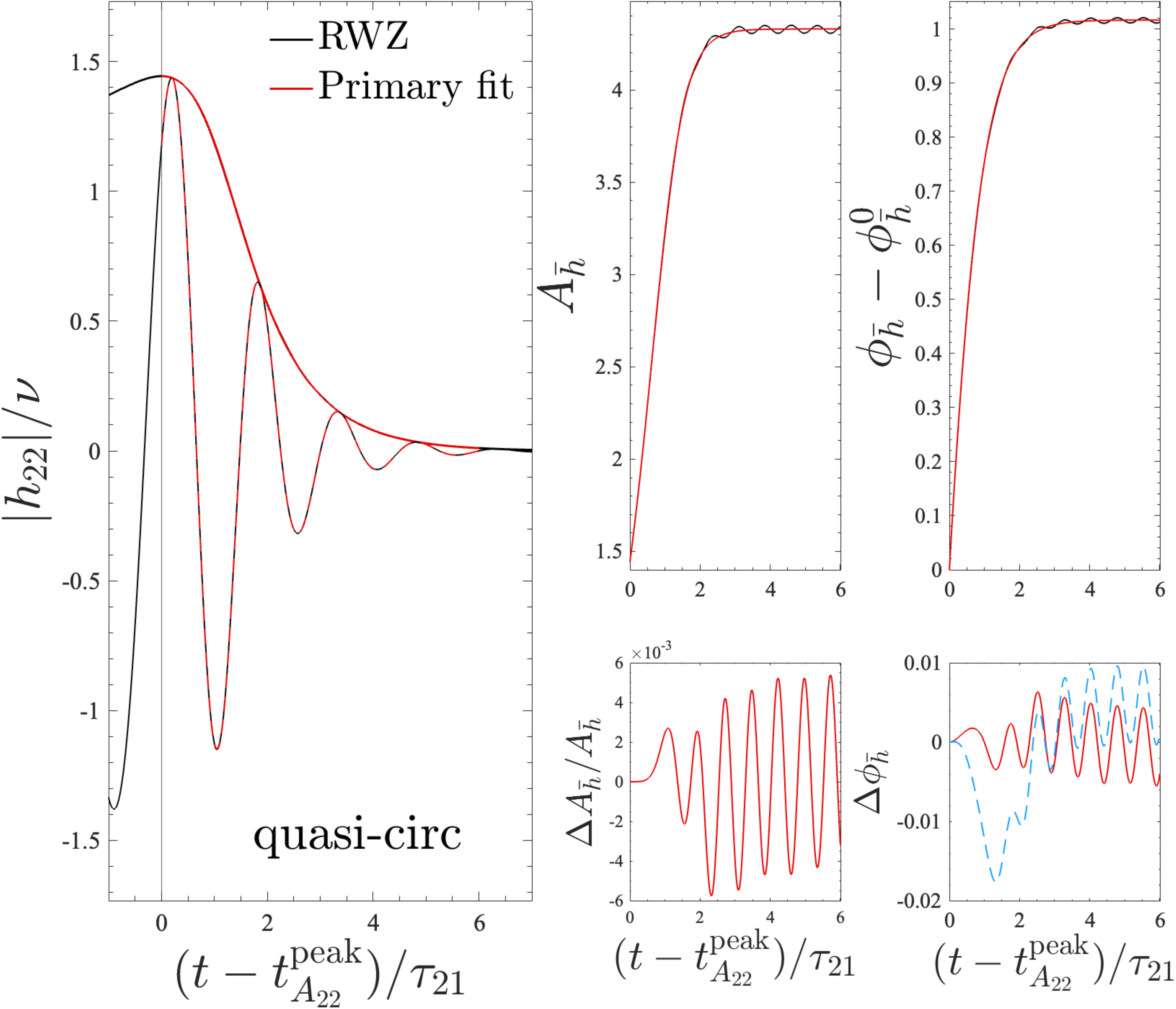}
	 \includegraphics[width=0.32\textwidth]{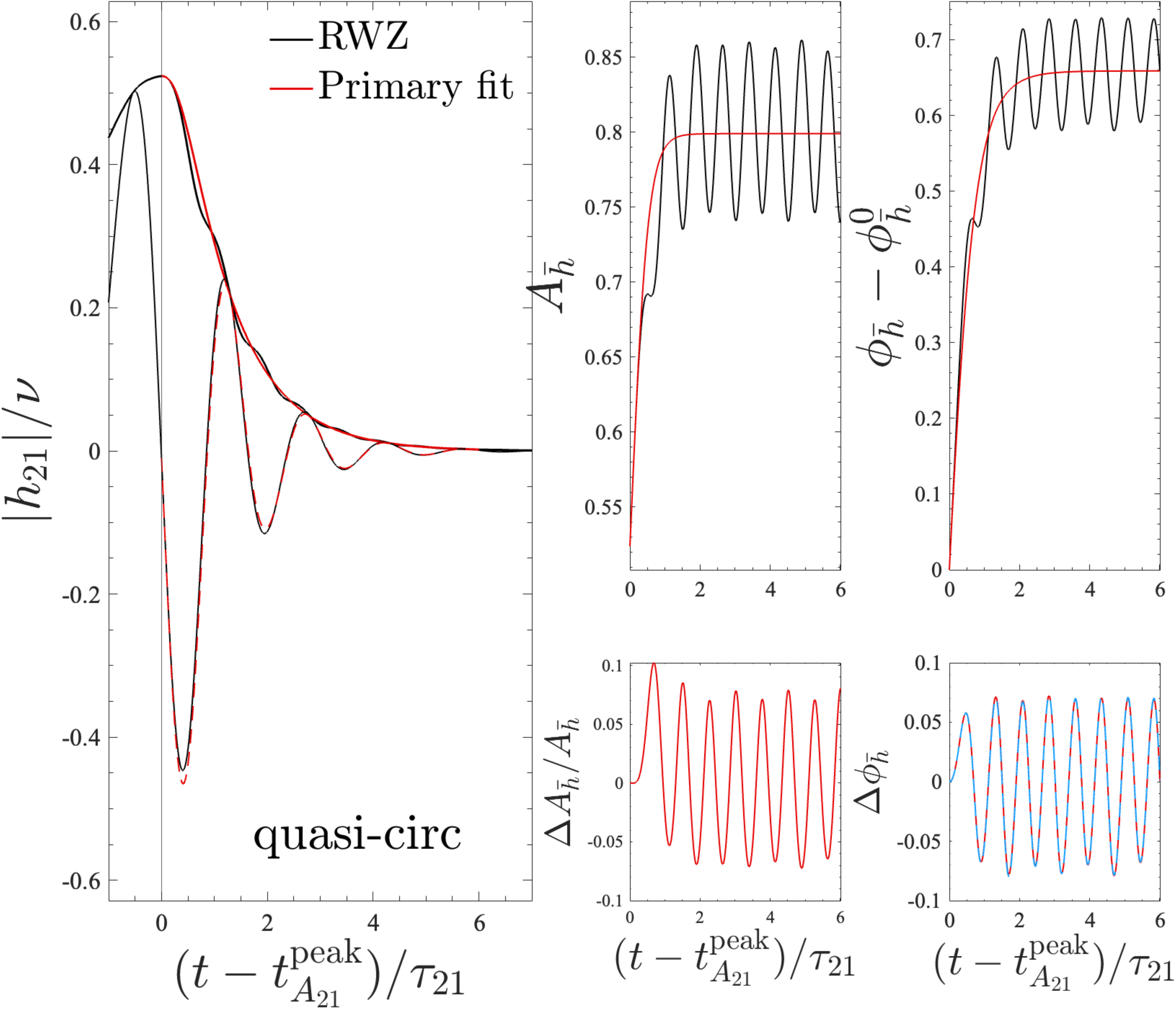}
	 \includegraphics[width=0.32\textwidth]{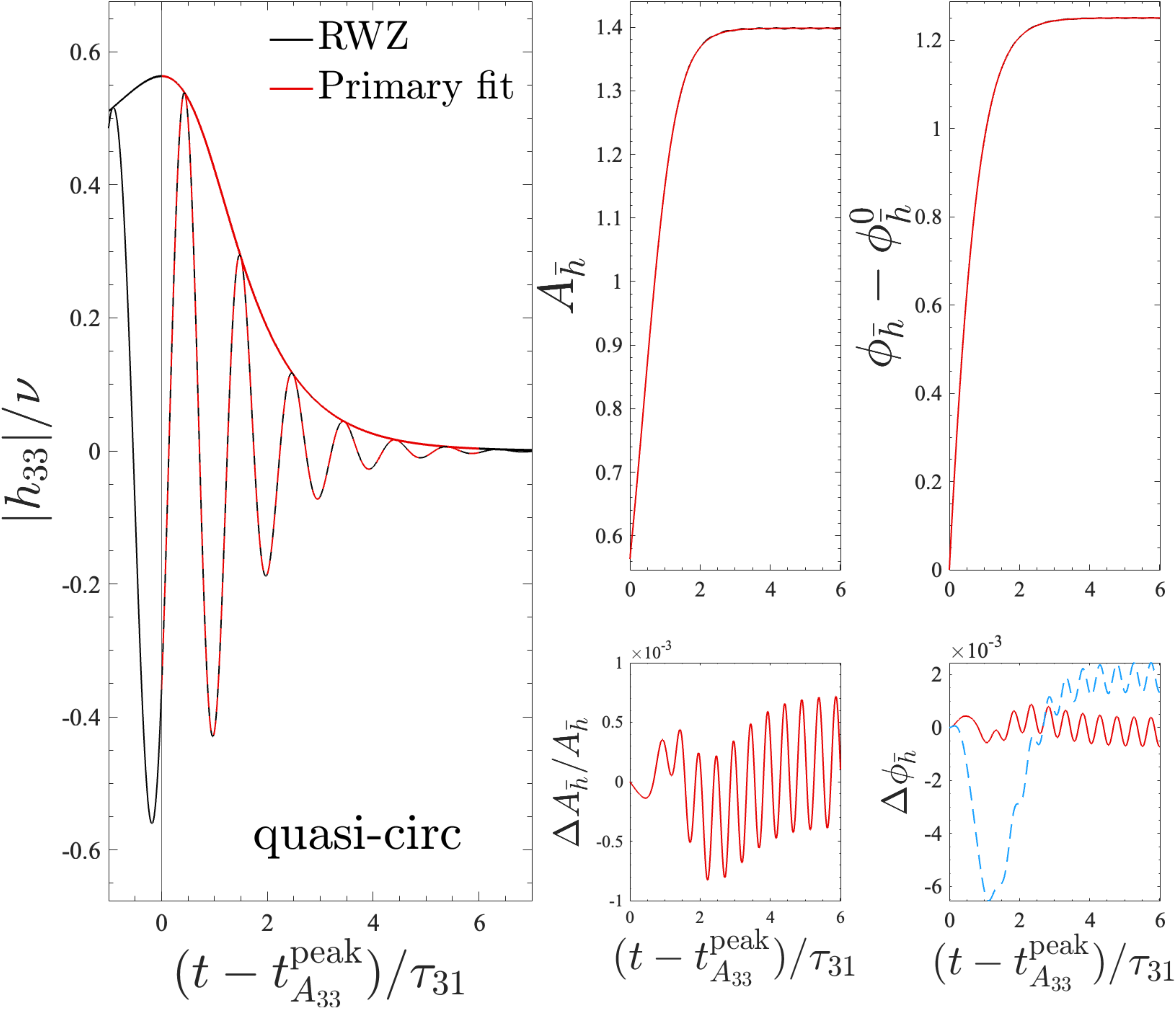}\\
	 \vspace{1cm}
	 \includegraphics[width=0.32\textwidth]{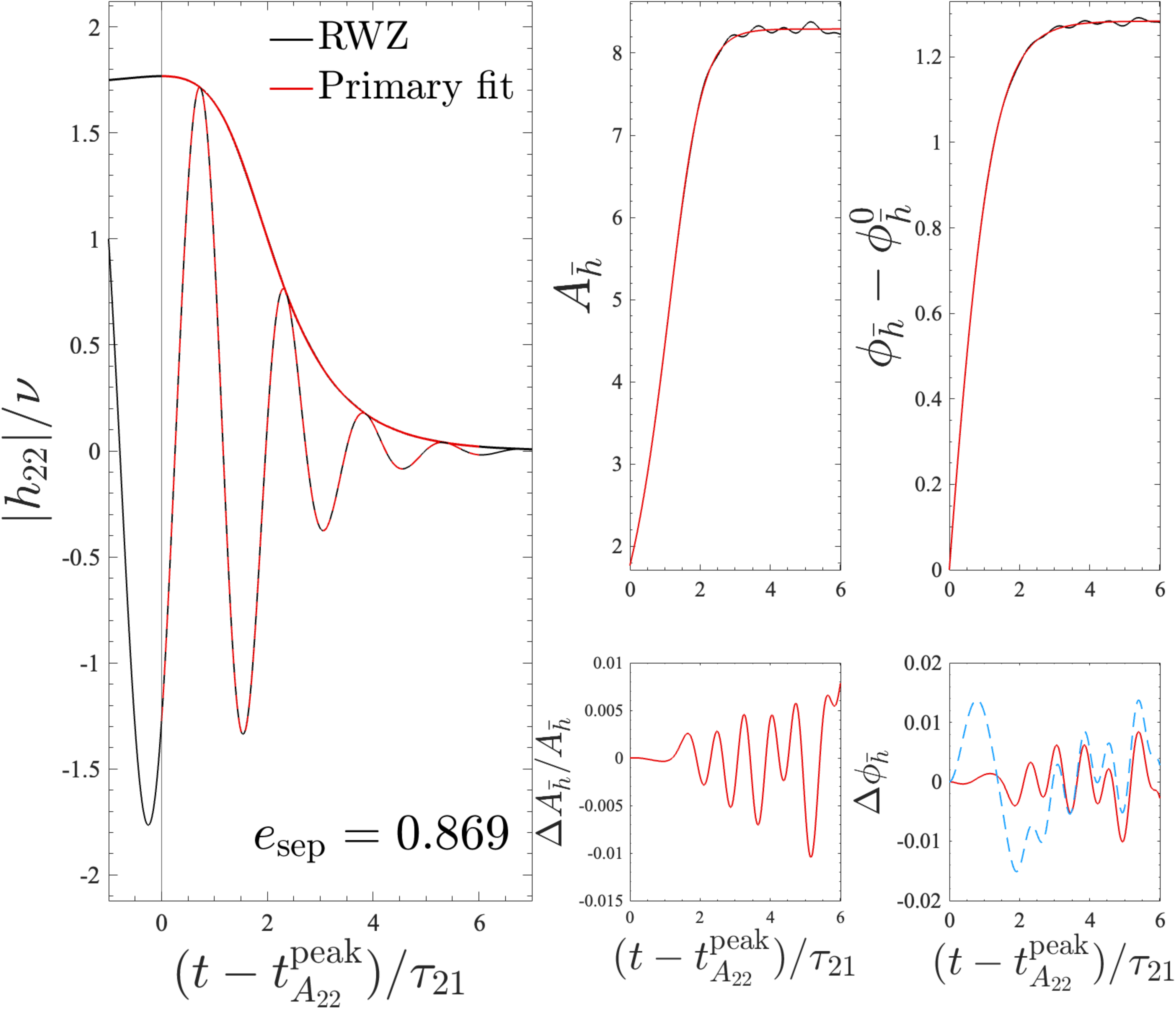}
	 \includegraphics[width=0.32\textwidth]{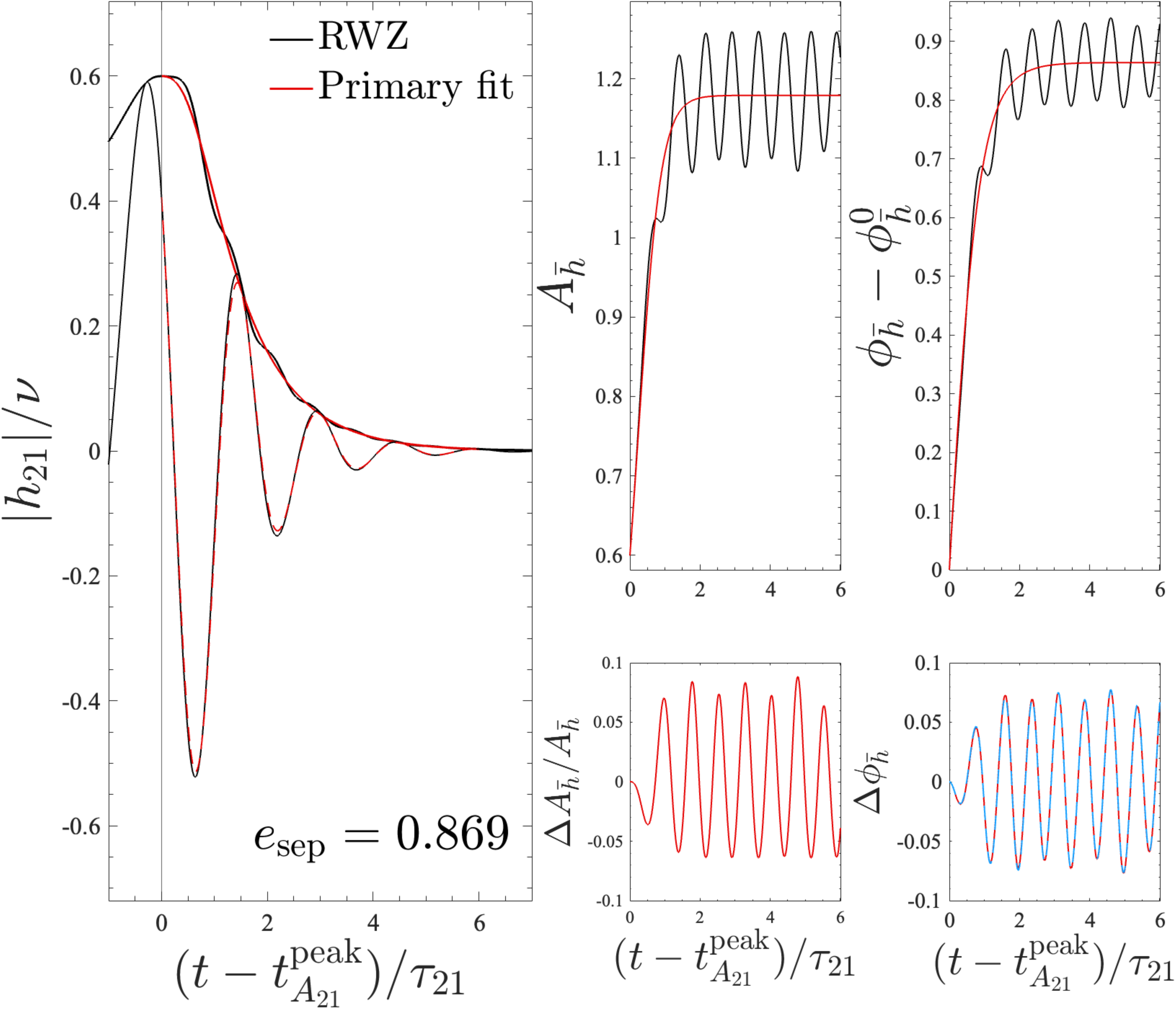}
	 \includegraphics[width=0.32\textwidth]{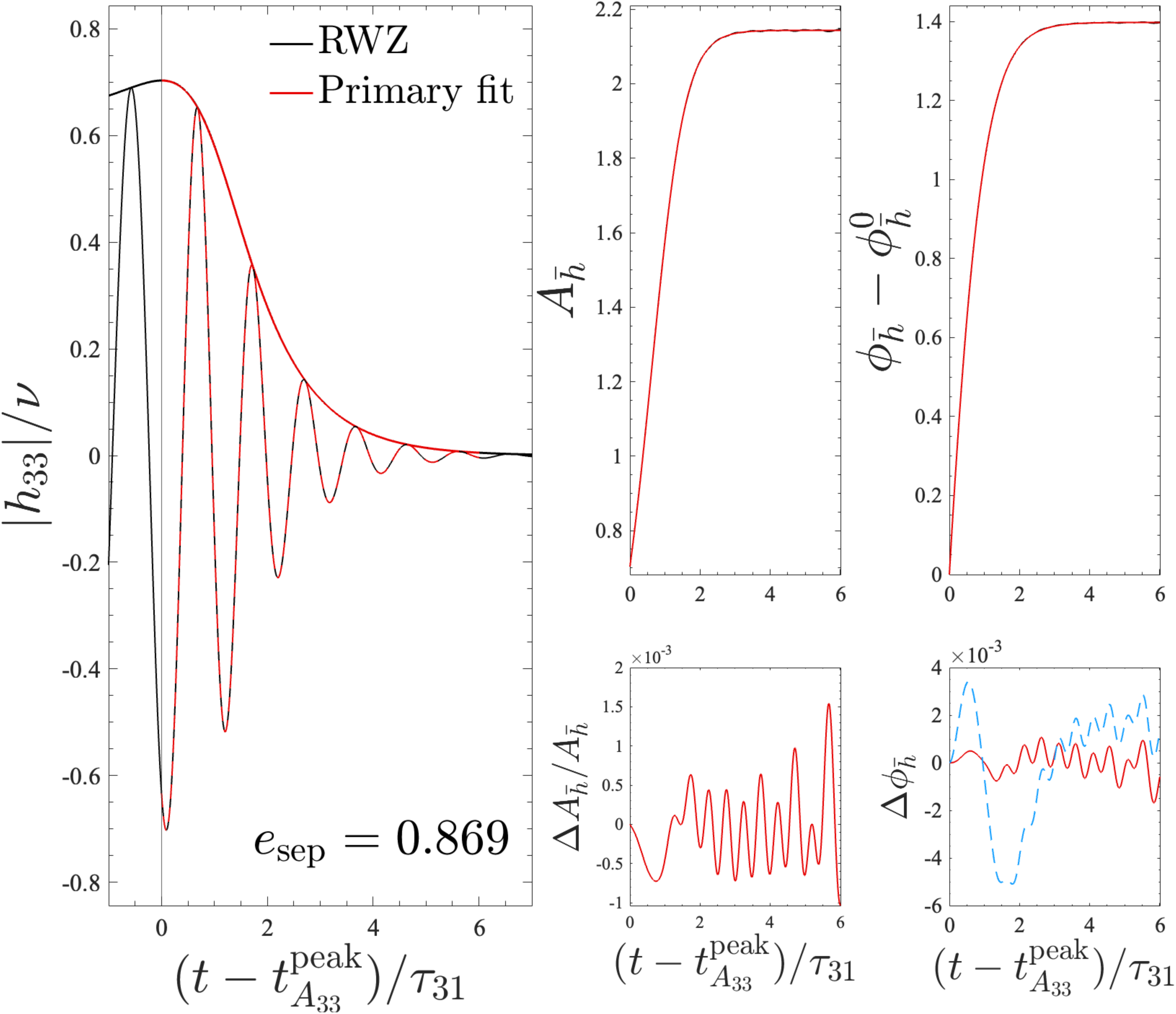}
	\caption{\label{fig:primary_fit}  Numerical waveform (black) and postpeak primary fits (red) for 
		the (2,2), (2,1) and (3,3) modes for the quasi-circular configuration (top row) and the configuration
		with $e_0=0.9$ (bottom row).
       We show the waveform $h_\lm$ and its amplitude together with the QNM-rescaled amplitude $A_{\bar{h}}$ 
       and phase $\phi_{\bar{h}}$. 
       In the two small bottom panels we show the fit/numerical relative difference for the rescaled amplitude 
       and the difference for the rescaled phase (red). Here $\phi_{\bar{h}}^0$ is the phase of $\bar{h}$ at $\tau=0$,
       and $\tau_{\l 1} = 1/\alpha_{\l 1}$ is the QNM-damping time of the fundamental mode. 
       We also show the RWZ/primary phase difference obtained imposing the condition $c_2^\phi=\Delta\alpha_{21}$
       in the primary fit (dashed blue).}
	\end{center}
\end{figure*}
During the ringdown, a relevant contribution to the waveform is given by the QNMs. 
If this is the only contribution, then each multipole can be written as
\begin{equation}
\label{eq:qnm_wave}
\Psi^\oe_\lm = \sum_{n=1}^{\infty} \left( C_{\lm n}^+ e^{-  \sigma_{\ell n}^+ \t} + C_{\lm n}^- e^{- \sigma_{\ell n}^- \t} \right) ,
\end{equation}
where $\sigma_{\ell n}^\pm = \alpha_{\ell n} \pm i \omega_{\ell n}$ are the complex QNM frequencies, 
and the $C_{\lm n}^\pm$ are complex constant coefficients\footnote{Note that we denote the fundamental QNM mode with the index $n=1$ so that the first 
overtone has index $n=2$, while in other recent works the fundamental mode is denoted with 0.}.
While the latter depend on the type of 
the perturbation, the frequencies depend only on the mass of the Schwarzschild black hole.
However, it has been shown that for comparable mass binaries the pure QNM description can be used only at later 
times~\cite{Baibhav:2023clw}. In Sec.~\ref{sec:qnm_phenom} we will argue that this is the case
also for the postpeak waveform generated by a particle falling into a Schwarzschild black hole. 
Therefore, we comply to the phenomenological ideas introduced in Ref.~\cite{Damour:2014yha},
based on the idea of factorizing away the contribution of the fundamental quasi-normal mode.
We thus introduce the QNM-rescaled waveform 
\begin{equation}
\label{eq:barh}
\bar{h}_\lm(\tau) = e^{\sigma_{\l 1}^+\tau + i \phi_\lm^{\rm peak}} h^{\rm rng}_\lm(\tau),
\end{equation}
where $\tau=t-\tAlm$, $\phi^{\rm peak}_\lm$ is the value of the phase at the amplitude peak, and
$h^{\rm rng}_{\lm}$ is the ringdown waveform.
The QNM-rescaled waveform $\bar{h}_{\lm}(\tau)$ is then written using two templates for the amplitude and
the phase, $\bar{h}(\tau) = A_{\bar{h}}(\tau) e^{i \phi_{\bar{h}}(\tau)}$. 
In Ref.~\cite{Albanesi:2021rby}, we pointed out that the original waveform template for the amplitude can become
incorrect when the mass ratio is large. For this reason we introduced a different fitting function.
Our templates thus read
\begin{align}
\label{eq:templateA}
A_{\bar{h}} (\tau) &=\left(\frac{c_1^A}{1 + e^{-c_2^A \tau + c_3^A}}+c_4^A \right)^{\frac{1}{c_5^A}},\\
\phi_{\bar{h}}(\tau)& = -c_1^\phi \ln \left( \frac{1+c_3^\phi e^{-c_2^\phi \tau} + c_4^\phi e^{-2c_2^\phi \tau}}{1+c_3^\phi + c_4^\phi} \right)\label{eq:templatePhi},
\end{align}
where we have dropped the $\lm$-indices for notation simplicity.
The sets of parameters $c^A$ and $c^\phi$ are constrained by requiring the continuity of the 
waveform at $\tau=0$ with the inspiral waveform.
Requiring the continuity of the amplitude, its first two time derivatives, and the frequency, we get 
\begin{align}
c_1^A &= \frac{c_5^A \alpha_1}{c_2^A} (\Amx)^{c_5^A} e^{-c_3^A} (1 + e^{c_3^A})^2,\\
c_4^A &= (\Amx)^{c_5^A} - \frac{c_1^A}{1 + e^{c_3^A}}, \\
c_5^A &= - \frac{\d2Amx}{\Amx \alpha_1^2} + \frac{c_2^A}{\alpha_1} \frac{e^{c_3^A} - 1}{1 + e^{c_3^A}} , \\
c_1^\phi &= \dfrac{1 + c_3^\phi + c_4^\phi}{c_2^\phi(c_3^\phi + 2 c_4^\phi)} \Deltaomgmx,
\end{align}
where $\d2Amx$ is the second time-derivative of the amplitude evaluated at the peak, 
and $\Deltaomgmx \equiv \omega_1 - \omgmx$.
Note that in previous works also the condition $c_2^\phi=\Delta\alpha_{21}\equiv \alpha_2 - \alpha_1$ was imposed, here instead
we leave $c_2^\phi$ as a free parameter.
The phase difference obtained with the constrained
$c_2^\phi$ using the damping times is shown in blue in the bottom-right panels of 
Fig.~\ref{fig:primary_fit}. The free $c^\phi_2$ improves the phase, especially for $\l=m$ modes. 

With these templates we are able to fit the numerical postpeak waveform for each multipole 
and every eccentricity considered in this work. 
In the first row of Fig.~\ref{fig:primary_fit} we show the primary fits for the (2,2), (2,1) and (3,3)
modes for the quasi-circular inspiral. The rescaled amplitude 
$A_{\bar{h}}$ and the rescaled phase $\phi_{\bar{h}}$ reach a plateau after $\simeq 2 \tau_{\ell 1}$,
where $\tau_{\ell 1} \equiv 1/\alpha_{\ell 1}$ is the QNM-damping time of the fundamental mode.
This means that, at this stage of the evolution, the only relevant contribution to the waveform
is given by the fundamental QNMs. Note that leaving $c_2^\phi$ as a free parameter strongly improves the
phase agreement for the (2,2) and (3,3) modes (cfr. red and blue lines in the phase difference of each plot),
while it is not relevant for the (2,1) mode.  
Similar considerations hold for the configuration with $e_0=0.9$, that is shown in the bottom
row of Fig.~\ref{fig:primary_fit}, and all the other eccentric configurations considered in this work. 

\subsection{Modeling the mode-mixing}
\label{sec:ringdown_modemixing}
The templates discussed above catch all the main features of the numerical waveform, except the mode-mixing 
generated by the negative-frequency QNMs. This effect can be already seen in the (2,2) mode, 
but become particularly relevant in the $m=1$ modes, as
shown for the (2,1) mode in Fig.~\ref{fig:primary_fit}. We will discuss this effect
in more detail in Sec.~\ref{sec:omgfit_iter}, here we just mention that it can be simply included in the
ringdown model doing the substitution
\be
\label{eq:modemix_in_hrng}
h^{\rm rng}_\lm (\tau) \rightarrow h^{\rm rng}_\lm(\tau) \left( 1 + \ha_{\lm 1} e^{2 i \omega_{\l,1} \tau } \sigma(\tau;\t^{\rm mm}_0) \right),
\ee
where $\hat{a}_{\ell m 1} = C_{\ell  m 1}^-/C_{\ell m 1}^+\equiv \hat{A}_{\lm 1} e^{i \theta_{\lm 1}}$ and 
$\sigma(t;\t^{\rm mm}_0)=1/(1+e^{-(\tau-\t^{\rm mm}_0)})$ is a sigmoid that activates the mode-mixing correction.
The $\ha_{\lm 1}$ coefficients can be extracted fitting the late ringdown frequency with a fundamental-QNM ansatz 
as outlined in Ref.~\cite{Bernuzzi:2010ty}; in this work we will follow a more refined procedure 
that we will discuss in Sec.~\ref{sec:omgfit_iter}. Note that, despite the fact that 
$C_{\lm 1}^\pm$ depend on the nature of the perturbation,
the modulus of their ratio, $\hat{A}_{\lm 1}$, does not seem to change with the eccentricity, as
shown for example for the (2,2) mode in Fig.~\ref{fig:ecc_together}.
The values of $\hat{A}_{\lm 1}$ can be found in Table~\ref{tab:Alm1}, while
the value of $\t^{\rm mm}_0$ is chosen in order to introduce the oscillations in the analytical
waveform only when they are also present in the numerical wave. 
For all the eccentricities, we use $\t^{\rm mm}_0=25$ for the (2,2) multipole, $\t^{\rm mm}_0=20$ 
for the $\l=m$ higher modes, $\t^{\rm mm}_0=8$ for (2,1), (3,2), (4,3), (4,2) and $t_0=3$ for (3,1) and (4,1).

\subsection{Global fits}
\label{sec:gfits}
In Sec.~\ref{sec:ringdown_primary} and Sec.~\ref{sec:ringdown_modemixing} we have discussed a 
phenomenological model that can be used to faithfully describe the postpeak waveform. This model
depends on different free parameters that are found fitting RWZ postpeak waveforms. However, in order 
to describe any eccentric case, we need to provide global fits of these parameters as
function of some system-characterizing quantity. While the eccentricity would be an intuitive 
choice\footnote{The eccentricity at the separatrix-crossing was used in the global fits performed in Ref.~\cite{Albanesi:2021rby}.}, it is not a gauge invariant quantity and it is not defined through the whole
evolution of the system. We thus use the quantity $b\equiv p_\varphi/\hat{E}$ evaluated at the peak of the 
orbital frequency (i.e.~at the light-ring crossing) and shifted with the corresponding quasi-circular value,
\be
\label{eq:bmrg}
\bmrg = b_\Omgpk - b_\Omgpk^{\rm QC},
\ee
where $b_\Omgpk^{\rm QC} = 3.6693$. Note that this parameter is gauge invariant since it 
is a combination of energy and angular momentum, and it vanishes in the quasi-circular case.
The latter feature is useful because we impose that the fits reduces to the exact values 
in the quasi-circular case.

We thus proceed to perform the global fits for each multipole. 
Note that in our global fits we only use the simulations in Table~\ref{tab:ID}
with odd identification number ($\#$) (i.e.~the simulations with "round" initial eccentricity), 
so that the eccentric simulations with even identification number can be though as a test-set 
for the analytical model.
We need to fit the free parameters of amplitude and
phase templates, $\left\lbrace c_2^A, c_3^A, c_2^\phi, c_3^\phi, c_4^\phi  \right\rbrace$, the quantities 
$\left\lbrace \Amx, \d2Amx, \omgmx \right\rbrace$ that are needed to compute the constrained parameters,
and the phase $\theta_1$ of the mode-mixing complex factor $\ha_{\lm 1}$ 
(the modulus does not depends on the eccentricity).
The primary fits are reported in Appendix~\ref{sec:gfit_tables}, and in particular in 
Tables~\ref{tab:gfits_primary1},~\ref{tab:gfits_primary2}. The fits for the mode-mixing
are instead reported in Table~\ref{tab:gfits_th1}.

\section{Effective-one-body waveform}
\label{sec:modeling}

\subsection{Inspiral EOB waveform}
\label{sec:inspiralwave}
The quasi-circular EOB inspiral waveform for is obtained factorizing and resumming the post-Newtonian (PN) expanded
multipoles~\cite{Damour:2008gu},
\begin{equation}
h_{\lm}= h_{\lm}^{(N, \epsilon)_{\rm c}} \hat{h}^{(\epsilon)_{\rm c}}_{\lm} = h_{\lm}^{(N, \epsilon)_{\rm c}} \hat{S}^{(\epsilon)} \hat{h}_\lm^{\rm tail} (\rho_{\lm})^{\l} , \label{eq:eobwave_qc}
\end{equation}
where $\epsilon$ denotes the parity of the multipole, $h_{\lm}^{(N, \epsilon)_{\rm c}}$ is the Newtonian circular 
contribution and $\hat{h}^{(\epsilon)_{\rm c}}_{\lm}$ is the circular PN correction. The term $\hat{S}^{(\epsilon)}$ 
is the effective-source term, i.e.~the energy if $\epsilon=0$ or the Newtonian-normalized
angular momentum if $\epsilon=1$, $\hat{h}_\lm^{\rm tail}=T_{\lm} e^{i \delta_{\lm}}$ is 
the tail factor and the $\rho_{\lm}$  are the residual amplitude corrections. 

The waveform~\eqref{eq:eobwave_qc} can be generalized to non-circular dynamics including corrections 
that are known up to 2PN~\cite{Chiaramello:2020ehz,Khalil:2021txt,Placidi:2021rkh,Albanesi:2022xge}.
In particular, Ref.~\cite{Chiaramello:2020ehz} proposed to generalize the waveform of Eq.~\eqref{eq:eobwave_qc}
generic orbits by simply replacing the Newtonian quasi-circular prefactor with its general expression, i.e.
\be
\label{eq:genericNP}
h_{\lm}^{(N, \epsilon)_{\rm c}} \rightarrow \hat{h}_\lm^{(N,\epsilon)_{\rm c}} \hat{h}_\lm^{(N,\epsilon)_{\rm nc}}.
\ee
For the (2,2) mode, the non-circular correction reads
\be
\label{eq:NP22}
\hat{h}_{22}^{(N,0)_{\rm nc}} = 1 - \frac{\ddot{r}}{2 r \Omega^2} - \frac{\dot{r}^2}{2 r^2 \Omega^2} + \frac{2 \mathrm{i} \dot{r}}{r \Omega} + \frac{\mathrm{i} \dot{\Omega}}{2 \Omega^2},
\ee
where $r$ is the radius and $\Omega$ is the orbital frequency. 
The time-derivatives in the generic Newtonian prefactor are computed using a 4th order centred stencil scheme, i.e.
no PN-expanded equations of motion are used to compute them.
While this correction is clearly crucial for eccentric inspirals, 
we will see that it is needed to improve the analytical/numerical agreement during the plunge also 
in the quasi-circular case.
The 2PN non-circular corrections to the Newtonian-factorized waveform provide a better analytical/numerical agreement for
the phase during the inspiral~\cite{Placidi:2021rkh,Albanesi:2022xge}, even if the main correction
is given by the Newtonian term~\eqref{eq:genericNP}. In this work we will consider the noncircular hereditary
term $\hat{h}_{22}^{\rm tail_{nc}}$ written in terms of $\dot{p}_{r_*}$ (see Eq.~C1 of Ref.~\cite{Placidi:2021rkh}) 
and the instantaneous corrections $\hat{h}_{22}^{\rm inst_{nc}}$ introduced in Ref.~\cite{Albanesi:2022xge}, 
so that the quadrupolar waveform reads 
\be
\label{eq:wave_2PN}
h_{22} = \hat{h}_{22}^{(N,\epsilon)_{\rm c}} \hat{h}_{22}^{(N,\epsilon)_{\rm nc}} \hat{h}^{(\epsilon)_{\rm c}}_{22} 
\hat{h}_{22}^{\rm inst_{nc}} \hat{h}_{22}^{\rm tail_{nc}}.
\ee
The 2PN non-circular corrections are switched-off at the beginning of the plunge using a sigmoid function, 
$ \sigma(t) = 1/[1+e^{-\alpha (t - \tplunge)}]$ with $\alpha=0.2$, both for the eccentric and quasi-circular cases.
We do not consider 2PN non-circular corrections for the higher-modes. The relevance of 
these corrections in the quasi-circular inspiral will be discussed in Sec.~\ref{sec:model_test}.

\subsection{Next-to-Quasi-Circular corrections}
\label{sec:nqc}
Even if the generic Newtonian prefactor of Eq.~\eqref{eq:NP22}
is useful to improve the waveform during the plunge, we still need to correct the plunge waveform
using numerical-informed correction known as Next-to-Quasi-Circular (NQC) corrections~\cite{Damour:2007xr}, 
especially for the higher modes. 
The complete EOB waveform is thus given by
\be
\label{eq:complete_eob_wave}
h_\lm = \theta(\tmatch-t) h_\lm^{\rm inspl} \hat{h}_\lm^{\rm NQC} + \theta(t-\tmatch) h_\lm^{\rm rng} ,
\ee
where $h_\lm^{\rm inspl}$ is the inspiral EOB waveform discussed in Sec.~\ref{sec:inspiralwave}, 
$h_\lm^{\rm rng}$ is the ringdown discussed in Sec.~\ref{sec:ringdown_primary}, $\hat{h}^{\rm NQC}_\lm$ is the NQC waveform correction,
$\theta(x)$ is the Heaviside step function, and $\tmatch$ is the matching time.
The NQC correction is written as 
\begin{equation}
\label{eq:hnqc}
\hat{h}^{\rm NQC}_\lm = \left( 1 + \sum_{i=1}^{3}  a_i^\lm n_i \right) \exp{ \left( i  \sum_{j=1}^{3} b_j^\lm n_{j+3} \right) },
\end{equation}
where $n_i$ are functions that are combinations of quantities negligible during the quasi-circular inspiral
but relevant during the plunge. To satisfy this requirement, it is natural to write
them in terms of time-derivatives of the radius or in terms of $p_{r_*}$. 
For all the higher modes we use the basis
\begin{subequations}
\begin{align}
n_1 & = \frac{p_{r_*}^2}{r^2\Omega^2}, \\
n_2 & = \frac{\ddot{r}}{r \Omega^2}, \\
n_3 & = n_1 \, p_{r_*}^2, \\
n_4 & = \frac{p_{r_*}}{r \Omega}, \\
n_5 & = n_4 \, \Omega, \\
n_6 & = n_5 \, p_{r_*}^2.
\end{align}
\end{subequations}
For the (2,2) mode, we use use the same $n_i$ with $i\leq4$, but 
we change $n_5 = n_4 \, r^2 \Omega^2$ and, consequently, $n_6  = n_5 \, p_{r_*}^2$.
Note that in \TEOBResumS{} only the first derivatives of $A$ and $\omega$ are considered,
so that $n_3$ and $n_6$ are not used. Moreover, in \TEOBResumS{} 
the $n_5$ function for the higher modes is different than the one considered here. 
The coefficients $a_i$ and $b_i$ are determined at a specific time $\tNQC$. 
If we consider $\tNQC>\tAlm$, then they are determined solving the linear system 
\begin{subequations}
\label{eq:nqc_syst}
\begin{align}
       A_\lm^{\rm EOB}(\tNQC)      & =            A_\lm^{\rm rng}(\tNQC),  \\
      \dot{A}_\lm^{\rm EOB}(\tNQC) & =       \dot{A}_\lm^{\rm rng}(\tNQC),  \\
     \ddot{A}_\lm^{\rm EOB}(\tNQC) & =      \ddot{A}_\lm^{\rm rng}(\tNQC),  \\
       \omega_\lm^{\rm EOB}(\tNQC) & =        \omega_\lm^{\rm rng}(\tNQC),  \\
 \dot{\omega}_\lm^{\rm EOB}(\tNQC) & =  \dot{\omega}_\lm^{\rm rng}(\tNQC),  \\
\ddot{\omega}_\lm^{\rm EOB}(\tNQC) & = \ddot{\omega}_\lm^{\rm rng}(\tNQC),
\end{align}
\end{subequations}
where on the left-hand side (lhs)~the amplitude, frequency and corresponding time-derivatives 
are computed from $h_\lm^{\rm inspl} \hat{h}_\lm^{\rm NQC}$, while on the right-hand side (rhs)~they
are computed from the ringdown model.  
If we want to use $\tNQC<\tAlm$, on the rhs we 
have to consider quantities extracted from numerical data at $\tNQC$.
Due to this reason, the choice $\tNQC>\tAlm$ is, in principle, preferable since it reduces 
the number of numerical-informed parameters in the model. However, as we will see in more detail later, 
choosing $\tNQC<\tAlm$ works better for the higher modes.

While the NQC correction is negligible during the quasi-circular inspiral by construction, 
they are not negligible in eccentric inspirals since $p_{r_*}$ is not small. 
For this reason we switch-off the NQC corrections during the eccentric inspiral using a sigmoid, 
\be
\hat{h}^{\rm NQC}_\lm  \rightarrow \hat{h}^{\rm NQC}_\lm \frac{1}{ 1 + e^{-\alpha^s\left(t-t_{\ddot{r}=0}\right)} }.
\ee
Given the discussion in Sec.~\ref{sec:RWZ}, it is natural to center the sigmoid in $t_{\ddot{r}=0}$ so that the
NQC corrections are switched-on in a region where the motion is indeed quasi-circular. Due to this choice, the relevance of
the precise value of $\alpha^s$ is not crucial; in this work we will use  $\alpha^s=0.2$. 

Finally, consider that to correctly evaluate the lhs of Eqs.~\eqref{eq:nqc_syst}, 
an interpolation on a refined time grid is needed, see Appendix~\ref{app:regridding} for more details.
\begin{figure*}[t]
	\begin{center}
	\includegraphics[width=0.23\textwidth]{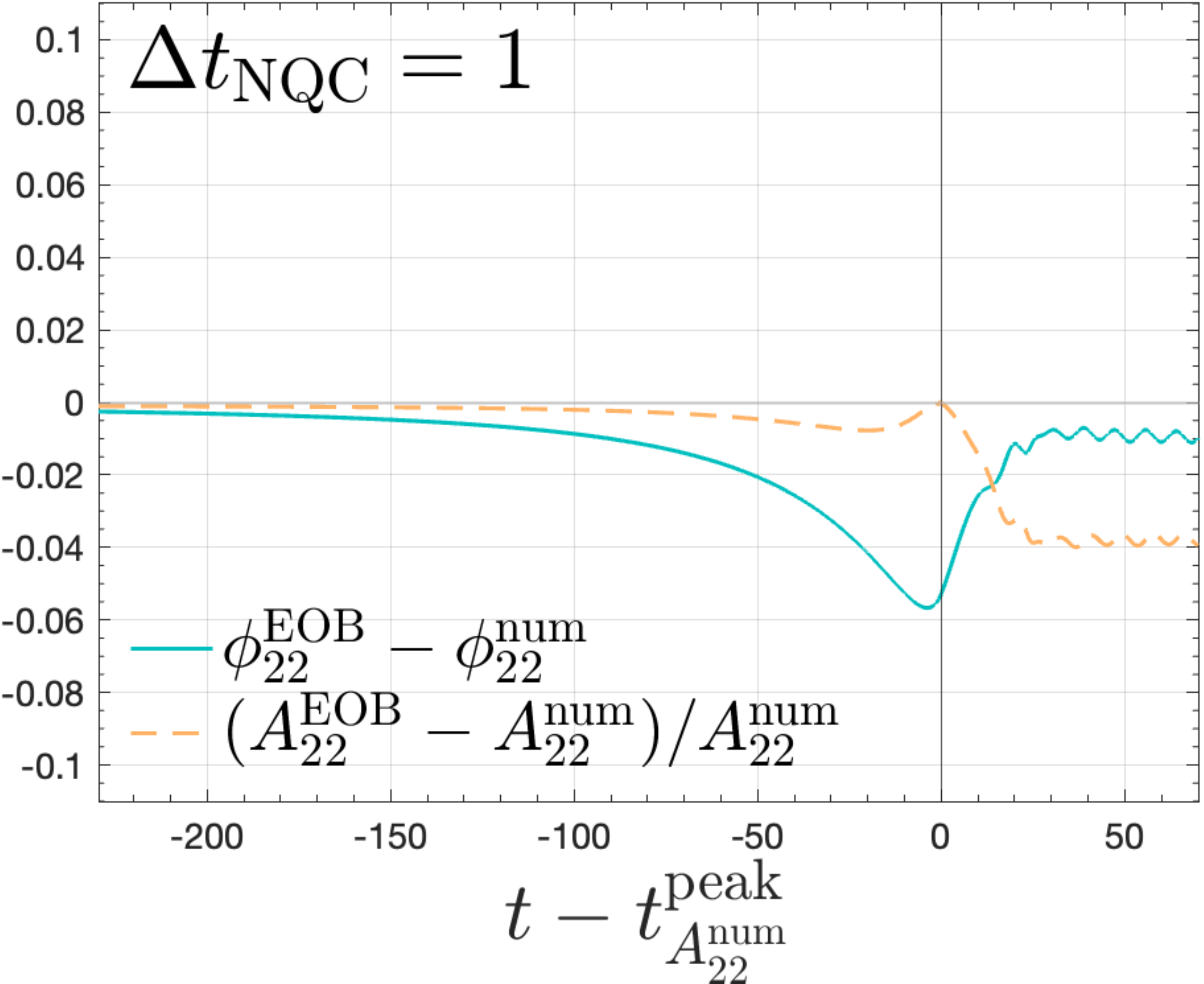}
	\includegraphics[width=0.23\textwidth]{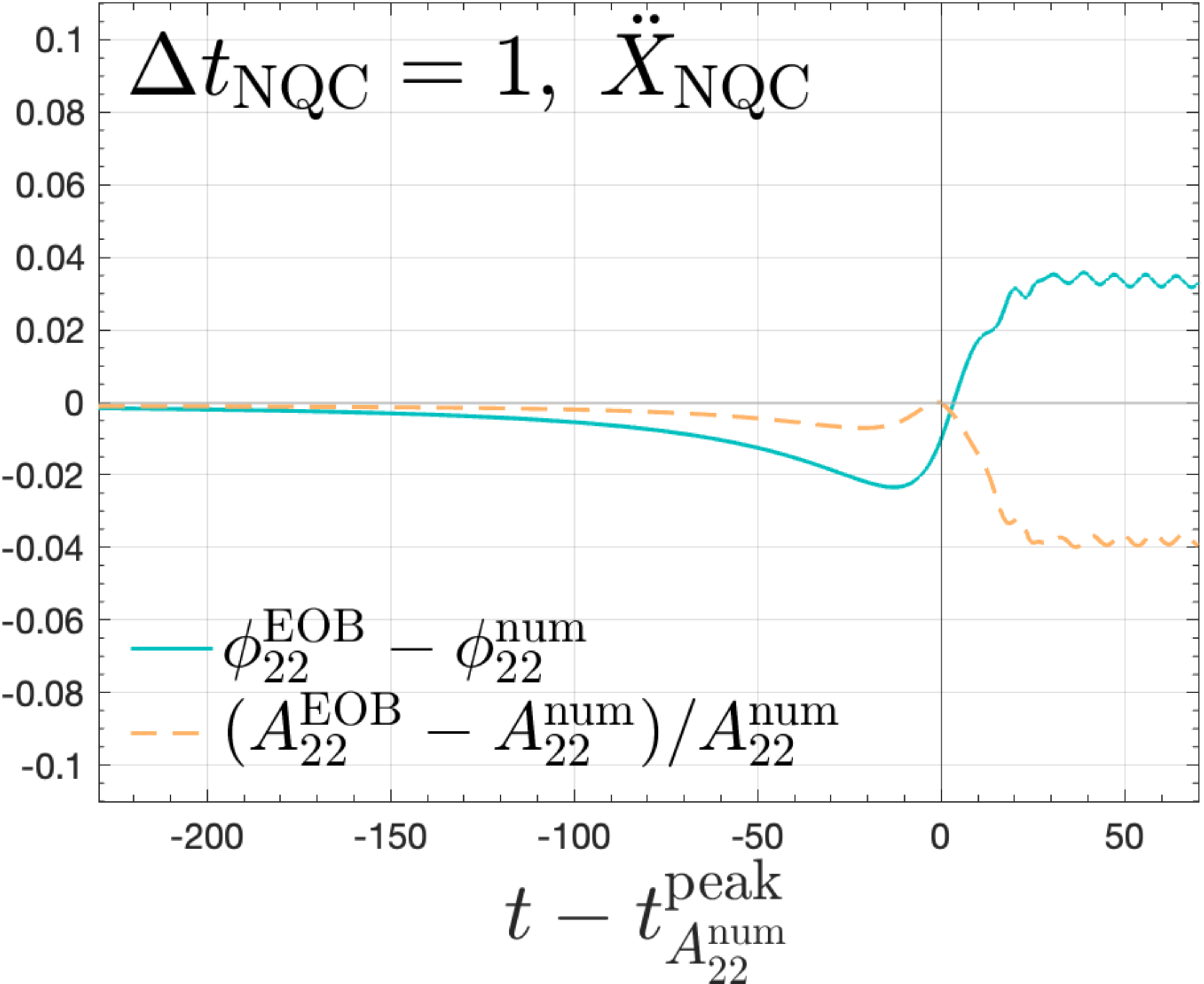}
	\includegraphics[width=0.23\textwidth]{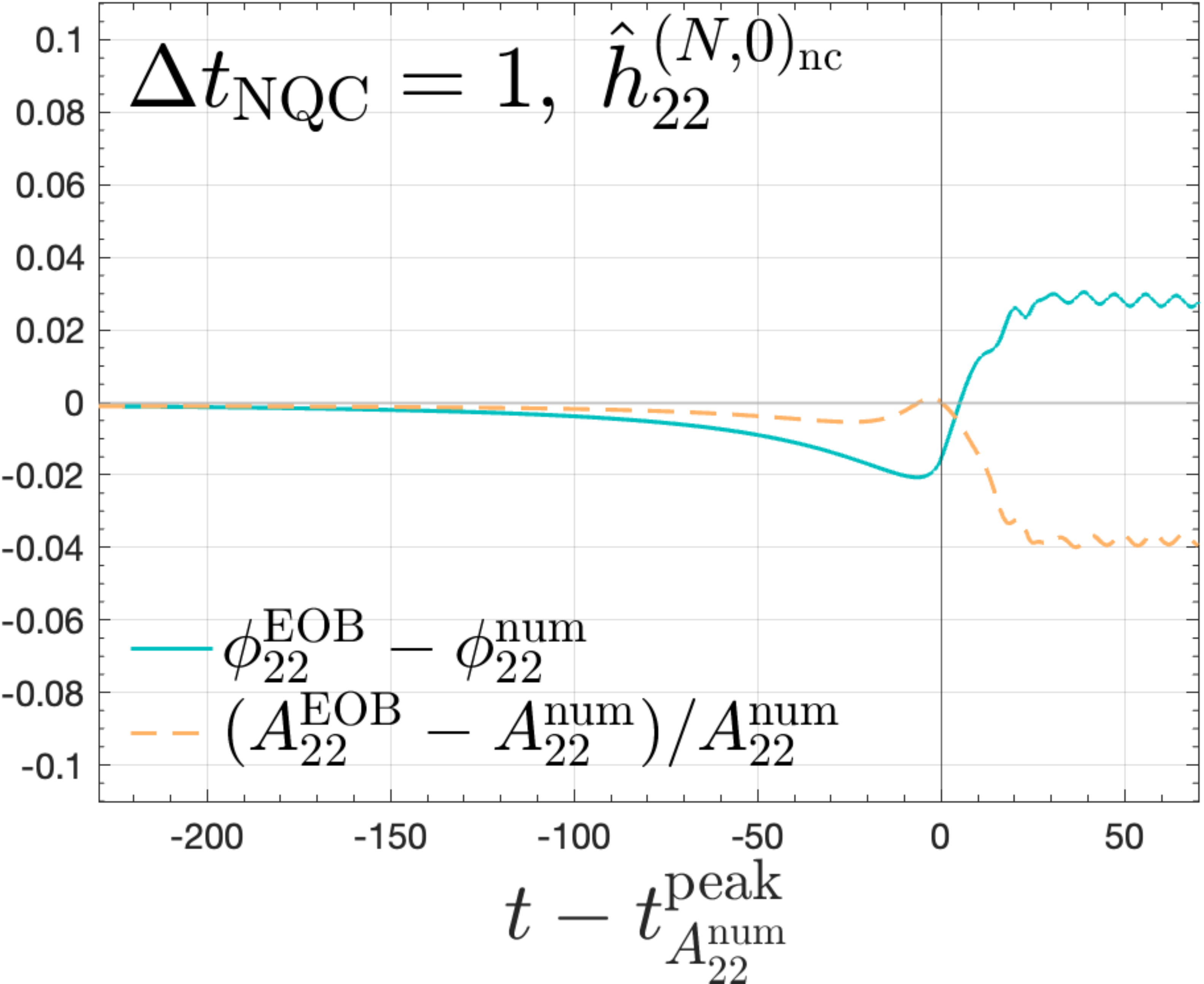}
	\includegraphics[width=0.23\textwidth]{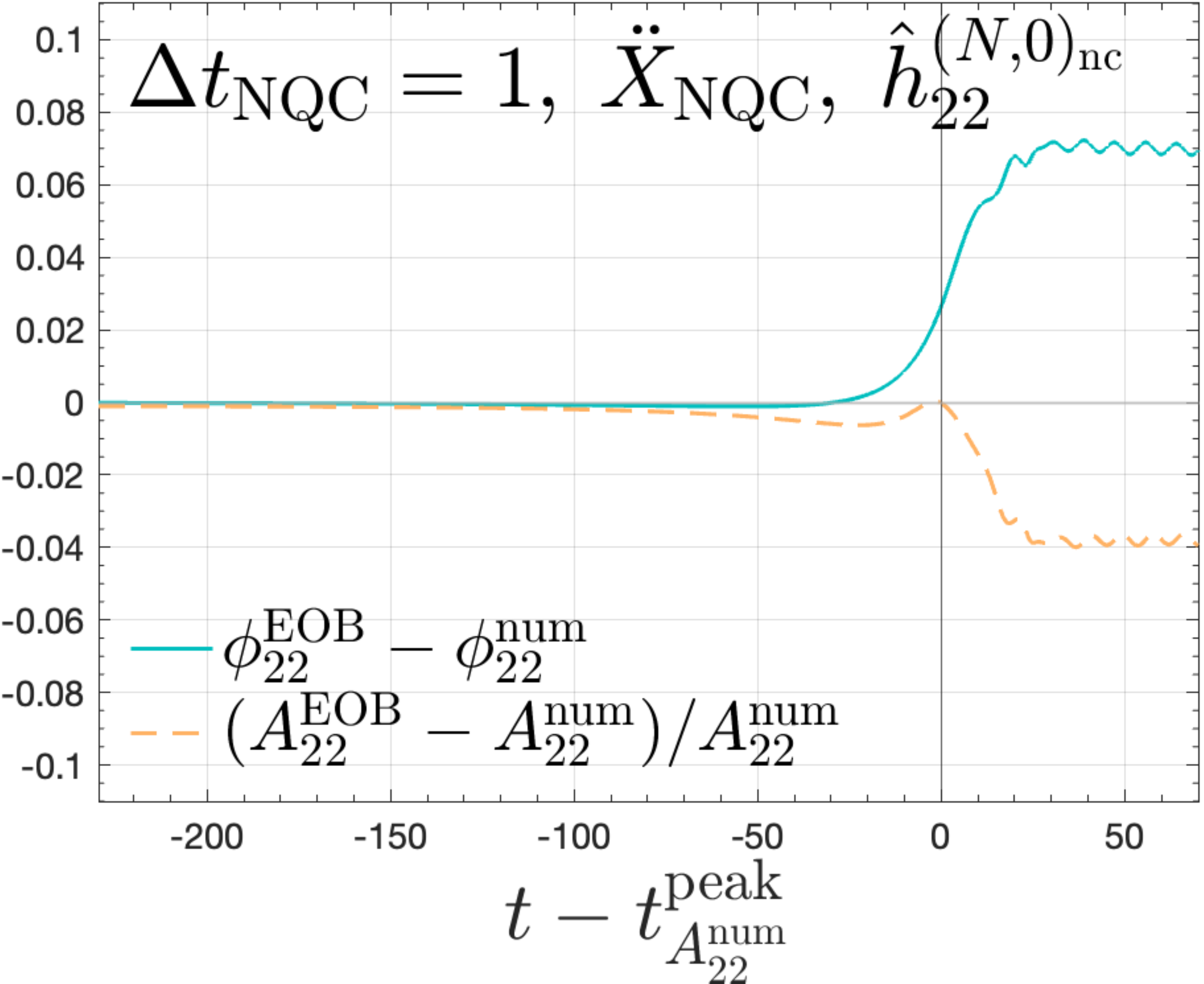}\\
	\includegraphics[width=0.23\textwidth]{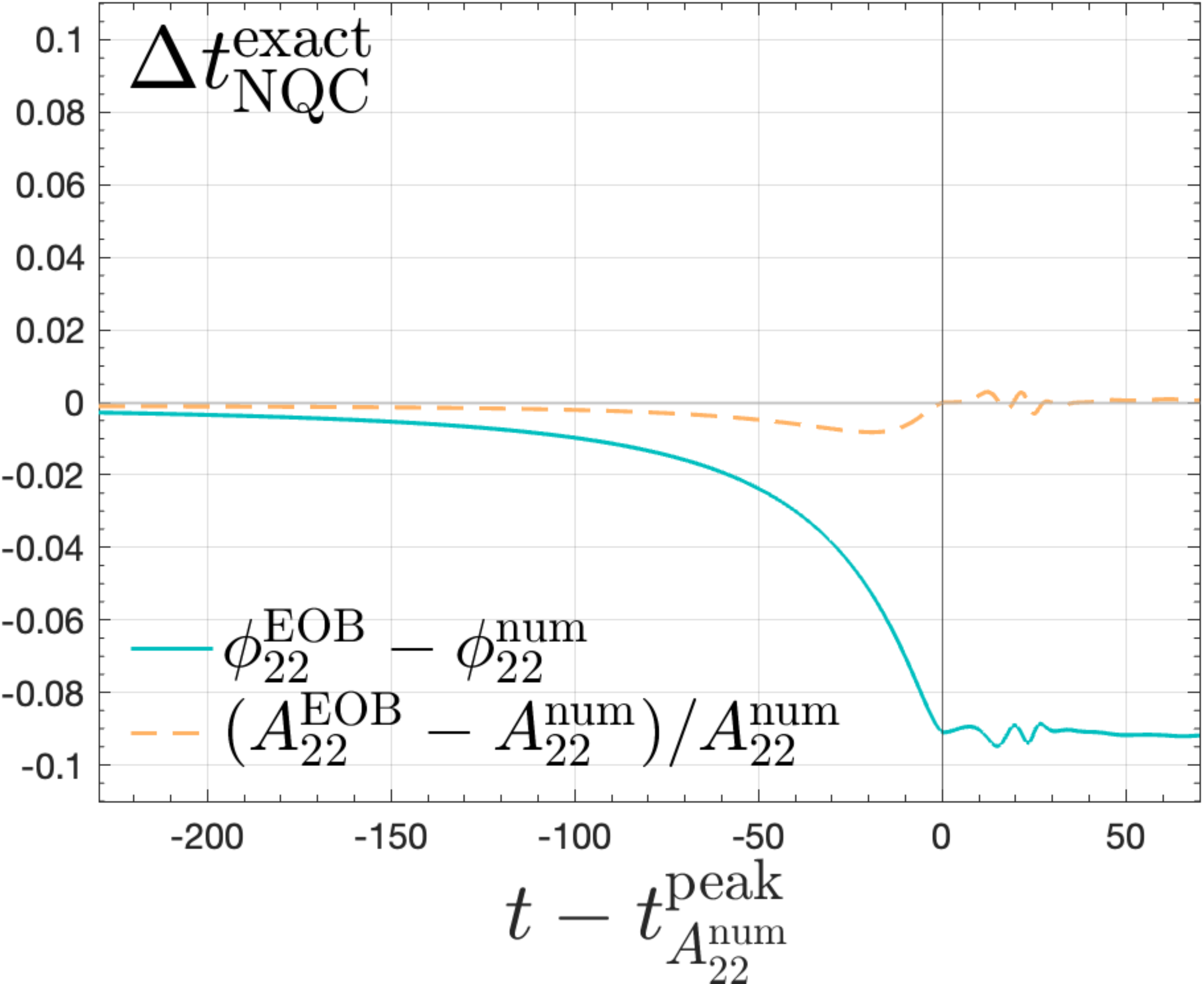}
	\includegraphics[width=0.23\textwidth]{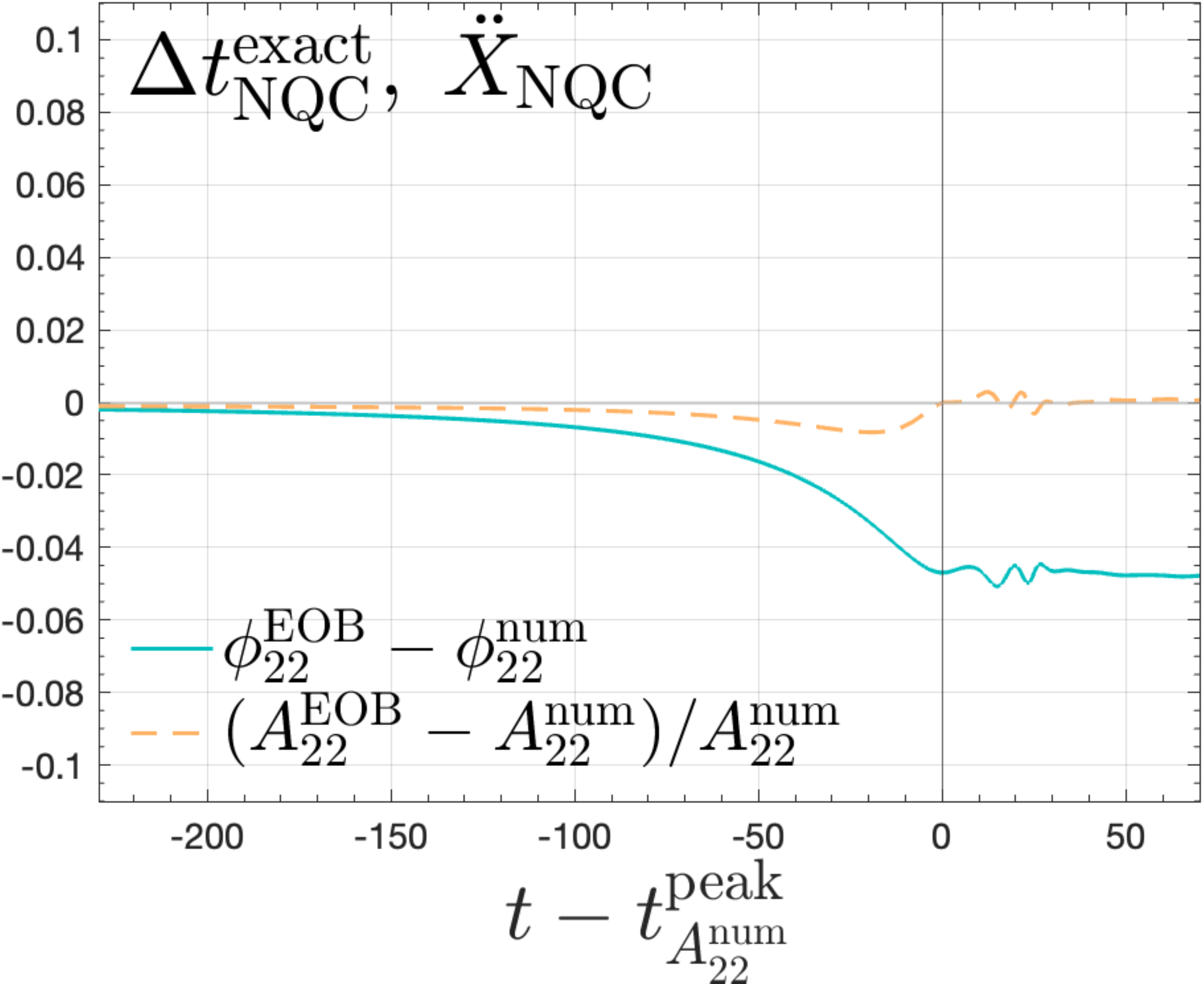}
	\includegraphics[width=0.23\textwidth]{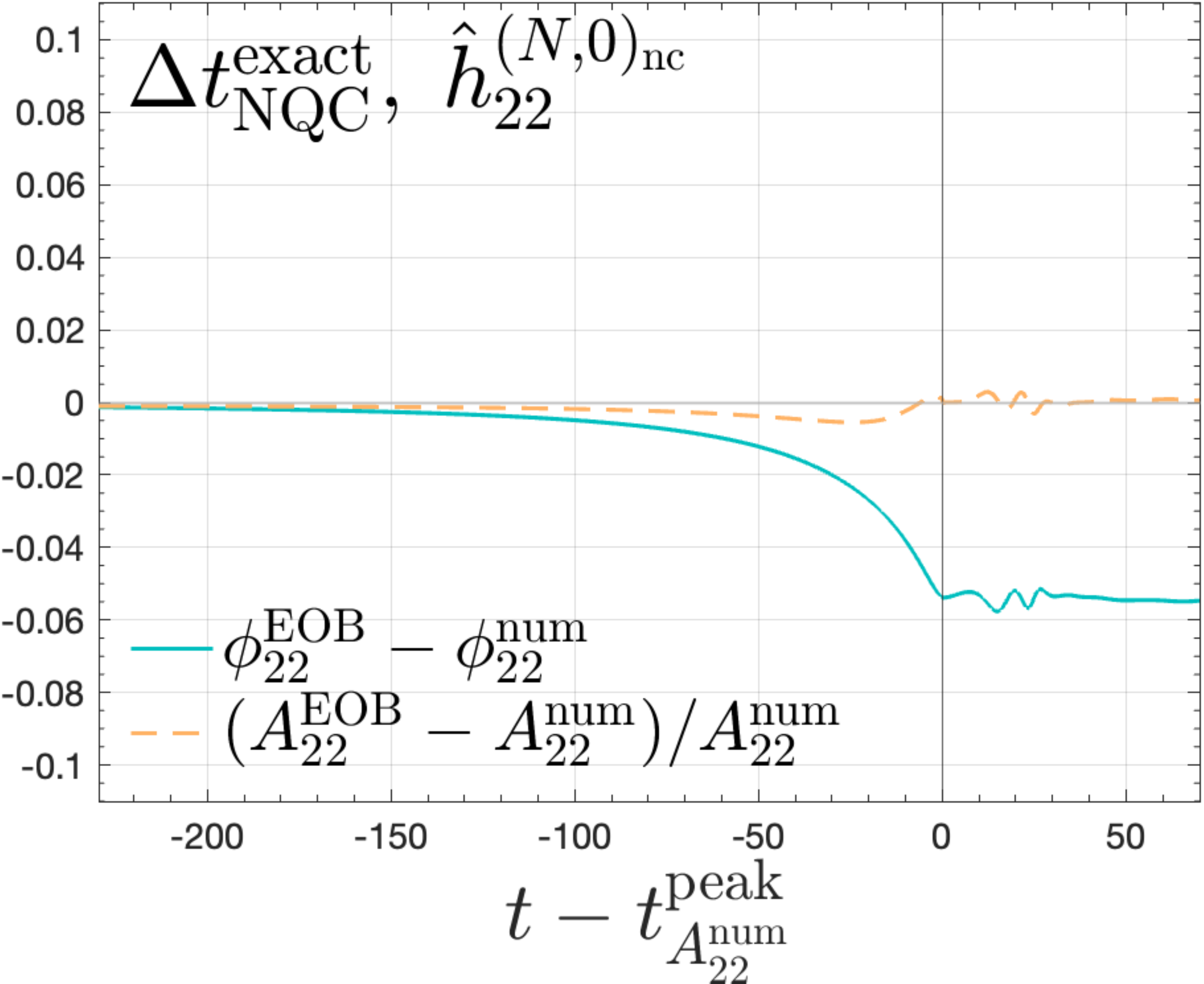}
	\includegraphics[width=0.23\textwidth]{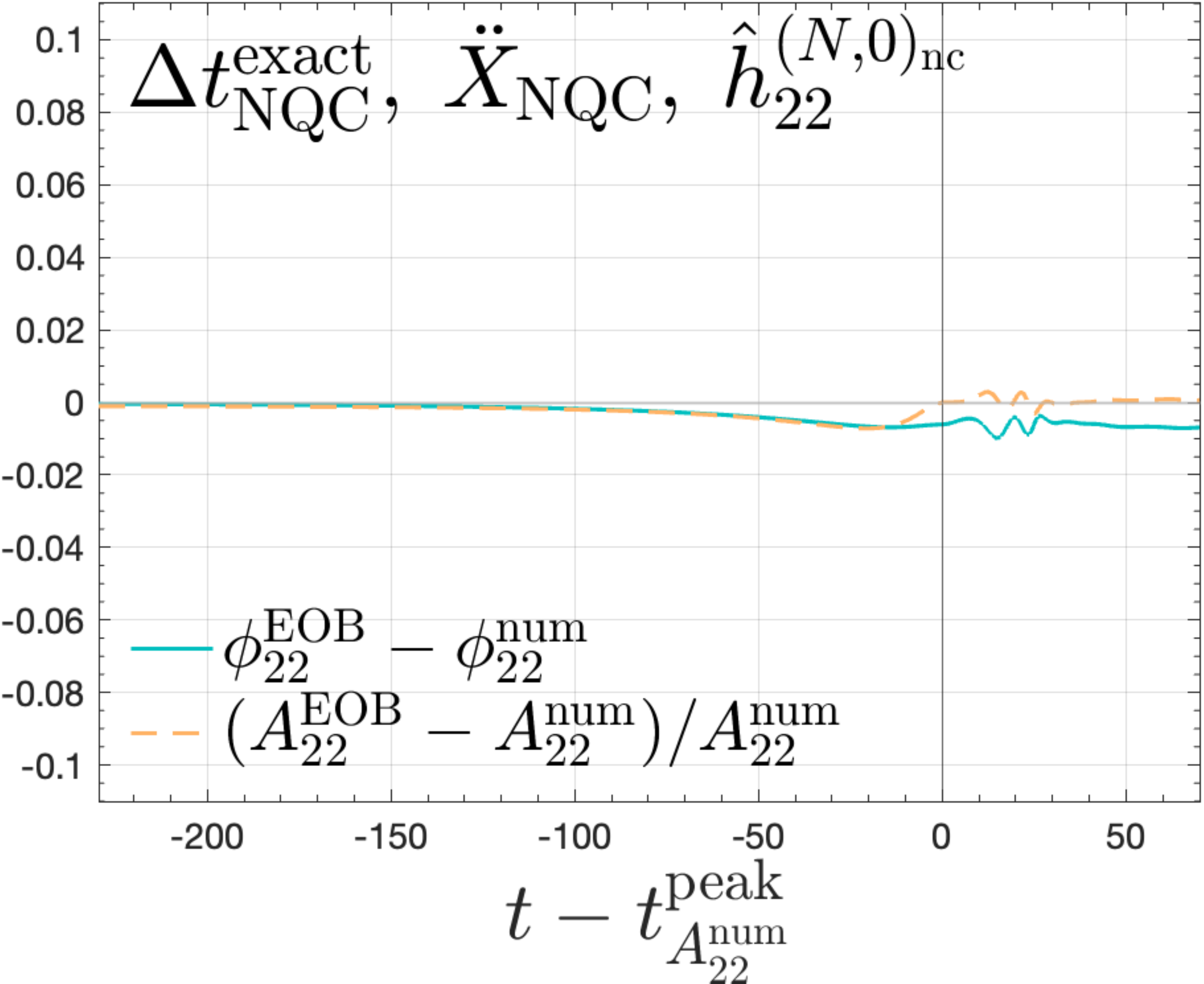}
	\caption{\label{fig:inspl_a0_e0_8pack} Analytical/numerical differences for the amplitude and the phase (in radians)
	 of the quadrupolar waveform for the quasi-circular inspiral-plunge in Schwarzschild. 
	 We consider different inspiral-plunge and ringdown
	 matching procedures. The differences are shown from the LSO crossing to $t_{A^{\rm num}_{22}}^{\rm peak}+70$. 
	 In the upper panels we consider the matching time
	 according to Eq.~\eqref{eq:old_tmrg}, while in the lower panels we use Eq.~\eqref{eq:new_tmrg}. Then, from left to right,
	 we progressively improve the model considering the second time derivatives in the NQC base, 
	 $\ddot{A}_{22}$ and $\ddot{\omega}_{22}$, and then the Newtonian noncircular correction, 
	 $\hat{h}_\lm^{(N,0)_{\rm nc}}$. The rightmost lower panel shows the differences for the state-of-the-art model.}
	\end{center}
\end{figure*}
%

\subsection{Matching point}
\label{sec:matchingpoint}
The match of the plunge and ringdown waveform is performed at 
$t_{\lm}^{\rm NQC}$ if $t_{\lm}^{\rm NQC}>\tAlm$, while it is performed at
$\tAlm$ if $t_{\lm}^{\rm NQC}\leq\tAlm$. The former prescription will be used for the
(2,2) mode, while the latter will be used for the higher modes.
In any case, in order to know the locations of $\tAlm$, we need to link them to dynamical quantities. 
In the \TEOBResumS{} model~\cite{Riemenschneider:2021ppj,Nagar:2023zxh} and also in Ref.~\cite{Albanesi:2021rby}, 
the heuristic used to find the peak of the quadrupolar amplitude was 
\be
\label{eq:old_tmrg}
\tA22 = \tOmgOrb - \Delta t_{\rm NQC} - 2, 
\ee
where $\tOmgOrb$ is the peak of the orbital frequency\footnote{Consider that for spinning binaries, $\tOmgOrb$ is the
peak of the {\it pure} orbital frequency, that is computed without considering the spin-orbit terms.} 
and $\Delta t_{\rm NQC}=1$. 
We keep $\Delta t_{\rm NQC}$ in the notation for continuity with previous works. 
The heuristic~\eqref{eq:old_tmrg} gives satisfactory results in the quasi-circular, since the exact value of
$\tA22$ extracted from the Zerilli waveform is $\Delta t_{\rm NQC}^{\rm exact} \simeq 0.559$. 
Most importably, Eq.~\eqref{eq:old_tmrg} has been shown to be reliable also for quasi-circular
binaries of comparable mass. However, when dealing with highly eccentric binaries, the approximation 
$\Delta t_{\rm NQC}=1$ is no longer valid and we thus perform a global fit as discussed in Sec.~\ref{sec:gfits}, 
finding 
\be
\label{eq:new_tmrg}
t_{A_{22}}^{\rm peak } = \tOmgOrb - \frac{ 2.559 + 7.574 \, \bmrg  - 18.830 \, \bmrg^2}{1 - 2.160 \, \bmrg}.
\ee
\begin{table}
  \caption{\label{tab:Dtlm}
  Time delays of the amplitude peaks $\Delta t_\lm$ for the higher modes with respect to
  the peak of the (2,2) amplitude, see definition in Eq.~\eqref{eq:Dtlm}.
  The global fitting template is 
  $\Delta t_\lm=\left( C^\lm_{\rm QC}+C^\lm_1\bmrg + C^\lm_2 \bmrg^2\right)/\left( 1 + D^\lm_1 \bmrg \right)$, 
  where $C^\lm_{\rm QC}$ is the quasi-circular value; see also discussion in Sec.~\ref{sec:gfits}. 
  }
  \begin{center}
\begin{ruledtabular}
\begin{tabular}{c | c | c | c | c }
 $(\l,m)$ & $C^\lm_{\rm QC}$ & $C^\lm_1$ & $C^\lm_2$ & $D^\lm_1$ \\
 \hline
(2,1) &   $ 11.960$ &  $ 51.831$ &   $\dots$  &    $ 2.704$      \\ 
(3,3) &   $ 3.563$  &  $ 5.507 $ &  $ 21.215$ &    $\dots$       \\ 
(3,2) &   $ 9.396$  &  $ 11.549$ &  $ 28.933$ &    $\dots$       \\ 
(3,1) &   $ 13.100$ &  $ 15.132$ &  $ 28.765$ &    $\dots$       \\ 
(4,4) &   $ 5.384$  &  $ 7.032 $ &  $ 23.973$ &    $\dots$       \\ 
(4,3) &   $ 9.766$  &  $ 11.101$ &  $ 27.870$ &    $\dots$       \\ 
(4,2) &   $ 12.090$ &  $ 13.150$ &  $ 28.396$ &    $\dots$       \\ 
(4,1) &   $ 13.280$ &  $ 15.033$ &  $ 28.667$ &    $\dots$       \\ 
(5,5) &   $ 6.679$  &  $ 7.803$  &  $ 24.818$ &    $\dots$       \\ 
\end{tabular}
\end{ruledtabular}
\end{center}
\end{table}

The peak amplitude of the other multipoles are delayed with respect to the quadrupolar one as
\be
\label{eq:Dtlm}
\tAlm = \tA22 + \Delta t_\lm,
\ee
with $\Delta t_\lm>0$ for all the higher modes. 
The values of $\Delta t_\lm$ the quasi-circular case, together with their global fits, are listed 
in Table~\ref{tab:Dtlm}. The fact that $\Delta t_\lm$ increases with $m$ at fixed $\l$
can be understood heuristically considering that all the $m$-modes have to reach the 
same final QNM frequency $\omega_{\l,1}$ (modulo mode-mixing), but the waveform 
frequency during the inspiral is given by $\omega_\lm^{\rm inspl} = m \Omega$ at leading order. 
Therefore, the modes with small $m$ will need more time to reach the final
frequency $\omega_{\l,1}$. 

We also point out that the mode-mixing becomes more relevant in low-$m$ higher modes (see e.g. Sec.~\ref{sec:omgfit_iter}),
so that the position of the amplitude peak for the higher modes can be contaminated by the mode-mixing. 
Once that the location of the amplitude peak is known, we can proceed to match, mode by mode, 
the inspiral waveform to the ringdown model. From a computational point of view, the matching 
is performed on a time grid that is finer than the one used to solve the dynamics, 
see Appendix~\ref{app:regridding} for details.

\section{Probing the effective one body analytical waveform}
\label{sec:model_test}
\subsection{Quasi-circular case}
\begin{figure}[t]
	\begin{center}
	\includegraphics[width=0.48\textwidth]{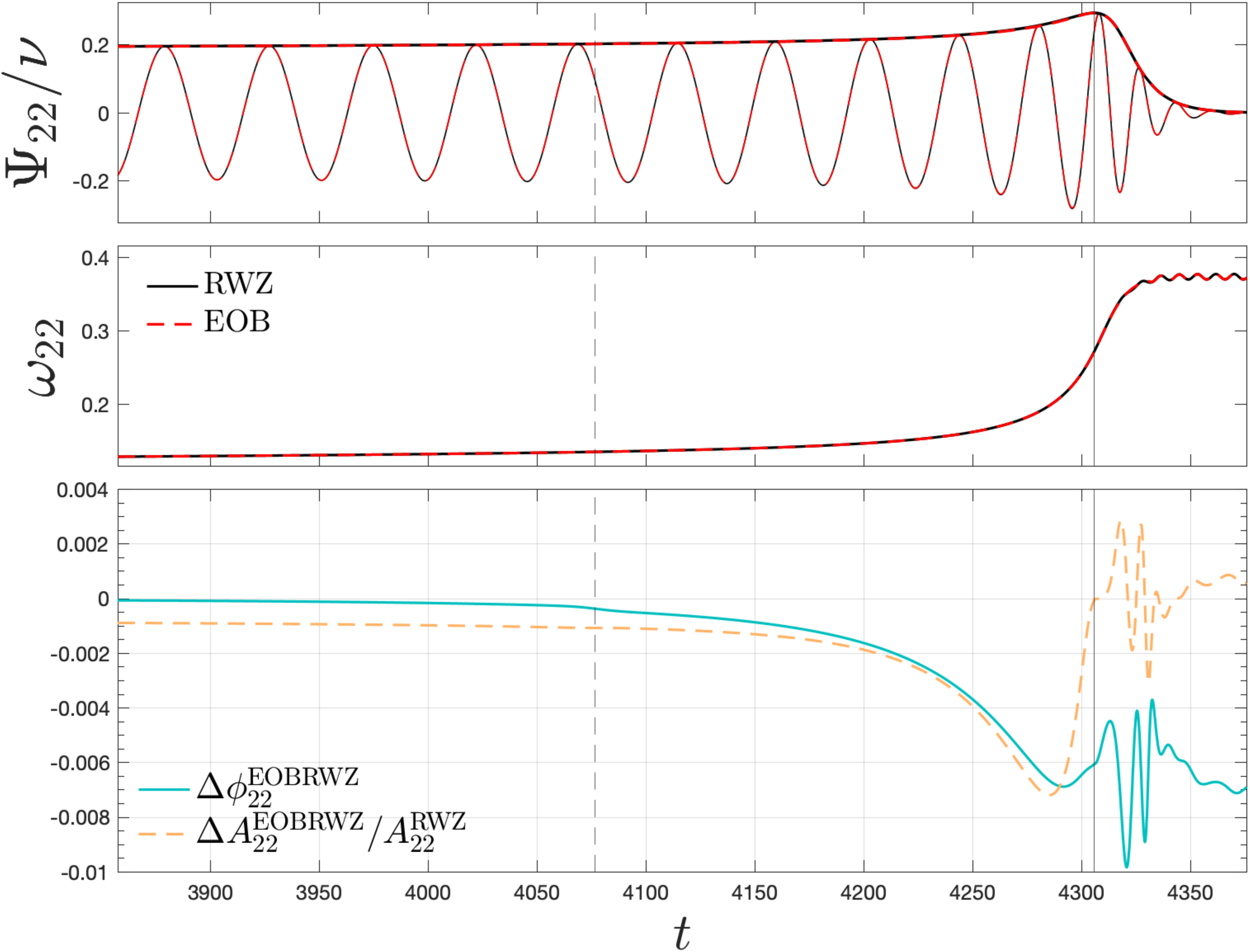}
	\caption{\label{fig:inspl_l2m2}Quasi-circular case, $\ell=m=2$ mode: RWZ waveform (black) compared to the
	complete EOB one (red, dashed). 
	The vertical lines mark the LSO crossing and the peak of $A_{22}$. 
	The bottom panel shows the relative amplitude difference (dashed orange) and the phase difference 
	in radians (solid light blue). Non-circular corrections to the waveform up to 2PN are included. }
	\end{center}
\end{figure}
We start by analyzing in detail the quasi-circular case, testing
the different prescriptions for the implementation of the NQC 
correction and discussing the accuracy of each waveform mode.
Focusing first on the $\ell=m=2$ mode, we have to discuss three aspects:
\begin{itemize}
\item[(i)]the impact of the precise location of the amplitude peak on the EOB temporal axis,  $\tA22$,
               as described by  Eq.~\eqref{eq:new_tmrg};
\item[(ii)]the impact of the NQC correction determined imposing also continuity between the EOB and RWZ
               second-time derivatives $\ddot{X}_{\rm NQC} = \left\lbrace \ddot{A}_{22}^{\rm NQC}, \ddot{\omega}_{22}^{\rm NQC}\right\rbrace$;
\item[(iii)]the impact due to the generic Newtonian prefactor, Eq.~\eqref{eq:NP22}. 
\end{itemize}
Note that here we are analyzing the plunge, therefore at this stage we do not consider the 2PN 
non-circular corrections introduced in Eq.~\eqref{eq:wave_2PN} .
Figure~\ref{fig:inspl_a0_e0_8pack} illustrates the analytical/numerical relative amplitude 
difference (dashed orange) and phase differences in radians (light blue) for all possible combinations. 
In the top row of the figure, the amplitude peak location is obtained according to 
Eq.~\eqref{eq:old_tmrg}, i.e.~the prescription that is adopted, for simplicity, in the comparable 
mass case within the \TEOBResumS{} model\footnote{See however Ref.~\cite{Damour:2012ky} for an 
early attempt to go beyond this simplifying choice.}.
In the bottom row of the figure we consider instead its exact location according to Eq.~\eqref{eq:new_tmrg}.
Then, moving from left to right, we add the second-time derivatives of amplitude and
frequency in the NQC-corrections, the analytical generic Newtonian prefactor, and
finally both effects together. In all the cases, the NQC corrections are obtained solving the
system given by Eqs.~\eqref{eq:nqc_syst} using $t_{22}^{\rm NQC} = \tA22 + 2$.

For the $\Delta t_{\rm NQC}=1$ case, we see that, as expected, the inclusion of both the improved
NQC corrections and of the generic Newtonian prefactor brings a considerable reduction of the phase 
difference up to merger. Moreover, the phase difference now grows monotonically, to saturate at
$\Delta\phi^{\rm EOBRWZ}_{22}\simeq 0.08$. As recently pointed out in Ref.~\cite{Nagar:2023zxh},
if such a behavior is reproduced for comparable-mass waveforms, generally indicates that one
will end up with excellent mismatches using actual detector power spectral density. This suggests
that the use of $\hat{h}_{22}^{(\rm N,0)_{nc}}$ and of $\ddot{X}_{\rm NQC}$ {\it also} for compable
mass binaries may result in a further reduction of the current EOB/NR disagreement ($\sim 0.2$~rad)
through merger and ringdown. By contrast, it is interesting to note that the amplitude difference
during the ringdown is nonegligible and remains substantially unchanged whatever choice is made.
When $\Delta t_{\rm NQC}^{\rm exact}$ is used we are thus not surprised to find a consistent reduction
of the amplitude difference during the ringdown (though evidently it remains unchanged up to merger).
By contrast, the progressive inclusion of additional physical elements (i.e.~$\hat{h}_{22}^{(\rm N,0)_{nc}}$ 
and $\ddot{X}_{\rm NQC}$) brings phase differences below $0.01$~rad through the full inspiral, merger
and ringdown.
The complete EOB/RWZ comparison for the final quadrupolar waveform, that incorporates also 2PN non-circular corrections,
is shown in Fig.~\ref{fig:inspl_l2m2} and 
complements the rightmost bottom panel of Fig.~\ref{fig:inspl_a0_e0_8pack} also showing the
EOB frequency. One appreciates that the phase difference reaches the $\sim 4\cdot 10^{-4}$~rad at LSO crossing, and 
remains always below the $0.01$~rad even at merger time.
The relative amplitude difference is $\sim 1\times 10^{-3}$ at LSO crossing to reach at most 
$\sim 7\times 10^{-3}$ around merger time.
\begin{figure}[t]
	\begin{center}
	\includegraphics[width=0.48\textwidth]{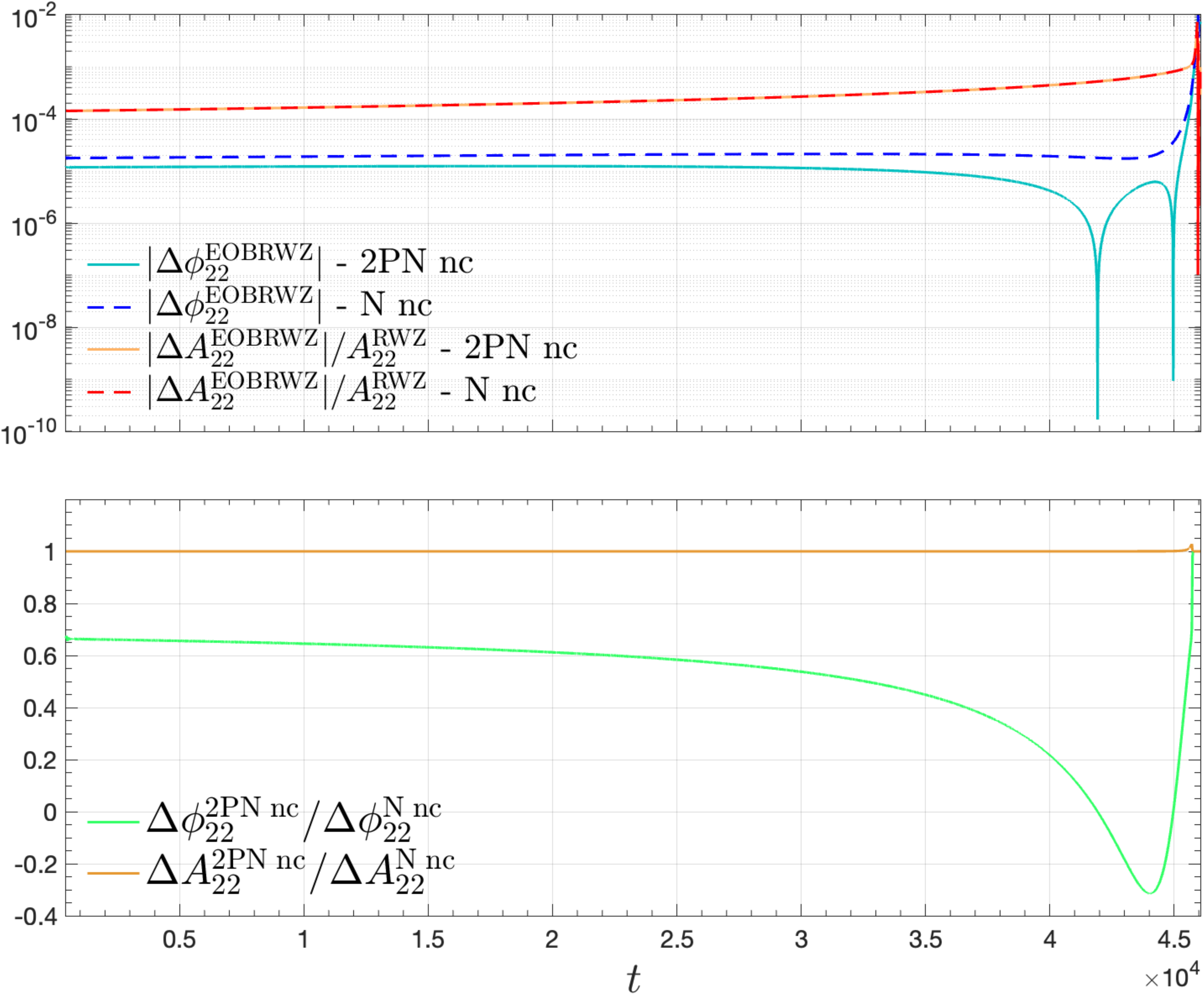}
	\caption{\label{fig:dphi22_2PN} 
	EOB/RWZ phase differences (radians) and relative amplitude differences for the (2,2) mode of 
	a quasi-circular inspiral starting from $r_0=9$. 
    In the upper panel we show the differences obtained with the waveform with non-circular 
    corrections up to 2PN (solid, light blue for the phase and orange for the amplitude),
    and the ones obtained considering only the generic Newtonian
	prefactor (dashed, blue for the phase and red for the amplitude). 
	In the bottom panel we show the ratios of these differences.
	}
	\end{center}
\end{figure}
Finally, we quantify the contribution of the 2PN non-circular corrections of Eq.~\eqref{eq:wave_2PN} in 
Fig.~\ref{fig:dphi22_2PN}, where we show the EOB/RWZ phase differences of the (2,2) mode for a quasi-circular inspiral
starting from $r_0=9$. The waveforms have been computed (i) considering the complete waveform, as discussed above 
and shown in Fig.~\ref{fig:inspl_l2m2}, (ii) considering only Newtonian non-circular corrections. As can be seen, the 
2PN non-circular corrections improve the phase agreement through the whole inspiral of the binary, but they
are not relevant for the amplitude.

%
\begin{figure*}
	\begin{center}
	\includegraphics[width=0.44\textwidth,height=6.81cm]{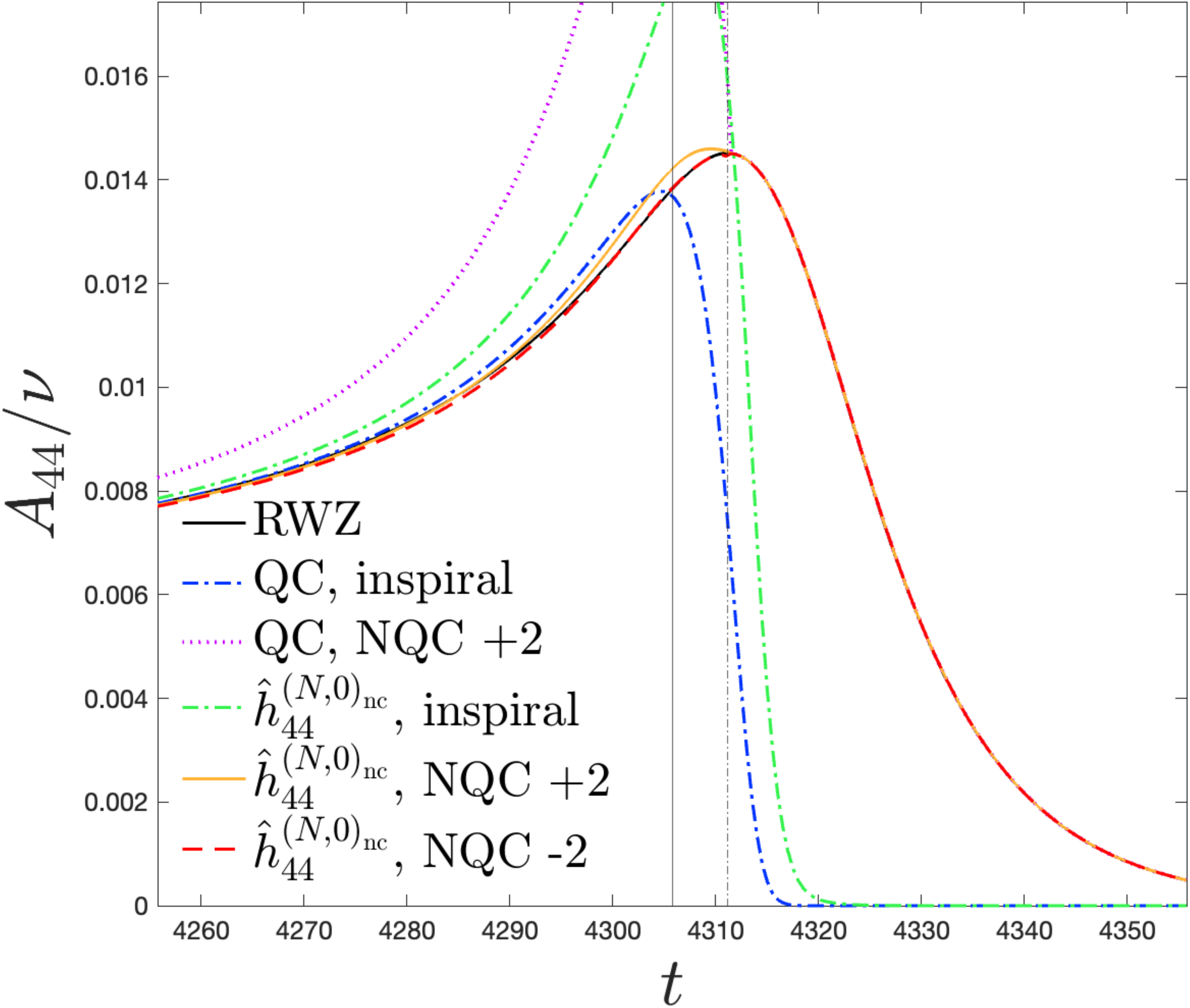}
	\hspace{0.12cm}
	\includegraphics[width=0.43\textwidth]{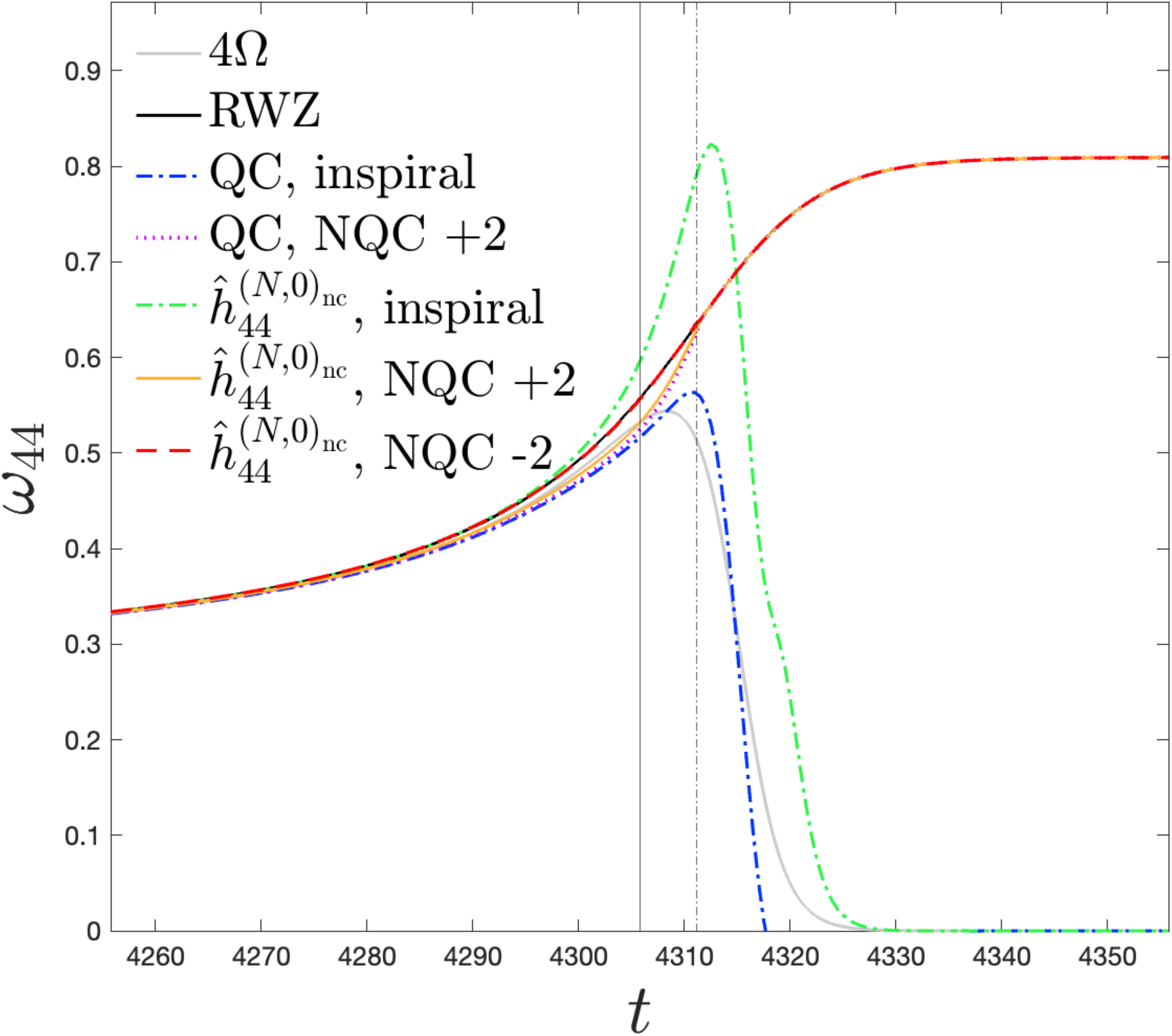}
	\caption{\label{fig:testHM_l4m4}Quasi-circular case, mode $\ell=m=4$: comparing various
	choices of analytical EOB waveform and different ways of determining  the NQC
	corrections.  Amplitude (left panel) and frequency (right panel)
	The best EOB/RWZ agreement is obtained by: (i) using the general, noncircular,
	Newtonian prefactor and (ii) when the NQC corrections are computed 
	at $t^{\rm NQC}_{44}=t_{A_{44}}^{\rm peak}-2$. We also show, as gray line,
	4 times the orbital frequency. From left to right, the two vertical line mark
	 $\tA22$ and $t_{A_{44}}^{\rm peak}$.}
	\end{center}
\end{figure*}
\begin{figure*}
	\begin{center}
	\includegraphics[width=0.32\textwidth,height=4.2cm]{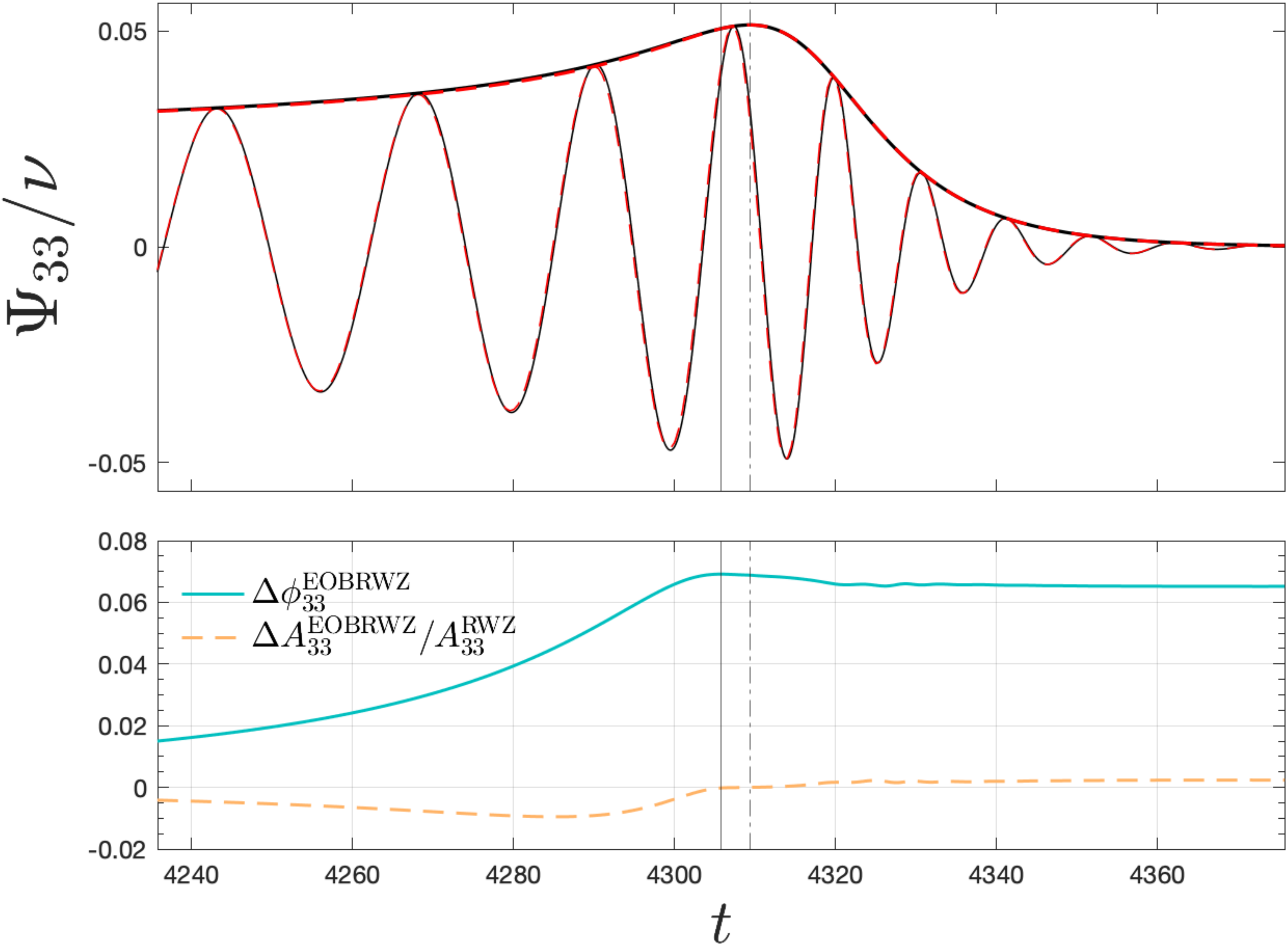}
	\includegraphics[width=0.32\textwidth,height=4.2cm]{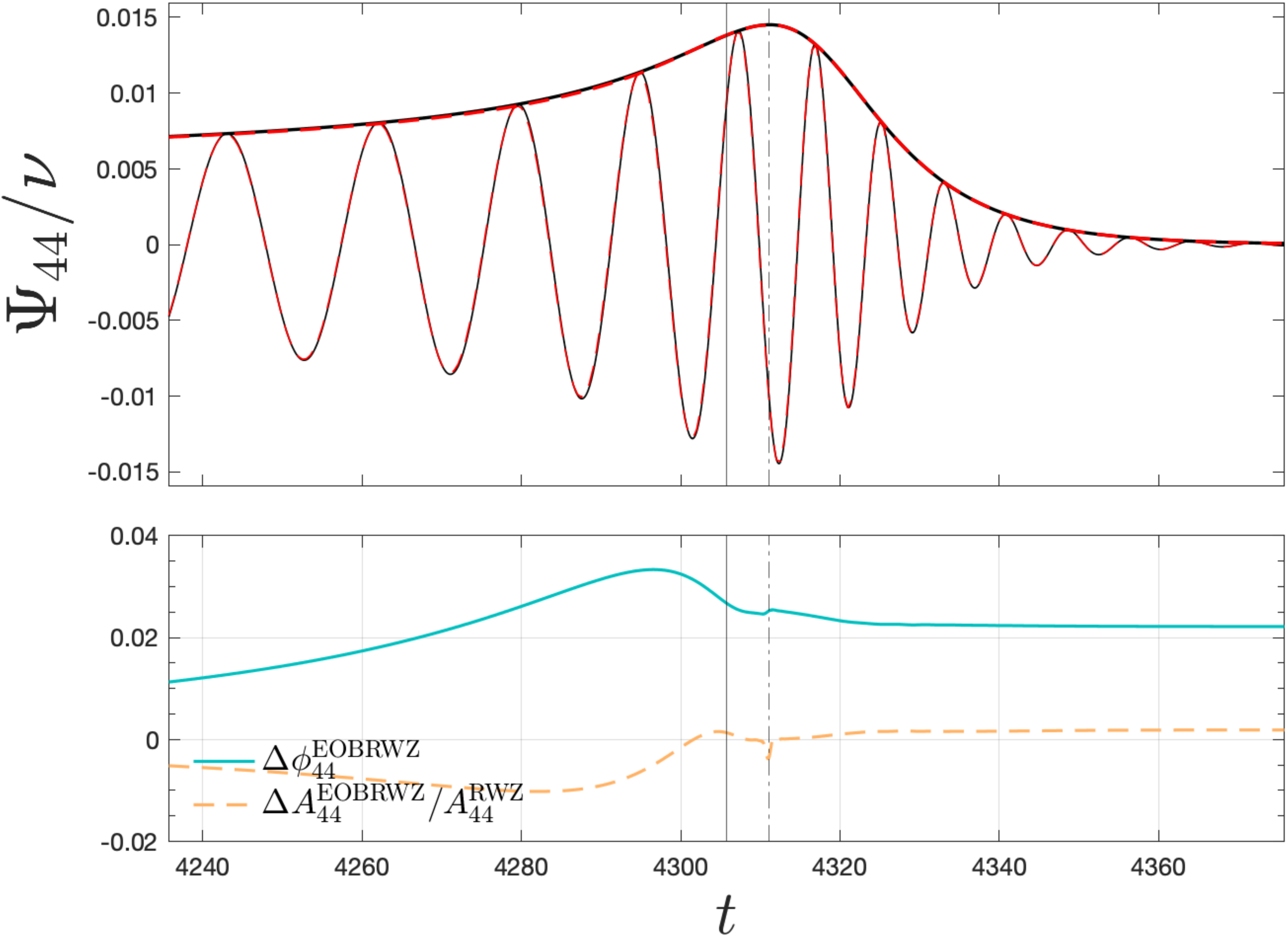}
	\includegraphics[width=0.32\textwidth,height=4.4cm]{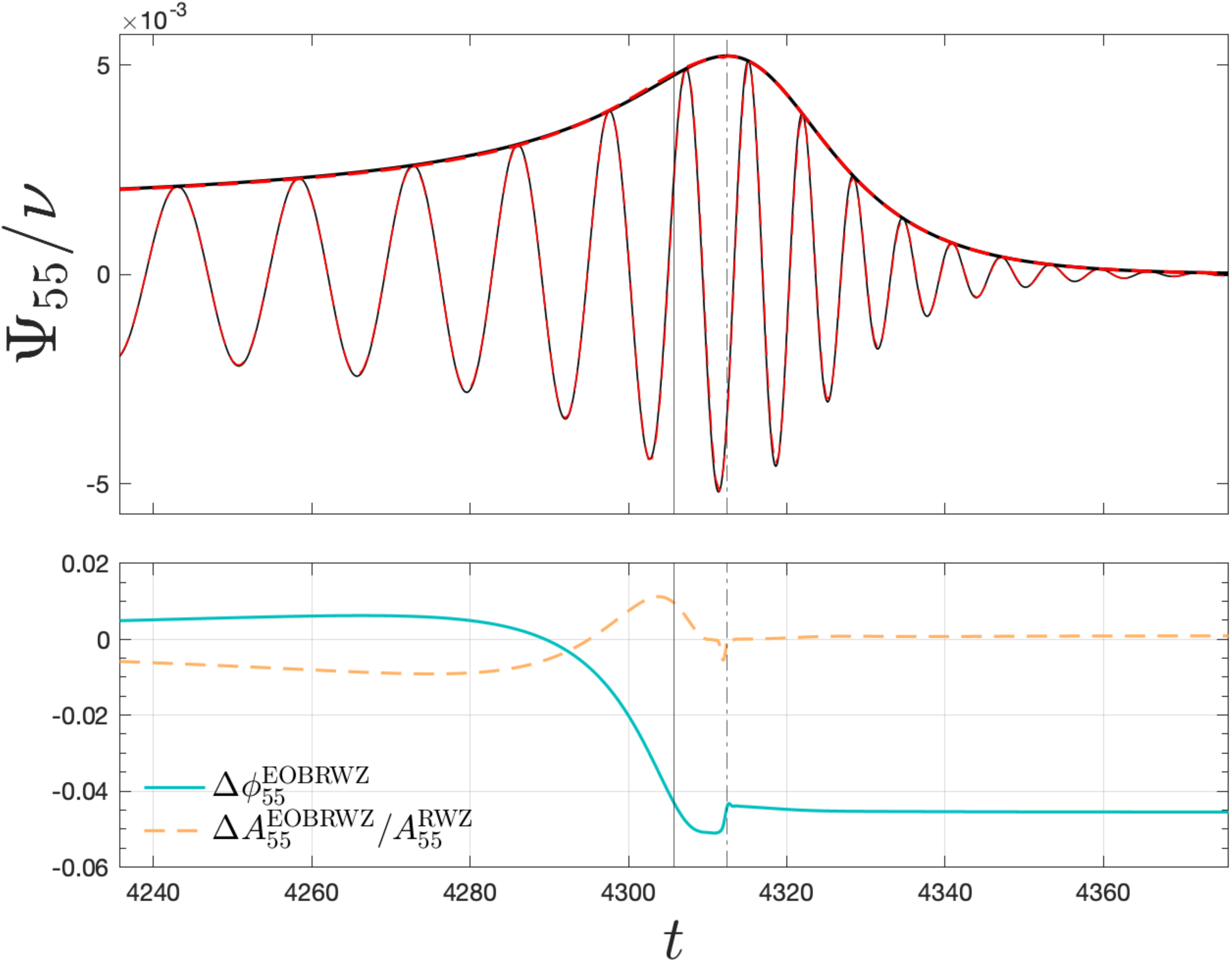} \\
	\vspace{0.1 cm}
	\includegraphics[width=0.32\textwidth,height=4.2cm]{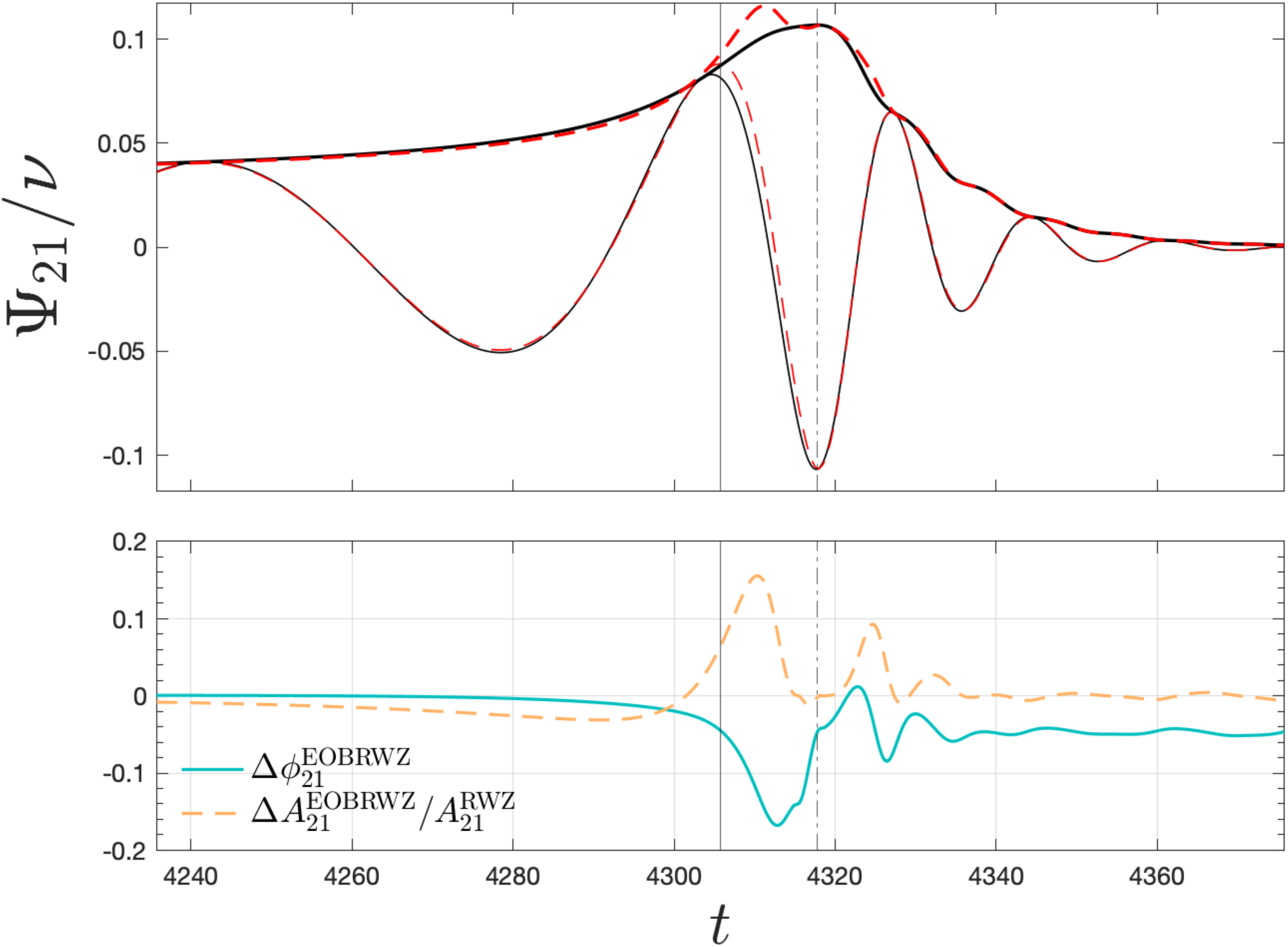}
	\includegraphics[width=0.32\textwidth,height=4.2cm]{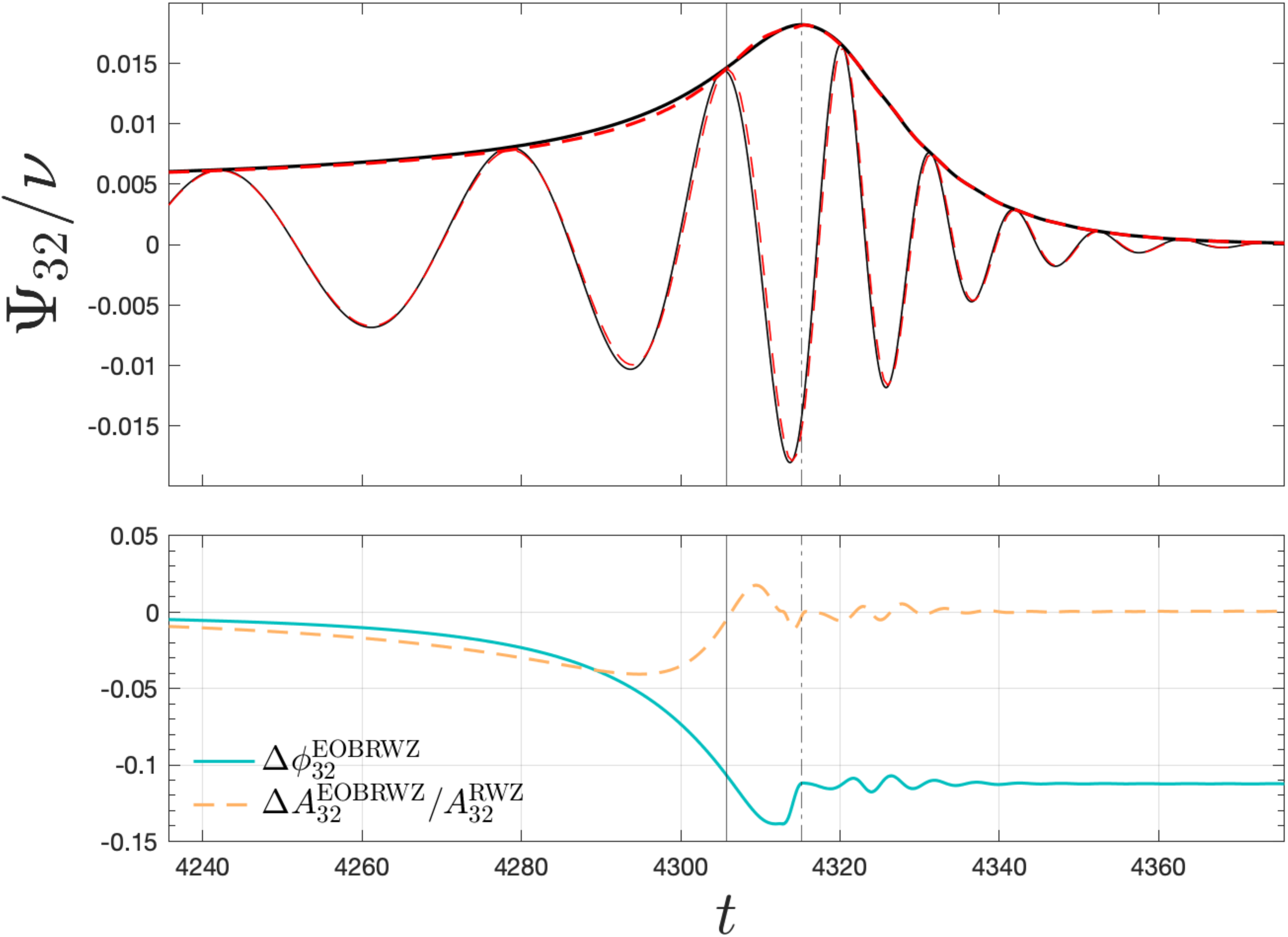}
	\includegraphics[width=0.32\textwidth,height=4.4cm]{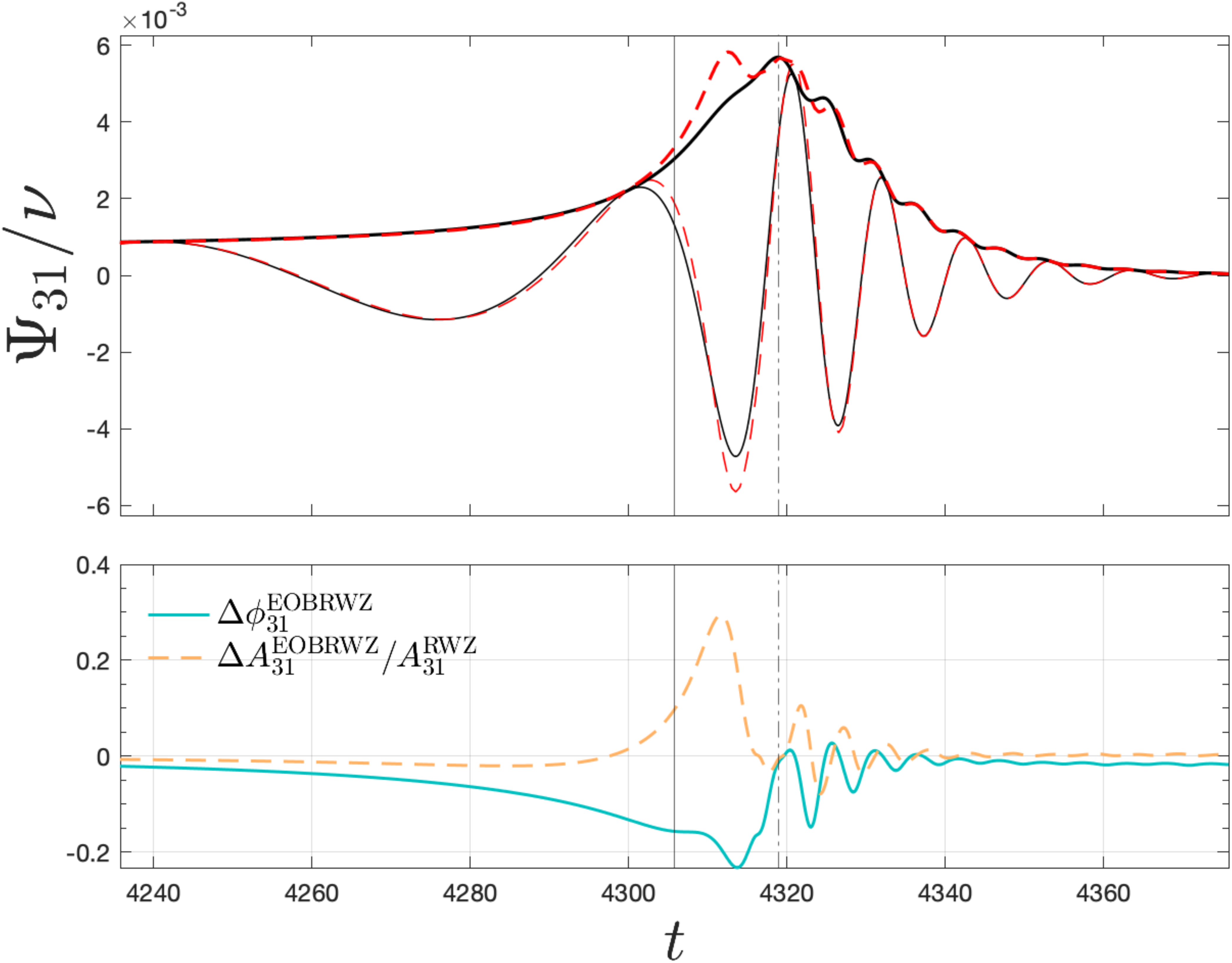}\\
	\vspace{0.1 cm}
	\includegraphics[width=0.32\textwidth,height=4.2cm]{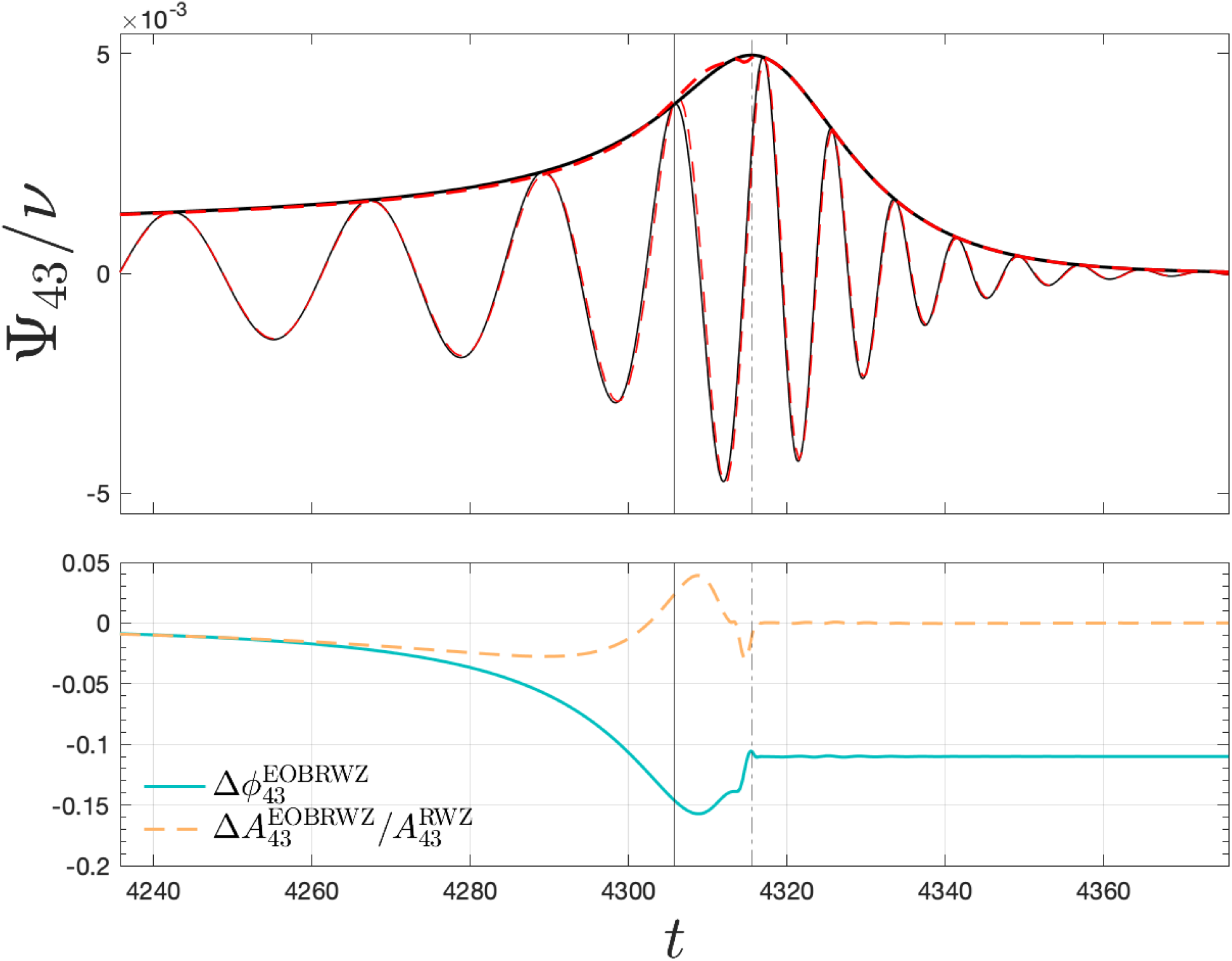}
	\includegraphics[width=0.32\textwidth,height=4.2cm]{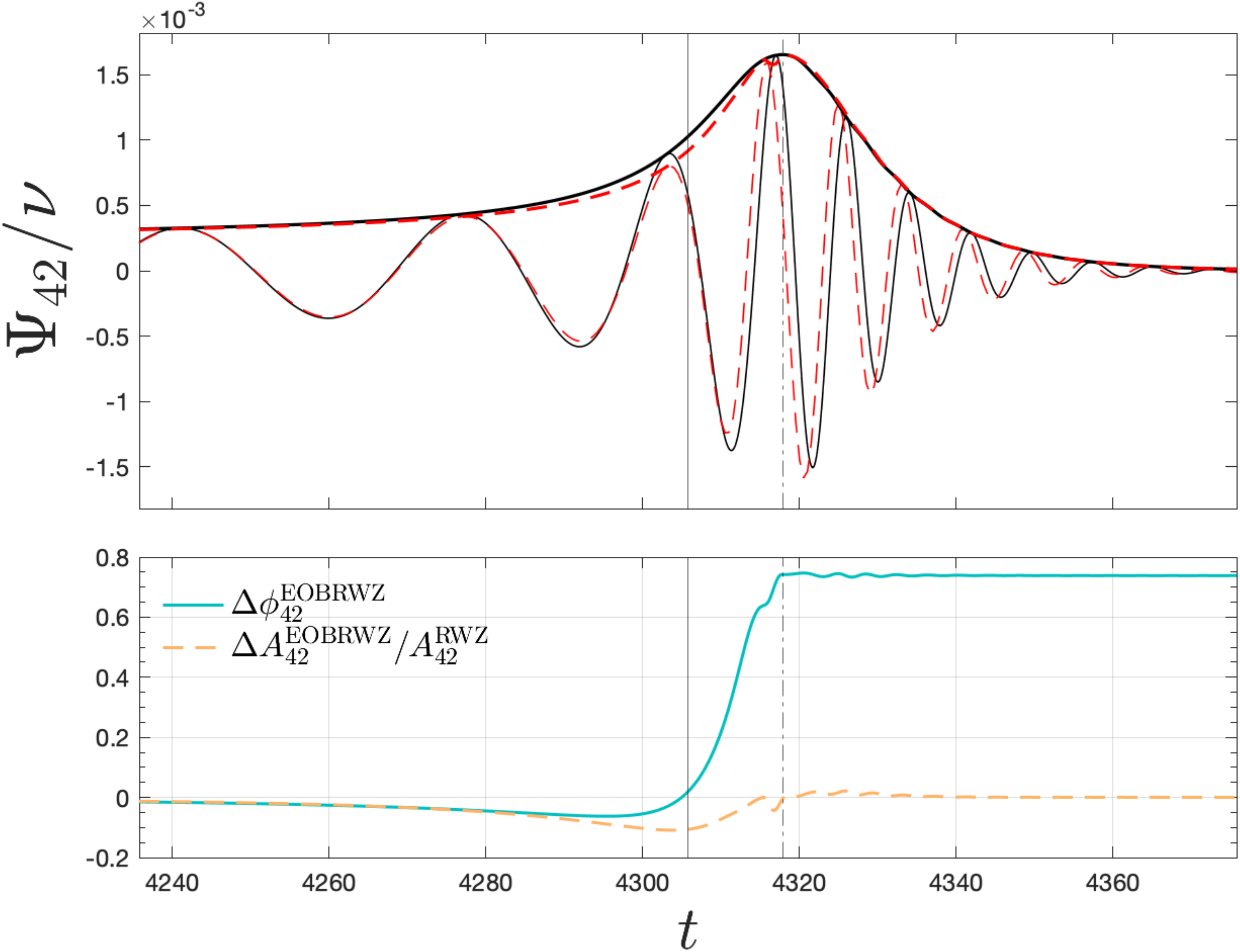}
	\includegraphics[width=0.32\textwidth,height=4.2cm]{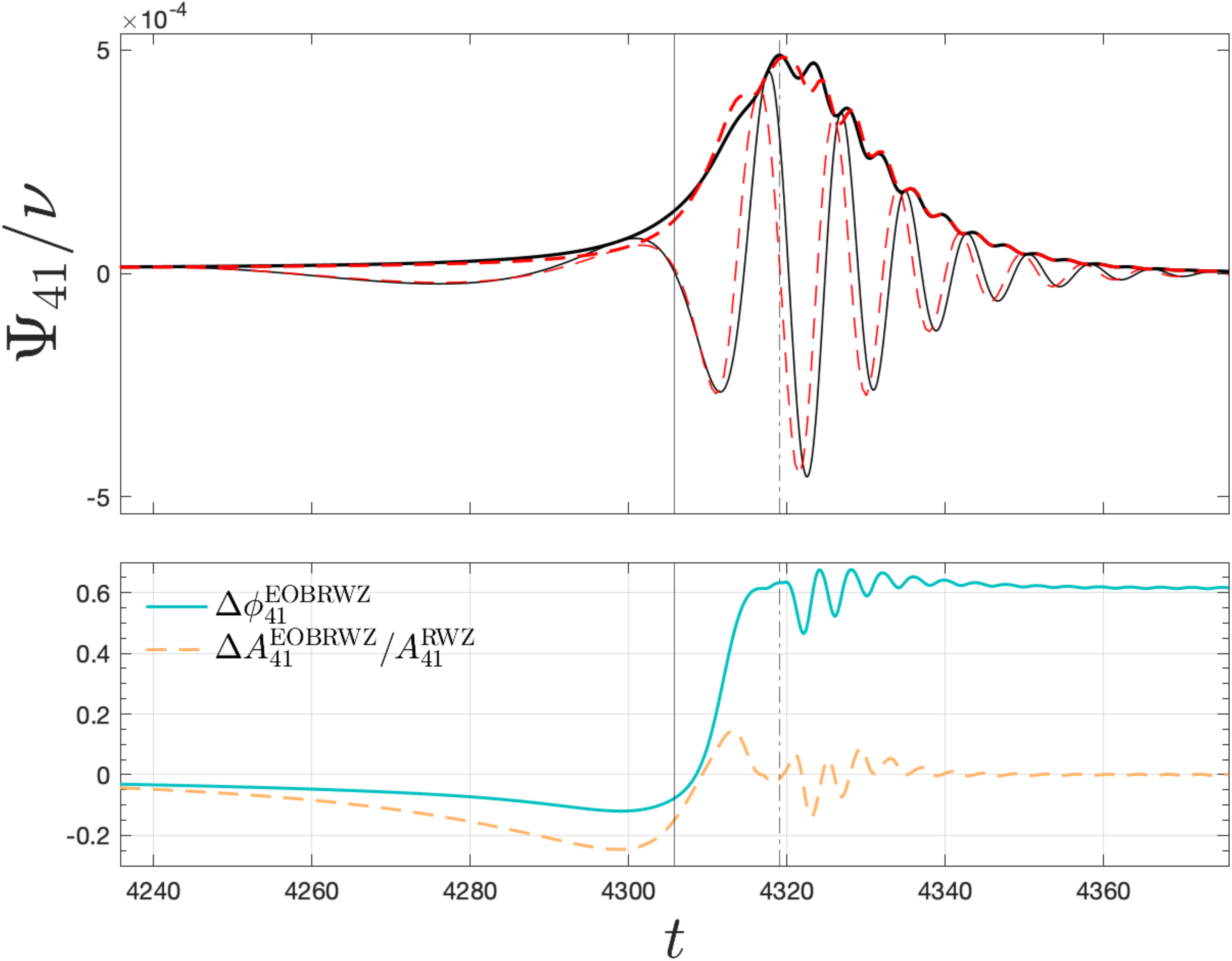}
	\caption{\label{fig:inspl_HM}Quasi-circular configuration, higher modes, EOB/RWZ comparisons.
	The vertical solid black lines mark $\tA22$, while the dash-dotted ones mark $\tAlm$.
	Bottom panels: relative amplitude difference (orange, dashed) and phase difference (light blue).
	For each mode, the NQC corrections are determined according to the best prescription selected
	in Fig.~\ref{fig:testHM_l4m4}.}
	\end{center}
\end{figure*}

Modeling higher modes correctly using NQC corrections determined using the standard
paradigm implemented in \TEOBResumS{} might be tricky. The main issue is that the amplitude 
peak of each mode is always delayed than the $(2,2)$ one~\cite{Bernuzzi:2010ty}, 
as reminded in Table~\ref{tab:Dtlm}. 
In the discussion above we have seen that the inclusion of the generic noncircular prefactor of Eq.~\eqref{eq:genericNP}
improves the EOB/RWZ agreement for the $(2,2)$ mode.
We now proceed to evaluate the relevance of this term also for the other multipoles considering
an illustrative higher mode, like the $(4,4)$. Figure~\ref{fig:testHM_l4m4} 
shows that the waveform with only the simple quasi-circular
factor (dash-dotted blue line) underestimates the waveform amplitude toward merger. 
Starting from this, it is not possible for the waveform NQC correction to improve 
the waveform behavior and assure a reliable matching to the ringdown, especially for the amplitude, 
as shown by the NQC-corrected quasi-circular waveform (dotted purple) in Fig.~\ref{fig:testHM_l4m4}. 
By contrast, one sees that when the noncircular factor of Eq.~\eqref{eq:genericNP} is used, the waveform
visibly overestimates the waveform amplitude toward merger. This situation is preferable and 
can be easily corrected by the NQC correction, as shown by the waveform obtained 
computing the NQC correction at $t^{\rm NQC}_{44}=t_{A_{44}}^{\rm peak}+2$ (yellow solid line).
We also find that, in order to considerably improve the NQC corrections, 
the system of Eqs.~\eqref{eq:nqc_syst} has to be evaluated at 
$\tNQC<\tAlm$. For all higher modes, we chose
\be
\label{eq:tnqc-2}
\tNQC = \tAlm-2.
\ee
Therefore, to compute the rhs~of Eqs.~\eqref{eq:nqc_syst}, we need to fit
$\left\lbrace 
A^{\rm NQC}_\lm, \dot{A}^{\rm NQC}_\lm, \ddot{A}^{\rm NQC}_\lm,
\omega^{\rm NQC}_\lm, \dot{\omega}^{\rm NQC}_\lm, \ddot{\omega}^{\rm NQC}_\lm
\right\rbrace$ from RWZ data.  
The global fits are discussed in Appendix~\ref{sec:gfit_tables} and reported in Table~\ref{tab:gfits_nqc1} and~\ref{tab:gfits_nqc2}.
The (4,4) multipole with the generic Newtonian prefactor
and the NQC evaluated according to Eq.~\eqref{eq:tnqc-2} is shown with a red dashed line 
in Fig.~\ref{fig:testHM_l4m4}.
As can be seen, both the amplitude and the frequency improves near the matching time with respect to the 
waveforms with NQC computed at $t^{\rm NQC}_{44}=t_{A_{44}}^{\rm peak}+2$.  

For $\l=m$ modes, we use the same prescriptions of the (2,2) mode, except for the fact that
the NQC corrections are computed before the peak amplitude, according to Eq.~\eqref{eq:tnqc-2}. 
The results for the (3,3), (4,4), and (5,5) modes are shown in the top row of Fig.~\ref{fig:inspl_HM}, where
we use the color black for the numerical waveform and frequency and red dashed lines for the complete 
EOB waveform.
The absolute value of phase difference, is always below $0.07$, $0.035$ and $0.05$ radians for the (3,3), 
(4,4), and (5,5) modes, respectively. 
The relative amplitude difference, instead, is around at most of the $1\%$ before the matching point for all the 
three cases. However, the amplitude difference in the late ringdown is around the $10^{-3}$ for the (3,3) and (4,4)
modes and even smaller ($8 \cdot 10^{-4}$) for the (5,5) mode. 

The $m<\l$ modes are shown in the middle and bottom rows of Fig.~\ref{fig:inspl_HM}.
In this case we do not consider the second-time derivative of the frequency in the NQC corrections. 
As for the higher modes with $\l=m$, the NQC are computed at $\tNQC=\tAlm-2$. However, for the $m<\l$
we also apply a downsampling and spline procedure in the interval $t\in [\tNQC,\tAlm]$ to improve the continuity
of the waveform. 
Indeed, since the NQC corrections are determined at $\tNQC$,
the waveform could be discontinuous at $\tAlm$, where the NQC-corrected plunge waveform is matched to the ringdown. 
The downsampling and subsequent spline-patching solves this issue.
While these higher modes are less accurate than the $\l=m$ ones, the phase agreement is still good and 
generally below the $0.2$ rad, with the exception of the (4,2) and (4,1) modes that are more de-phased.
The degradation of the accuracy is strictly linked to the higher delay of the matching point (i.e.~$\tAlm$);
see Table~\ref{tab:Dtlm}. 
However, these modes are not relevant as the others in 
the complete strain, that can be computed using Eq.~\eqref{eq:strain}. In the first
panel of Fig.~\ref{fig:inspl_strain} we show the strain 
for the observational direction $(\Theta, \Phi) = (\pi/4, 0)$ computed considering all the modes shown so far, 
both in the numerical and analytical strain.
During the inspiral, the relative amplitude difference reaches at most the $2\cdot 10^{-3}$,
while the absolute value of the phase difference never exceeds the $2\cdot 10^{-3}$ radians.
The differences oscillate more in the ringdown; the amplitude difference reaches at most the $8\%$ in the early 
ringdown, while the phase difference reaches at most $0.06$ radians. 
Note however that, on average, during the ringdown both the amplitude and the phase difference are much smaller.
\begin{figure*}
	\begin{center}
	\includegraphics[width=0.32\textwidth]{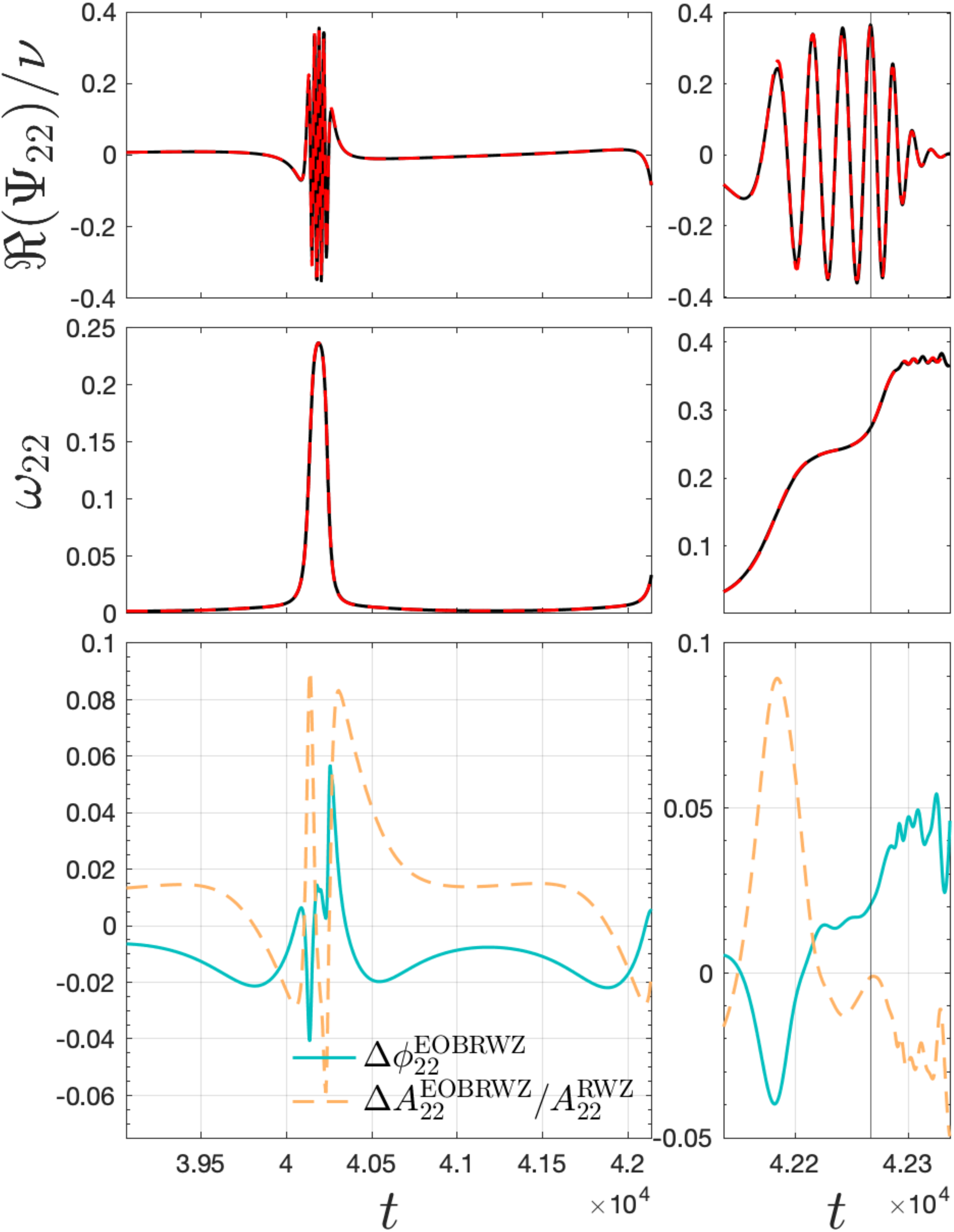}
	\hspace{0.1cm}
	\includegraphics[width=0.32\textwidth]{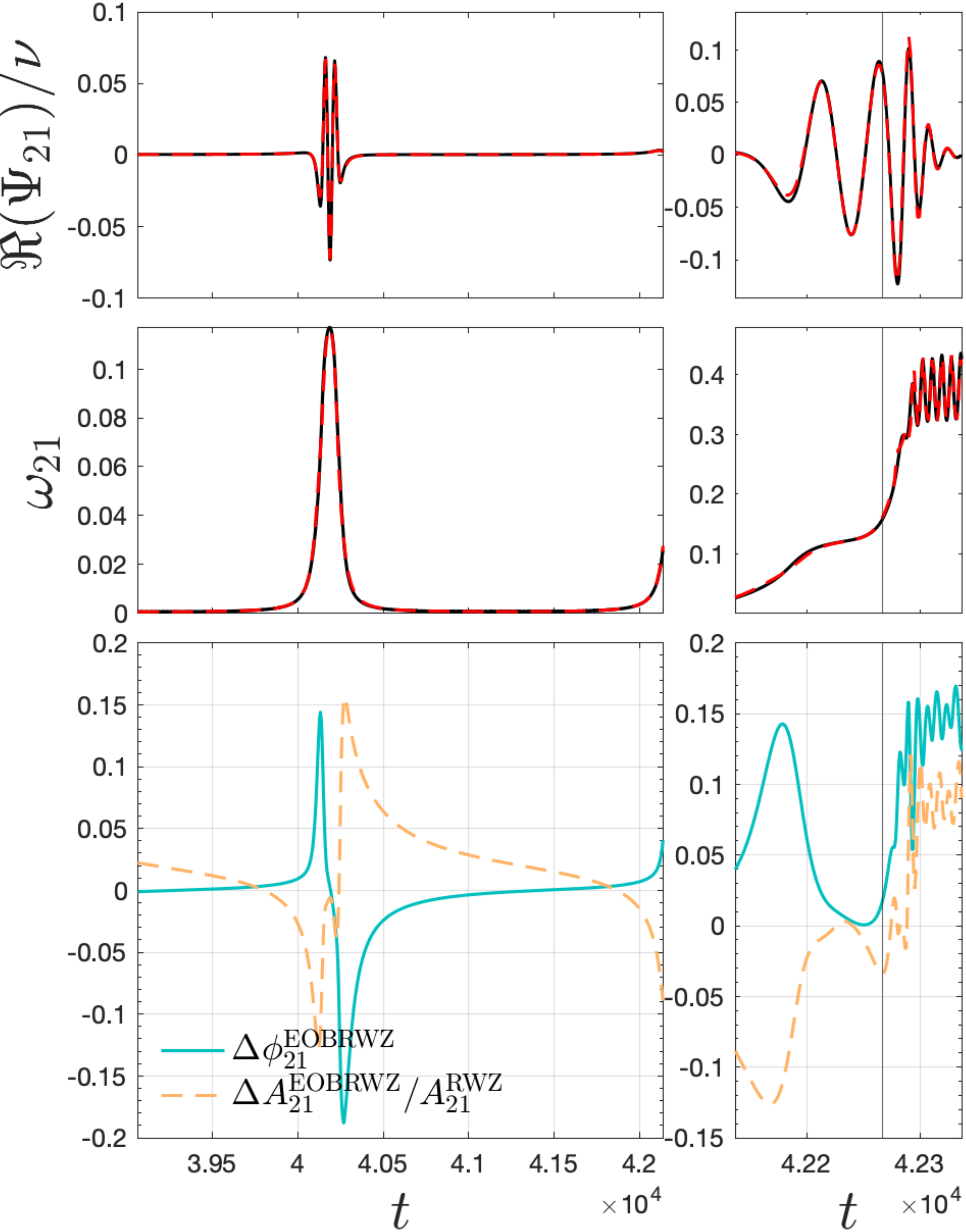}
	\hspace{0.1cm}
	\includegraphics[width=0.32\textwidth]{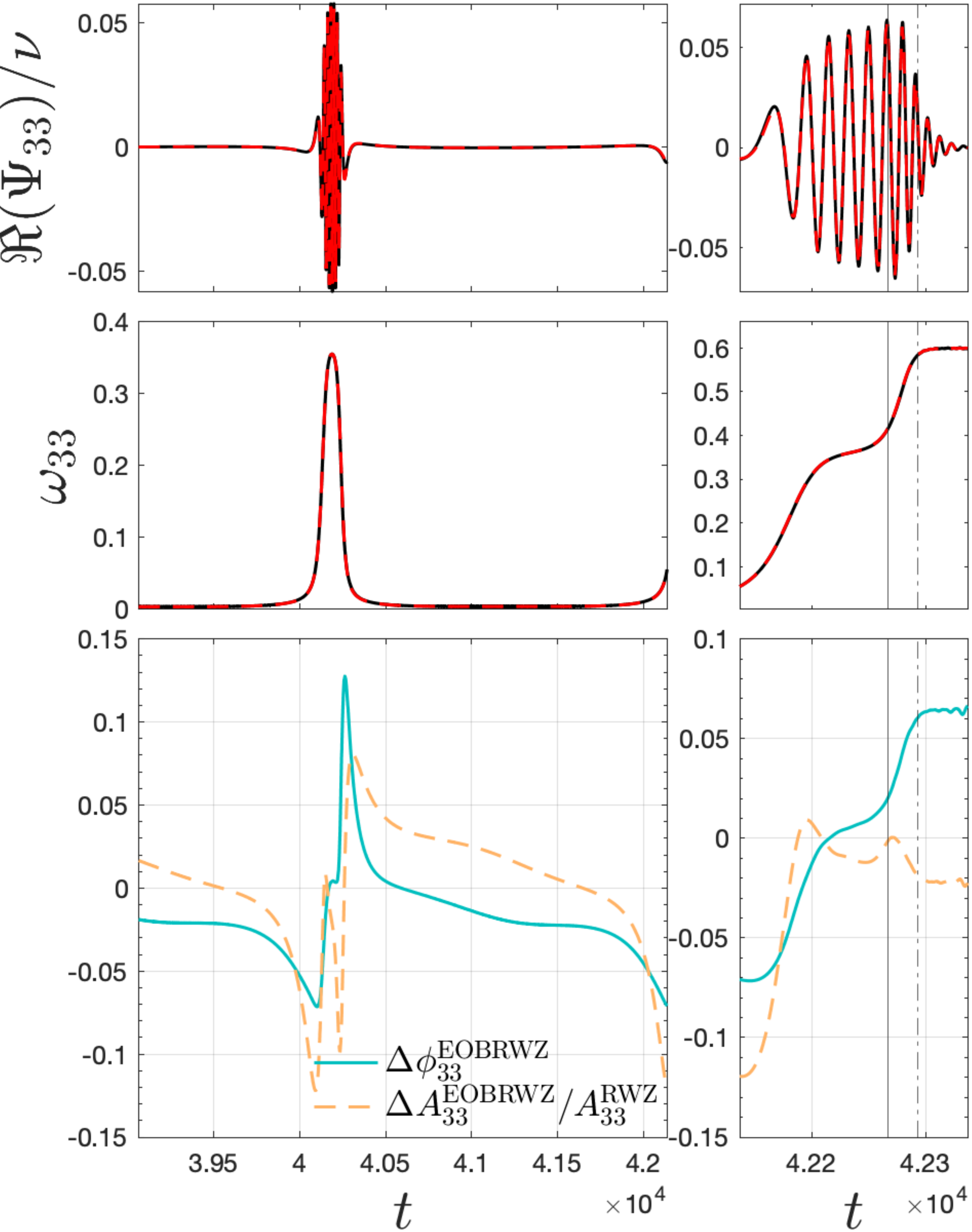}
	\caption{\label{fig:inspl_e095} Analytical/numerical comparisons for the (2,2), (2,1) and (3,3) multipoles
	for the configuration with initial eccentricity $e_0=0.95$.
	We show the real part and the frequency of the RWZ waveform (black) and the complete EOB waveform (red, dashed).
	In the bottom panels we show the relative amplitude difference (orange dashed)
	and the phase difference in radians (light blue). The vertical solid lines mark the peak of $A_{22}$, while
	the dash-dotted ones mark $\tAlm$.}
	\end{center}
\end{figure*}
\begin{figure*}
	\begin{center}
	\includegraphics[width=0.32\textwidth]{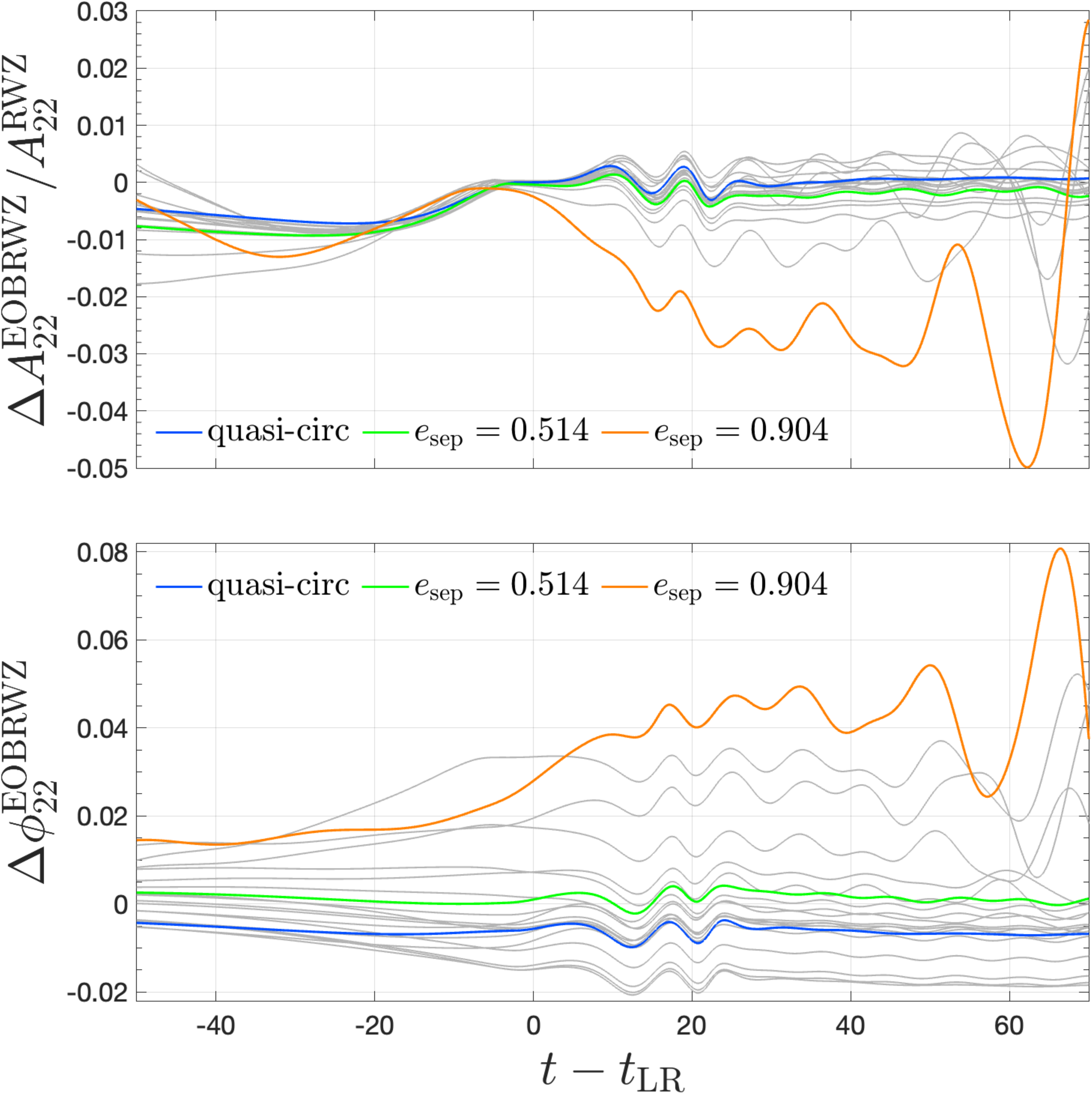}
	\hspace{0.1cm}
	\includegraphics[width=0.32\textwidth]{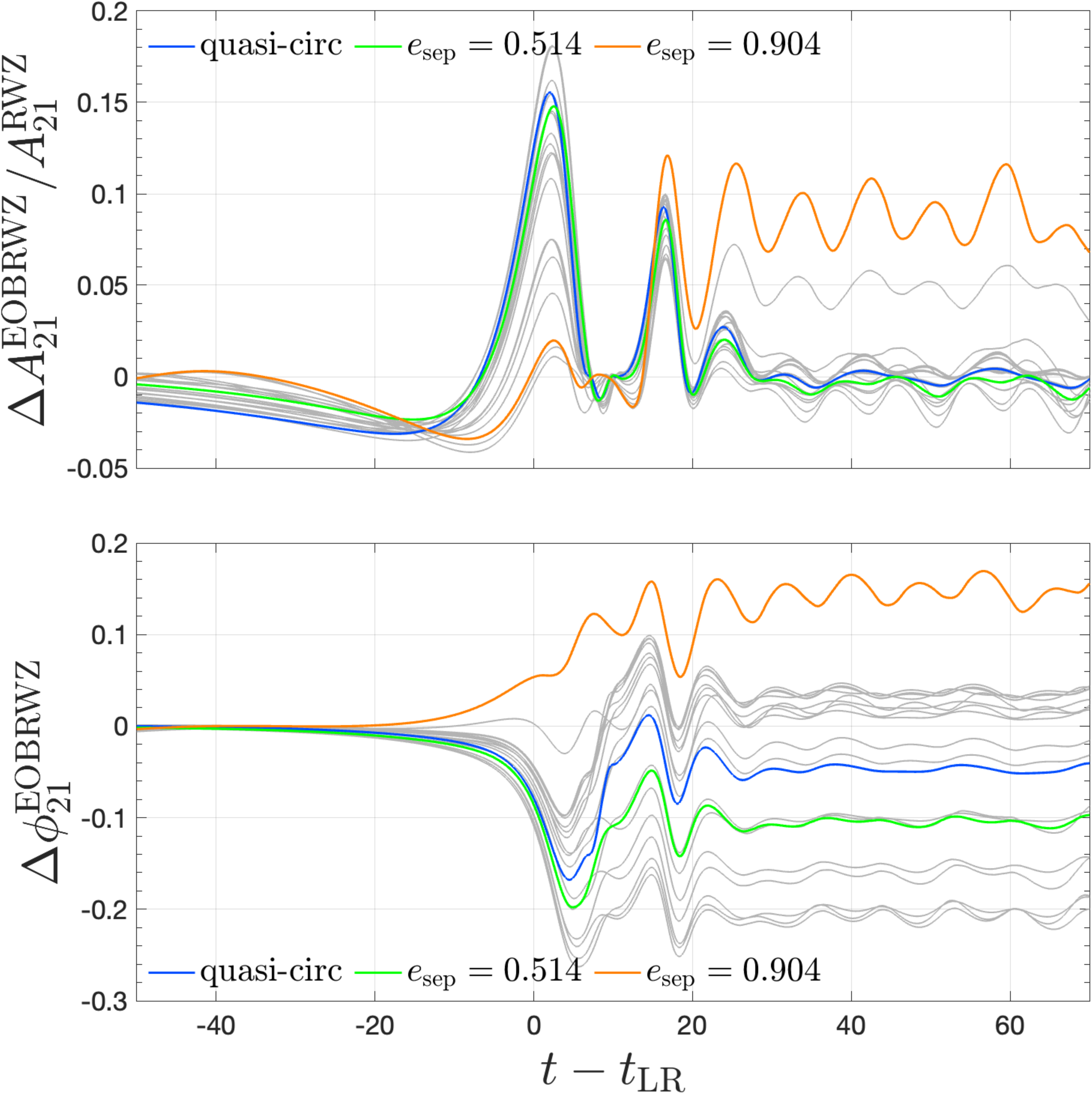}
	\hspace{0.1cm}
	\includegraphics[width=0.32\textwidth]{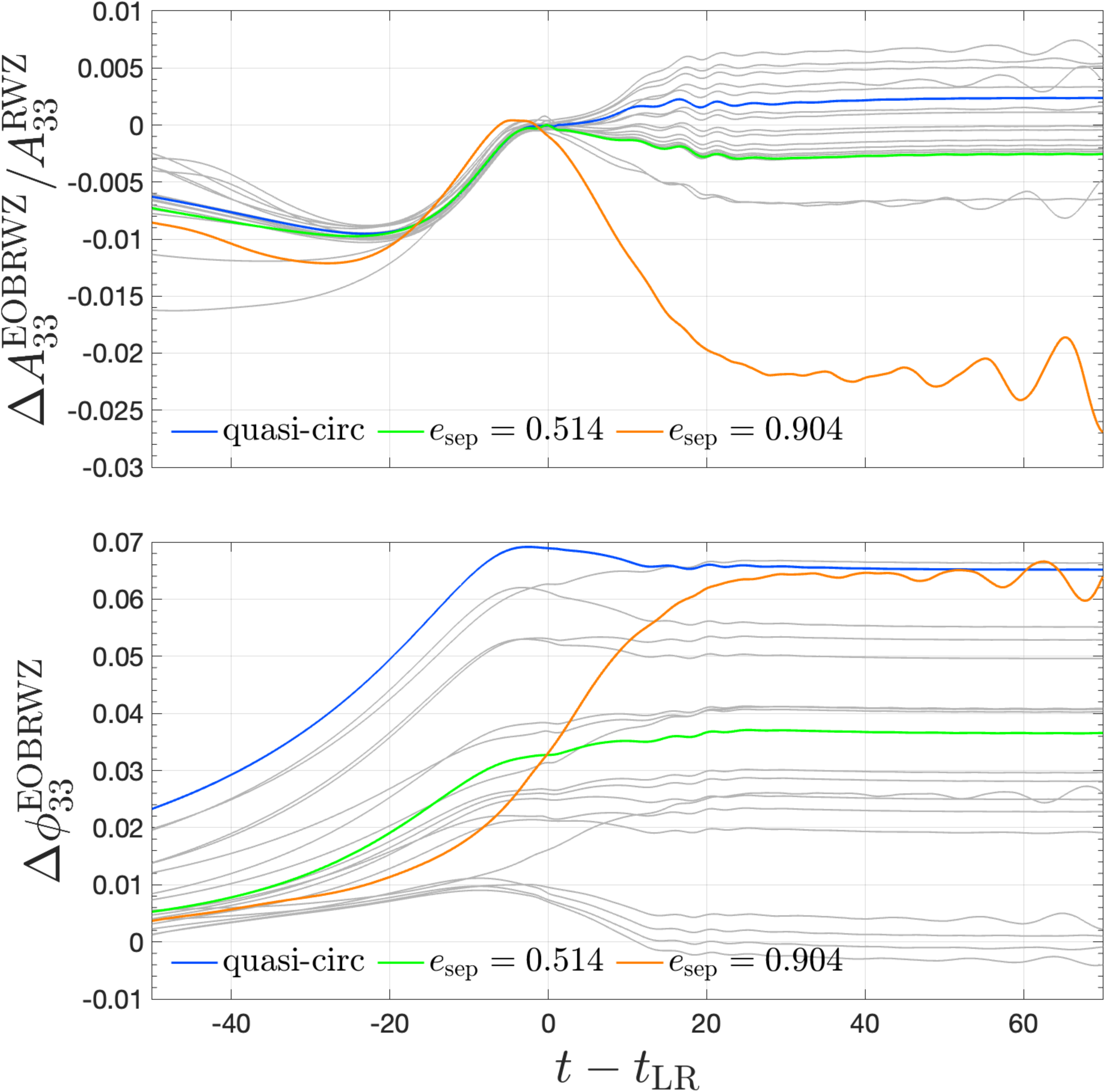}
	\caption{\label{fig:ecc_diffs} Analytical/numerical relative difference for the amplitude (upper panels) and 
	phase difference (bottom panels) for the plunge-ringdown of all the configurations considered in this work. 
	Modes (2,2), (2,1), and (3,3).
	We highlight the quasi-circular configuration (blue) and the ones with 
	$e_0=\left\lbrace 0.55,0.95 \right\rbrace$ (green and orange, respectively).}
	\end{center}
\end{figure*}
\begin{figure*}
	\begin{center}
	\includegraphics[width=0.32\textwidth]{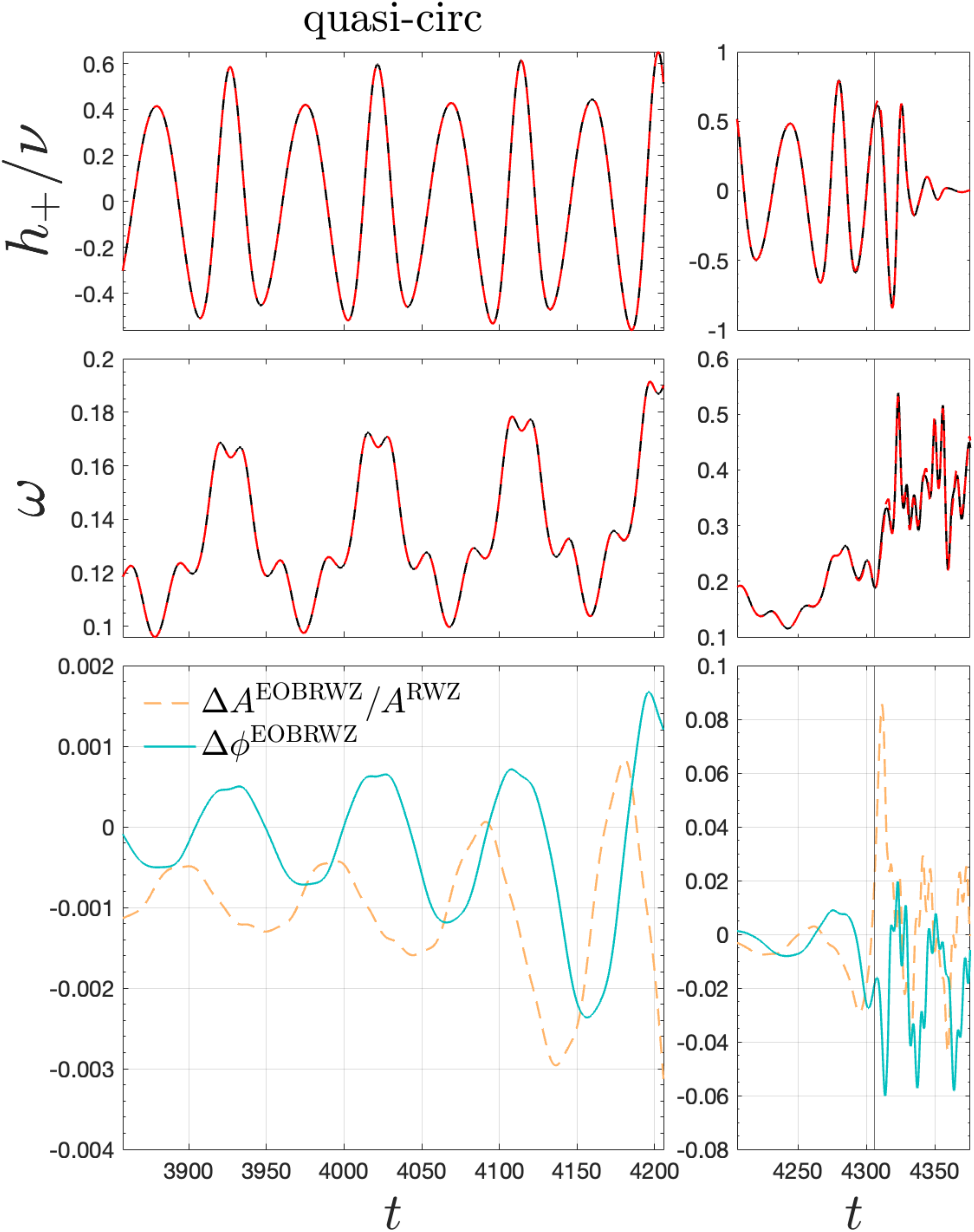}
	\includegraphics[width=0.32\textwidth]{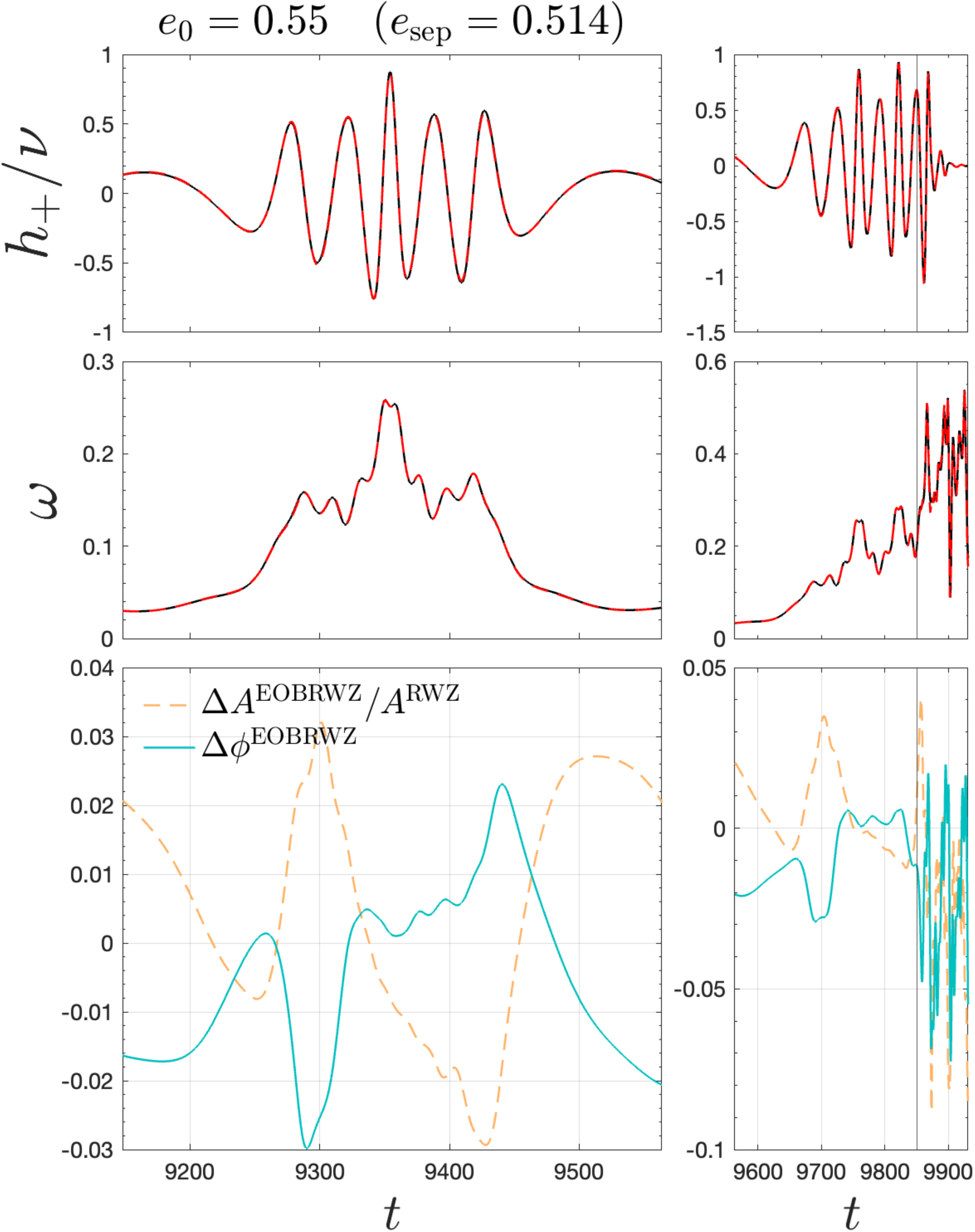}
	\includegraphics[width=0.32\textwidth]{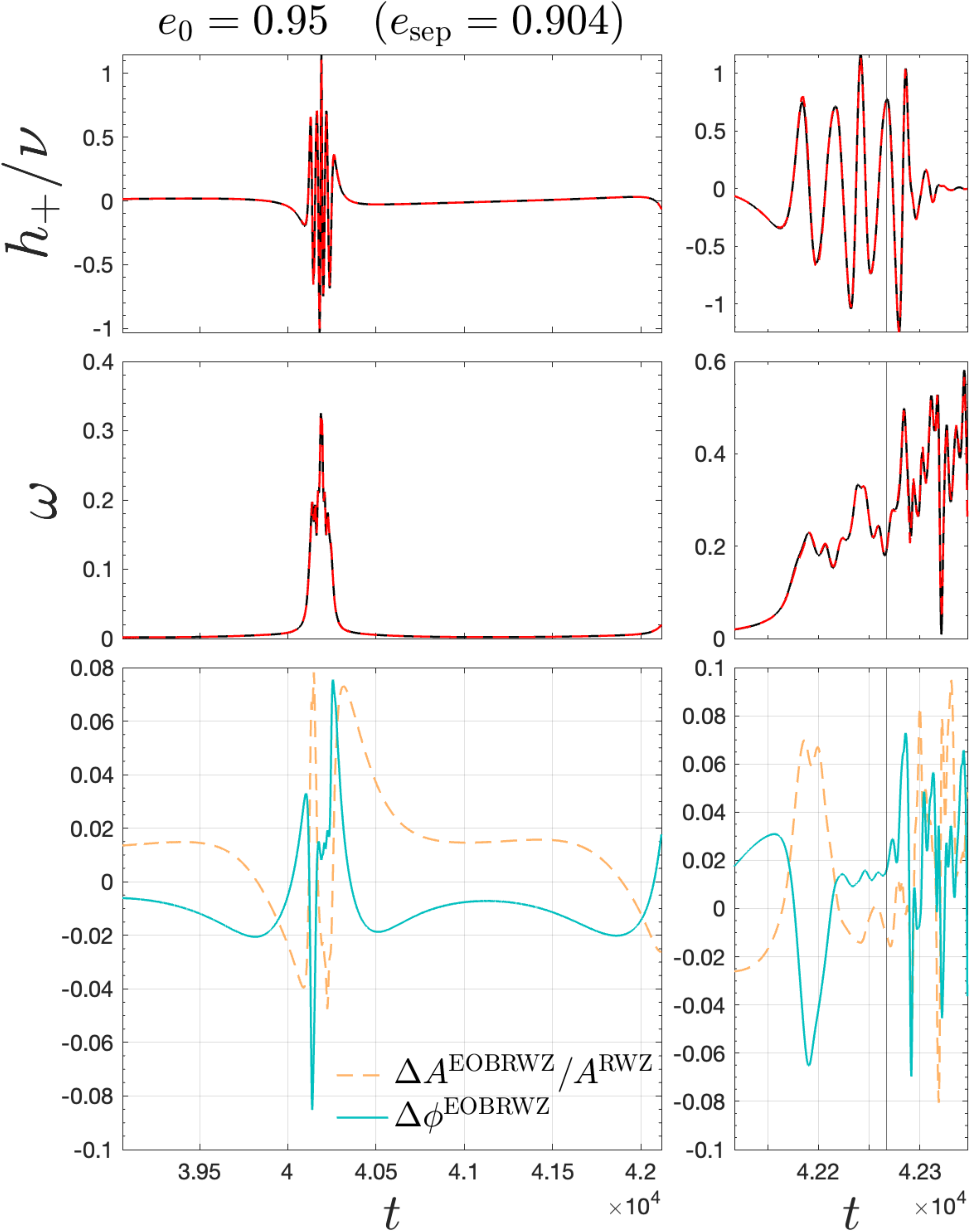}
	\caption{\label{fig:inspl_strain}EOB (red dashed) versus RWZ (black, solid) 
	strain comparison for  $e_0=\left\lbrace0, 0.55, 0.95\right\rbrace$. 
	The direction is  $(\Theta, \Phi) = (\pi/4, 0)$ 
	and all modes with $m>0$ are summed up to $\l=4$, plus the (5,5) one. 
	Bottom panels: relative amplitude difference and phase difference (in radians).  
	The vertical lines mark the peak of the quadrupolar amplitude. }
	\end{center}
\end{figure*}
%

\subsection{The eccentric case} 
\label{sec:ecc}
Let us move now to discussing the eccentric case. The same procedures considered optimal in
the quasi-circular case are retained also in the presence of eccentricity so to provide comparisons
with the eccentric configurations listed in Table~\ref{tab:ID}.
As an explicit example that efficiently summarizes the performance of the model all over the 
parameter space, Fig.~\ref{fig:inspl_e095} shows the EOB/RWZ performance for configuration
$\# 20$ of Table~\ref{tab:ID}, that corresponds to initial eccentricity $e_0=0.95$. 
Note that this configuration is not used in the global fits, as discussed in Sec.~\ref{sec:gfits}.
The figure includes modes $(2,2)$, $(2,1)$ and $(3,3)$.
The performance of the model all over the RWZ-covered points of the parameter space is assessed,
for the same waveform multipoles, in Fig.~\ref{fig:ecc_diffs}, that only reports relative amplitude differences
(top row plots) and phase differences (bottom row plots). Note that three specific configurations are
highlighted in color, so to point out that the performance of the model degrades (slightly) as eccentricity
is increased.
The simplest route to have a handle on the accuracy of all multipoles is simply to compare the strain for
EOB and RWZ. This is done in Fig.~\ref{fig:inspl_strain} for the quasi-circular case, for $e_0=0.55$ and
for $e_0=0.95$. 
It is interesting to note that, despite the EOB performance during the (eccentric) inspiral degrades with
eccentricity, as expected, due to the lack of high-order corrections (see Refs.~\cite{Placidi:2021rkh,Albanesi:2022xge}), 
the behavior during merger and ringdown is practically comparable among the three cases.

\subsection{Improved description of ringdown for $m\neq \ell $ modes} 
%
\begin{figure*}
	\begin{center}
	\includegraphics[width=0.32\textwidth]{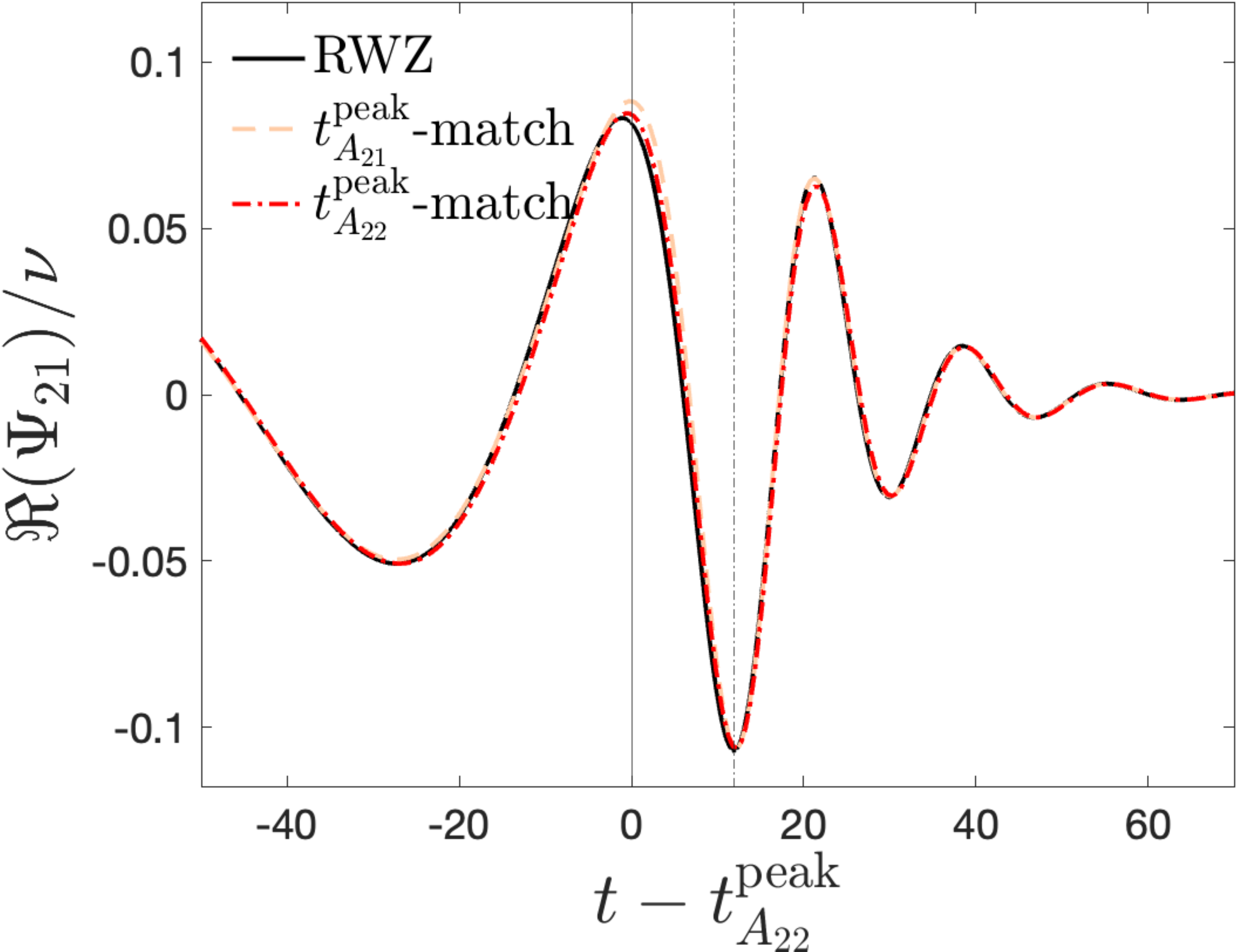}
	\includegraphics[width=0.32\textwidth]{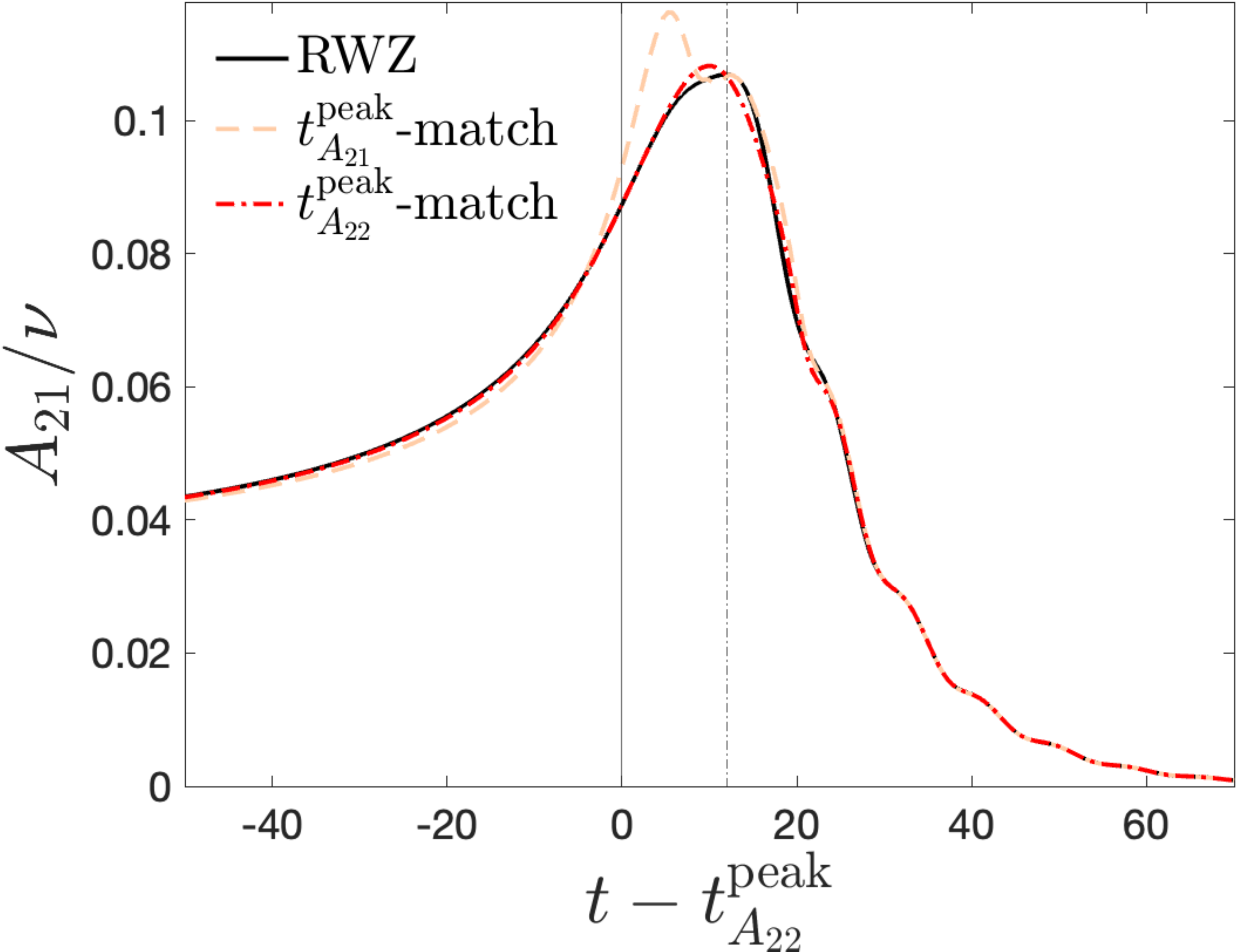}
	\includegraphics[width=0.32\textwidth]{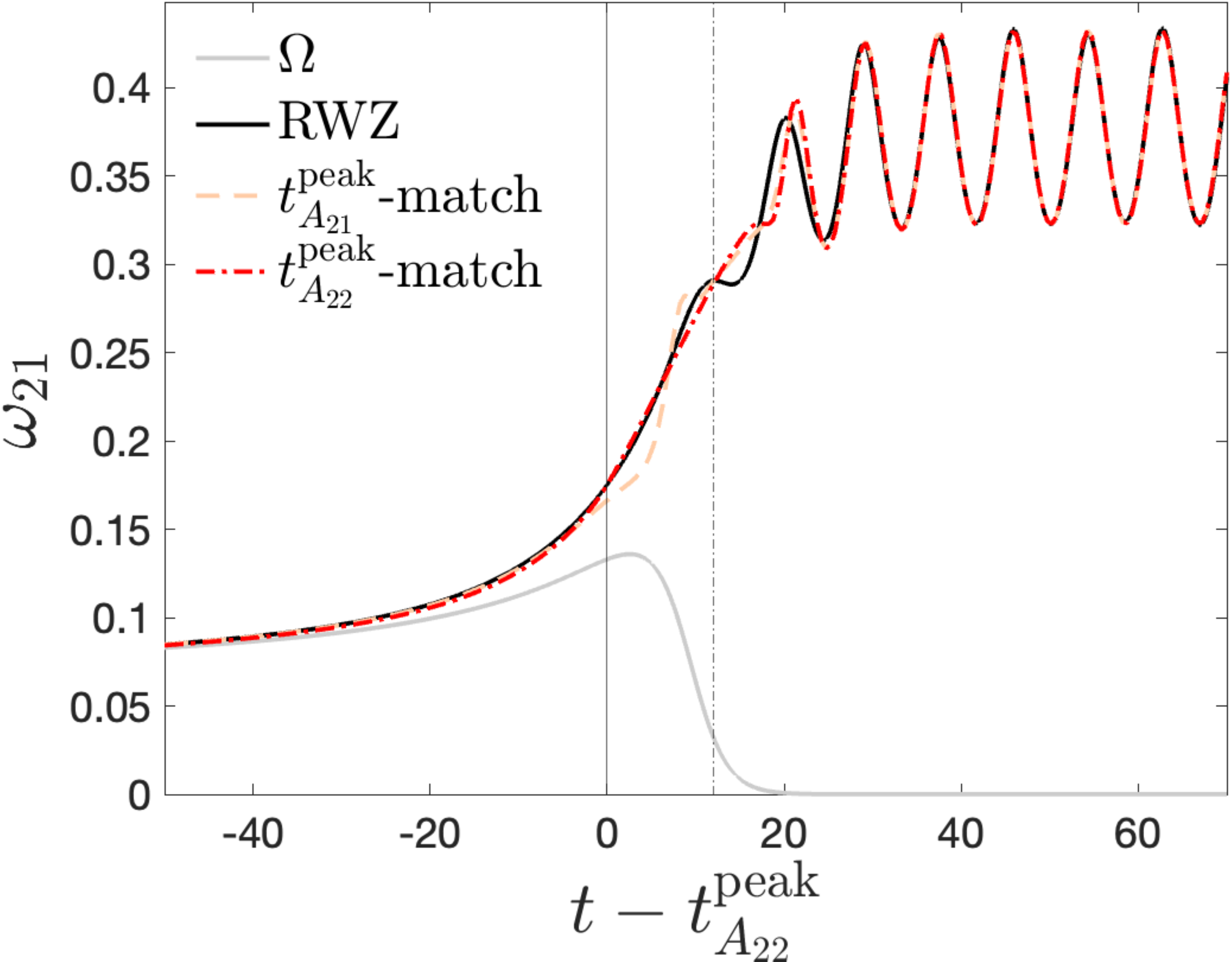}\\
	\includegraphics[width=0.32\textwidth]{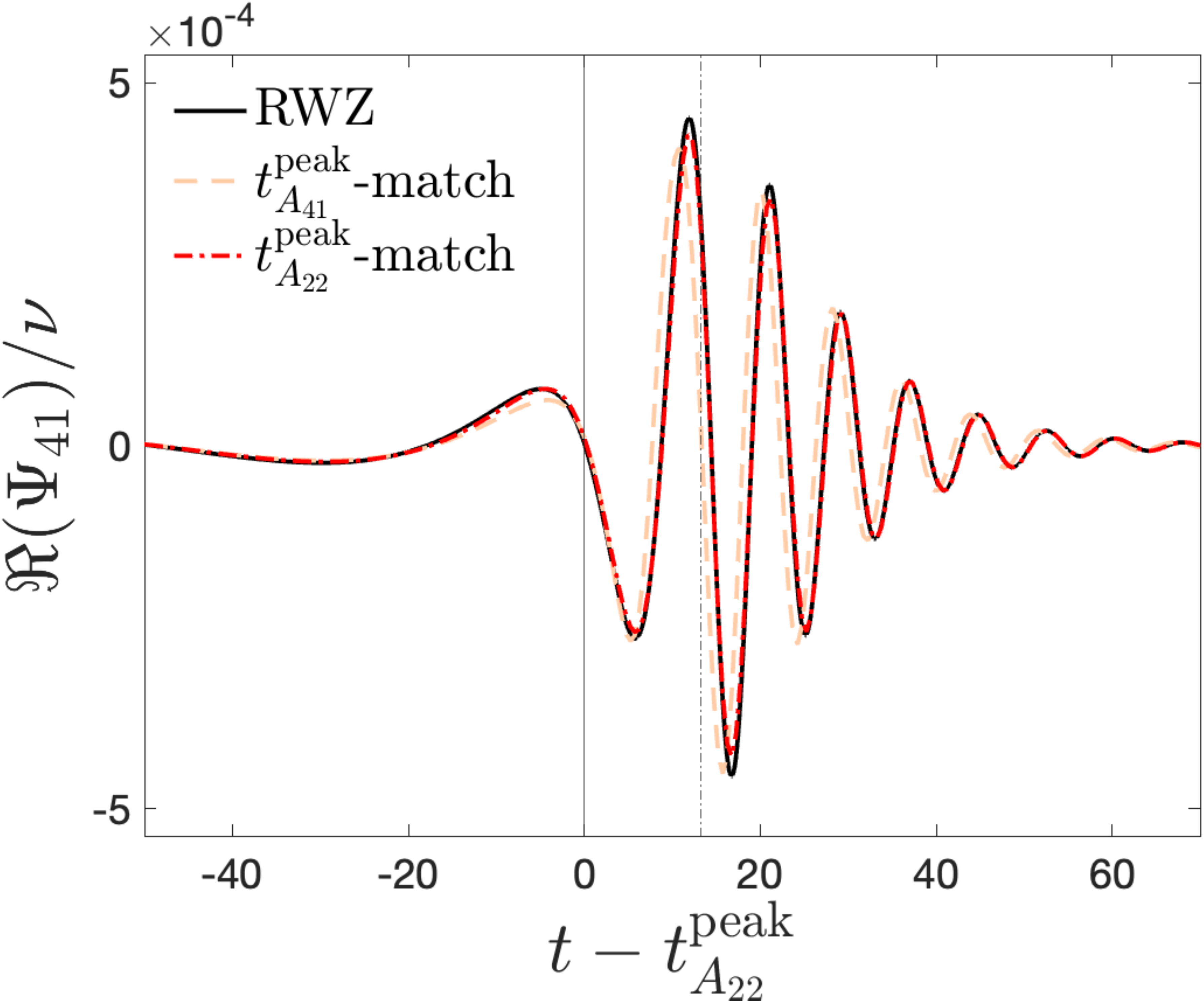}
	\includegraphics[width=0.32\textwidth]{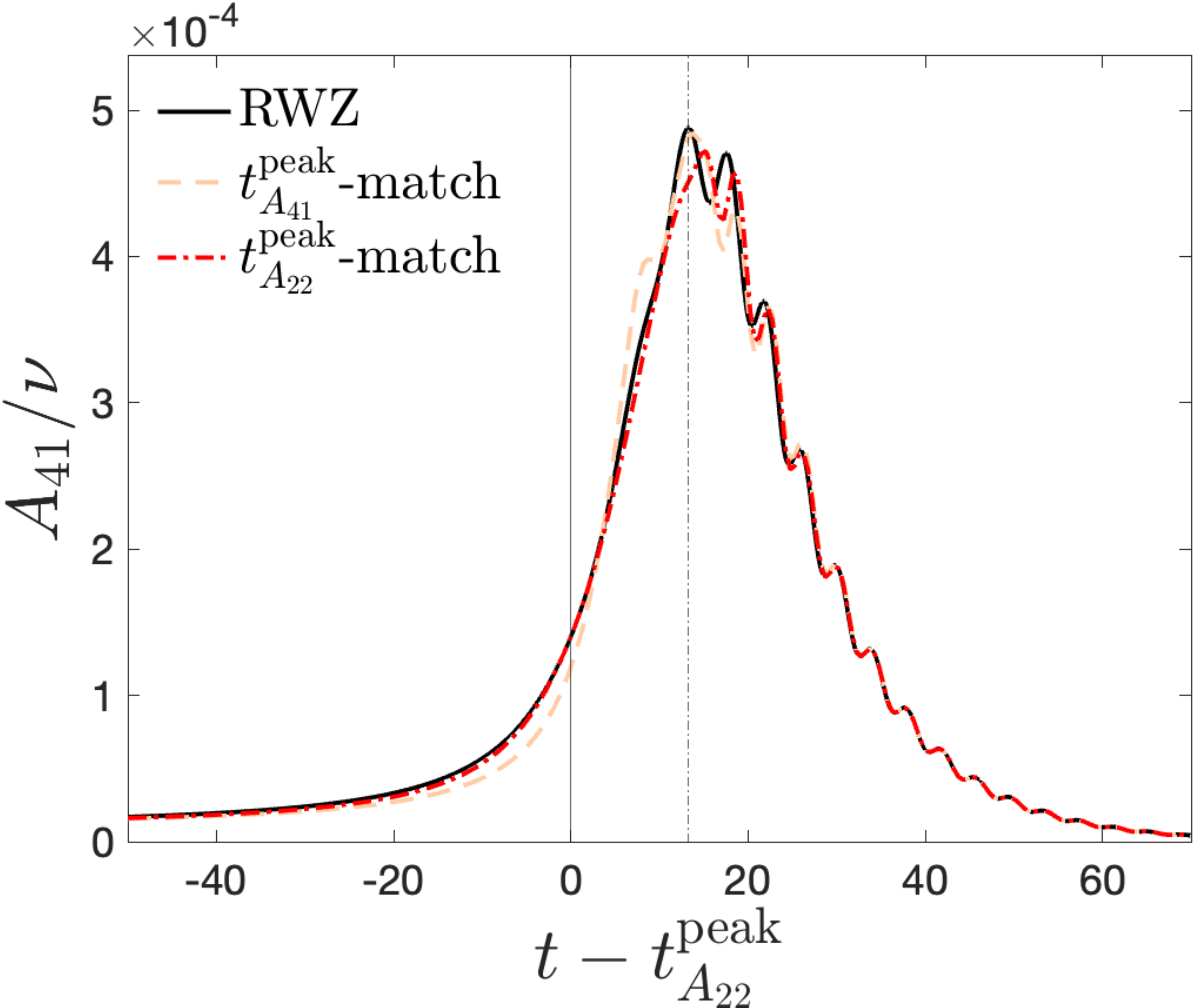}
	\includegraphics[width=0.32\textwidth]{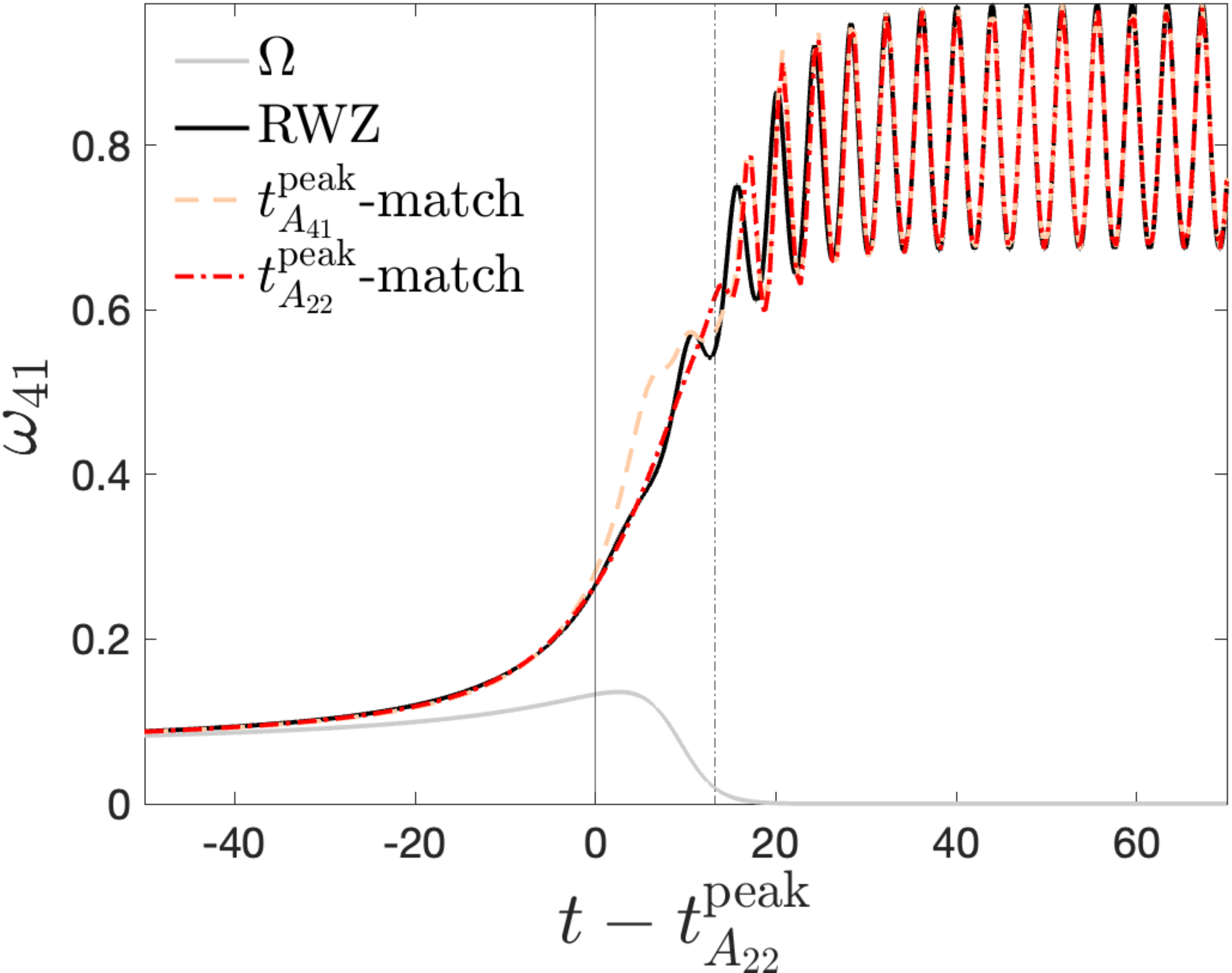}\\
	\caption{\label{fig:new_rng_schw}Nonspinning case. New ringdown (red dashed) contrasted with the standard
	one discussed above (orange, dashed) for the $(2,1)$ and $(4,1)$ modes. 
	The RWZ ringdown waveform is fitted from $\tA22$ (solid vertical lines) 
	rather than from $\tAlm$ (dash-dotted vertical lines).
	This ensures a more accurate waveform description around $\tAlm$.}
	\end{center}
\end{figure*}
%
%
\begin{figure}
	\begin{center}
	\includegraphics[width=0.48\textwidth]{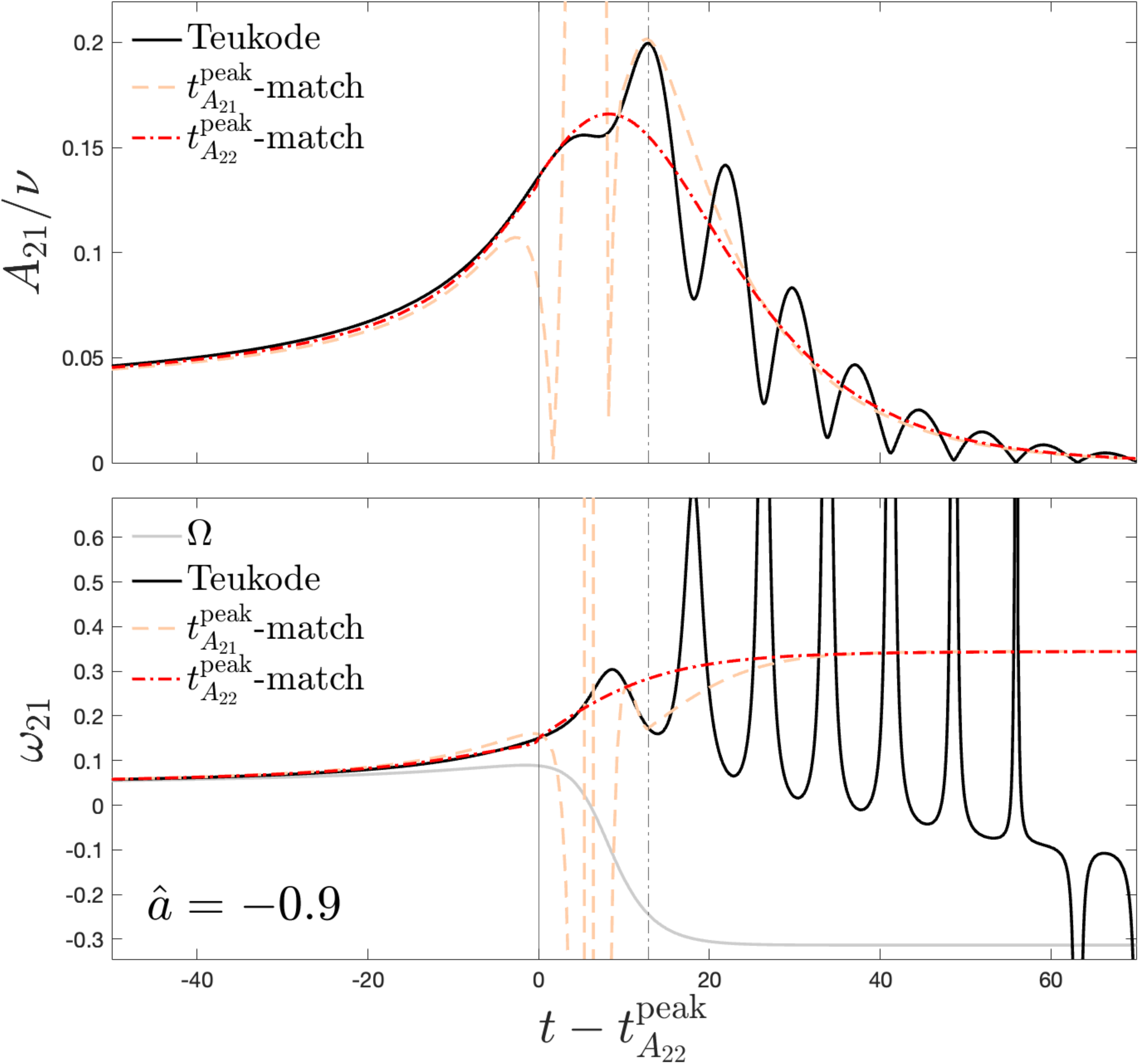}
	\caption{\label{fig:new_rng_kerr}Spinning case with Kerr parameter $\hat{a}=-0.9$. 
	The NQC and ringdown determined from  $\tA22$ instead of $\tAlm$ allow for an excellent
	EOB/Teukode agreement. Note that the mode mixing in this case is not implemented in
	the EOB waveform.}
	\end{center}
\end{figure}
So far, we have seen that the EOB waveform model gives more than satisfactory results
also for the higher modes. However, if one carefully inspects the $m=1$ EOB modes in the 
quasi-circular case (see Fig.~\ref{fig:inspl_HM}), one sees the ubiquitous presence of
a bump before the actual amplitude peak. Interestingly, this feature occurs in all modes 
and it is related to the NQC amplitude correction. In general, this is also true for other
modes with $m\neq \ell$, though the effect is less visible.
To overcome this difficulty, we decided to explore a different way to model the ringdown 
for higher modes. The procedure is substantially the same used  in the EOB 
models of the \texttt{SEOBNR}-family, in particular {\tt SEOBNRv4PHM}~\cite{Cotesta:2018fcv,Ossokine:2020kjp}
and {\tt SEOBNRv5PHM}~\cite{Pompili:2023tna,Ramos-Buades:2023ehm}, although the
fitting template is different from the one used there. This approach is crucial to obtain 
very reliable waveforms when $\tAlm$ is far from $\tA22$, i.e.~situations where the 
NQC corrections cannot guarantee a reliable match of the plunge waveform to the
merger ringdown one modeled after $\tAlm$. The main idea is to have the RWZ-informed
part of the waveform starting directly from $\tA22$ (and not from $\tAlm$) for all modes
with $\ell\neq m$.
To do so, we consider the QNM-rescaled waveform similar to the one of Eq.~\eqref{eq:barh}, 
but where the time is shifted using $\tA22$ for each $(\l,m)$ mode. It thus reads
\begin{equation}
\label{eq:barh_mod}
\bar{h}_\lm(\bar{\tau}) = e^{\sigma_{\lm 1}^+\bar{\tau} + i \phi_\lm^{0}} h^{\rm rng}_\lm(\bar{\tau}) \ ,
\end{equation}
where $\bar{\tau} \equiv t - \tA22$ and $\phi_\lm^{0}$ is the phase of the $(\l,m)$ multipole at $\tA22$. 
This rescaled waveform is then written as $\bar{h}^{\rm mod}_\lm(\bar{\tau}) = A_{\bar{h}} e^{i \phi_{\bar{h}}}$, where
the templates for $A_{\bar{h}}$ and $\phi_{\bar{h}}$ are given in Eqs.~\eqref{eq:templateA} and~\eqref{eq:templatePhi}.
We impose continuity conditions constraining 
$c_1^A$, $c_2^A$, $c_4^A$ and $c_1^\phi$:
\begin{align}
c_1^A    =& e^{-c_3^A} \left(e^{2 c_3^A}-1\right) c_5^A (\Amxq)^{c_5^A} (\alpha_1  \Amxq+ \cr 
& \dAmxq)^2  \left[ \alpha_1 ^2 (\Amxq)^2 c_5^A  +\Amxq (2\alpha_1  c_5^A \dAmxq +\right. \cr
& \left. \ddAmxq)+(c_5^A-1) (\dAmxq)^2 \right]^{-1}, \\
c_2^A    =& \frac{c_5^A}{c_1^A} e^{-c_3^A} \left(e^{c_3^A}+1\right)^2  \left(\Amxq \right)^{c_5^A-1} \times \cr 
& \left[\alpha_1 \Amxq+\dAmxq\right] ,\\
c_4^A    =& (\Amxq)^{c_5^A} -\frac{ c_1^A}{e^{c_3^A}+1},\\
c_1^\phi =& \dfrac{1 + c_3^\phi + c_4^\phi}{c_2^\phi(c_3^\phi + 2 c_4^\phi)} \left(\omega_1-\omega_\tA22\right),
\end{align} 
where $\Amxq$, $\dAmxq$, $\ddAmxq$ and $\omega_\tA22$ are, respectively, the amplitude of $h_\lm$, 
its first and second time-derivative and the frequency evaluated at $\tA22$; $\alpha_1+i \omega_1$ is the $\l$-fundamental
QNM frequency. The coefficients $\left\lbrace c_3^A, c_5^A, c_2^\phi, c_3^\phi, c_4^\phi \right\rbrace$
are determined performing the primary fits of $A_{\bar{h}}$ and $\phi_{\bar{h}}$ starting from $\tA22$. 
Finally, this fitted waveform is used to determine the NQC corrections at $\tA22$ and then is matched 
to the inspiral wave, always at $\tA22$.
Concerning the structure of the NQC correction to amplitude and phase, there is an additional subtlety. 
We realized that the standard NQC basis we used so far is not efficient when the NQC corrections are determined
at $\tA22$ for $m\neq \ell$, and it is thus better to resort to the NQC basis used by {\tt SEOBNRv5PHM}
that reads
\begin{align}
\label{eq:seobn1}
n_1  &= \dfrac{p_{r_*}^2}{(r \Omega)^2} , \\
\label{eq:seobn2}
n_2  &= \dfrac{n_1}{r} \ , \\
\label{eq:seobn3}
n_3  &= n_1 r^{-3/2} , \\
\label{eq:seobn4}
n_4  &= \dfrac{p_{r_*}}{r\Omega} , \\
\label{eq:seobn5}
n_5  &= n_4 p_{r_*}^2 .
\end{align}
The real part, amplitude and frequency of the final result for the $(2,1)$ and $(4,1)$ 
modes are shown in Fig.~\ref{fig:new_rng_schw}.
It is remarkable how the different NQC approach can visibly improve the
EOB/RWZ agreement. The only visible remaining differences between the two curves (mostly around
the waveform peak) are related to the fact that the mode mixing  only includes the fundamental mode 
and not the overtones.

Although the improvement discussed in the nonspinning case may be considered relatively marginal, 
it becomes absolutely essential when the central black hole is spinning and the spin is large and anti-aligned
with the orbital angular momentum. In this case, the orbital frequency has a zero and thus the NQC basis
becomes meaningless. This is clarified in Fig.~\ref{fig:new_rng_kerr}, that refers to the $(2,1)$ mode
for a particle inspiralling and plunging on a Kerr black hole with dimensionless spin $\hat{a}=-0.9$,
where the numerical waveform (black) has been obtained with the time-domain code \texttt{Teukode}~\cite{Harms:2014dqa}.
As the orbital frequency (gray online) passes through zero, the NQC corrected waveform determined 
using the standard approach oscillates unphysically. By contrast, the NQC correction determined at
$\tA22$, using the basis of Eqs.~\eqref{eq:seobn1}-\eqref{eq:seobn5} allows one to smoothly and 
reliably connect the inspiral waveform to the ringdown one. In this preliminary study, we are evidently
not considering the mode mixing during the Kerr ringdown~\cite{Taracchini:2014zpa}, 
so the EOB frequency and amplitude do not present any modulation.

\section{Phenomenology of quasi-normal-modes excitation}
\label{sec:qnm_phenom}
In the previous section we have provided an accurate and complete EOB waveform where the 
ringdown model was based on a phenomenological description. In doing so, we assumed that 
the fundamental QNM was excited, but we did not attempt any qualitative (nor quantitative) investigation
to understand the origin of this excitation. In this section we attempt to do this, still in a 
somehow phenomenological and heuristic way. Our main aim is to correlate the QNMs excitation
with the behavior of the source of the RWZ equations that is driven by the dynamics. The material 
presented here is inspired by and extends the (qualitative) discussion 
of Sec.~IIIB of Ref.~\cite{Damour:2007xr} (see also Fig.~4 therein).

\subsection{The RWZ source term during ringdown}
\label{sec:source}
We start by analyzing the source terms of the RWZ equations, Eq.~\eqref{eq:RWZ}.
Their functional form is~\cite{Nagar:2006xv}
\begin{align}
\label{eq:Slm}
S^\oe_{\ell m} = & \bar{G}^\oe_\lm(\tilde{r},t)\delta(\tilde{r}_*-r_*(t)) + \nonumber \\
                 & \bar{F}^\oe_\lm(\tilde{r},t)\de_{\tilde{r}_*}\delta(\tilde{r}_*-r_*(t)) \ ,
\end{align}
where the tilde denotes the field tortoise coordinate, while $r_*(t)$ is the tortoise coordinate of the particle. 
In order to understand the relevance of the source terms during the ringdown, we evaluate it on the particle
dynamics. More precisely, we neglect the term proportional to
$\de_{\tilde{r}_*}\delta(\tilde{r}_*-r_*(t))$ in Eq.~\eqref{eq:Slm}
and just evaluate $\bar{G}^\oe_\lm$.
This yields the expressions  
\begin{subequations}
\label{eq:Flm}
\begin{align}
		\label{eq:even_source}
		F_\lm^{\rm (e)}(t) & \equiv \frac{16 \pi \mu \, Y^*_\lm}{r \, \hat{H}\, \lambda[r(\lambda-2)+6]} 
		\Biggl\{ 2 i m \, A \, p_{r_*}  p_{\varphi} \cr
		&- A\Bigg[3 \Bigg(1+\frac{4 \hat{H}^2 r}{r(\lambda-2)+6}\Bigg)-\frac{r \,\lambda}{2} \cr
		&+\frac{p^2_\varphi}{r^2 (\lambda-2)}\Bigg(r(\lambda-2)(m^2-\lambda-1) \cr
		&+2(3m^2-\lambda-5)\Bigg)+\frac{2}{r^2}\Big(p^2_\varphi+r^2\Big)\Bigg]\Biggr\}, \\
			\label{eq:odd_source}
	F_\lm^{\rm (o)}(t)  & \equiv  \frac{16 \pi \mu \, \partial_{\Theta}Y^*_\lm }{r \, \lambda(\lambda-2)} 
	\Bigg[\frac{d}{dt}\Bigg(\frac{p_{r_*}  p_\varphi}{\hat{H}}\Bigg) \cr 
	&-2 \frac{p_\varphi \, A}{r} -i m \frac{A \, p_{r_*}  p_\varphi^2 }{r^2\hat{H}^2}\Bigg],
\end{align}
\end{subequations}
where $\lambda \equiv \ell(\ell+1)$. After having set $\Theta=\pi/2$,
we show  $|F_\lm^\oe(t)|$ for the (2,2) mode in Fig.~\ref{fig:Flm}.
Interestingly, $|F_{22}^{\rm (e)}(t)|$ reaches its maximum after the peak 
of $A_{22}$ and remains quite relevant also later on.
For example, at $t=\tA22+10$ we have 
$\hat{F}^{\rm (e)}_{22}\equiv |F^{\rm (e)}_{22}|/\max(|F^{\rm (e)}_{22}|)\simeq 0.38$,
and $\hat{F}^{\rm (e)}_{22}<10^{-2}$ only from $t = \tA22+ 18.2$.
For the odd modes we have that the maximum of the source is delayed with respect to the even ones.
Since the source term is quite relevant during the ringdown, 
we do not expect {\it a priori} that a pure QNMs description (i.e.~part of the solution of the homogeneous RWZ equations)
can be used for the whole postpeak waveform. However, from a sufficient late time $t>\tAlm$, the $(\l,m)$-mode can
be fully described in terms of QNMs using the ansatz of Eq.~\eqref{eq:qnm_wave} since the source term becomes negligible. 
\begin{figure}[t]
	\label{fig:Flm}
	\begin{center}
	    		\includegraphics[width=0.48\textwidth]{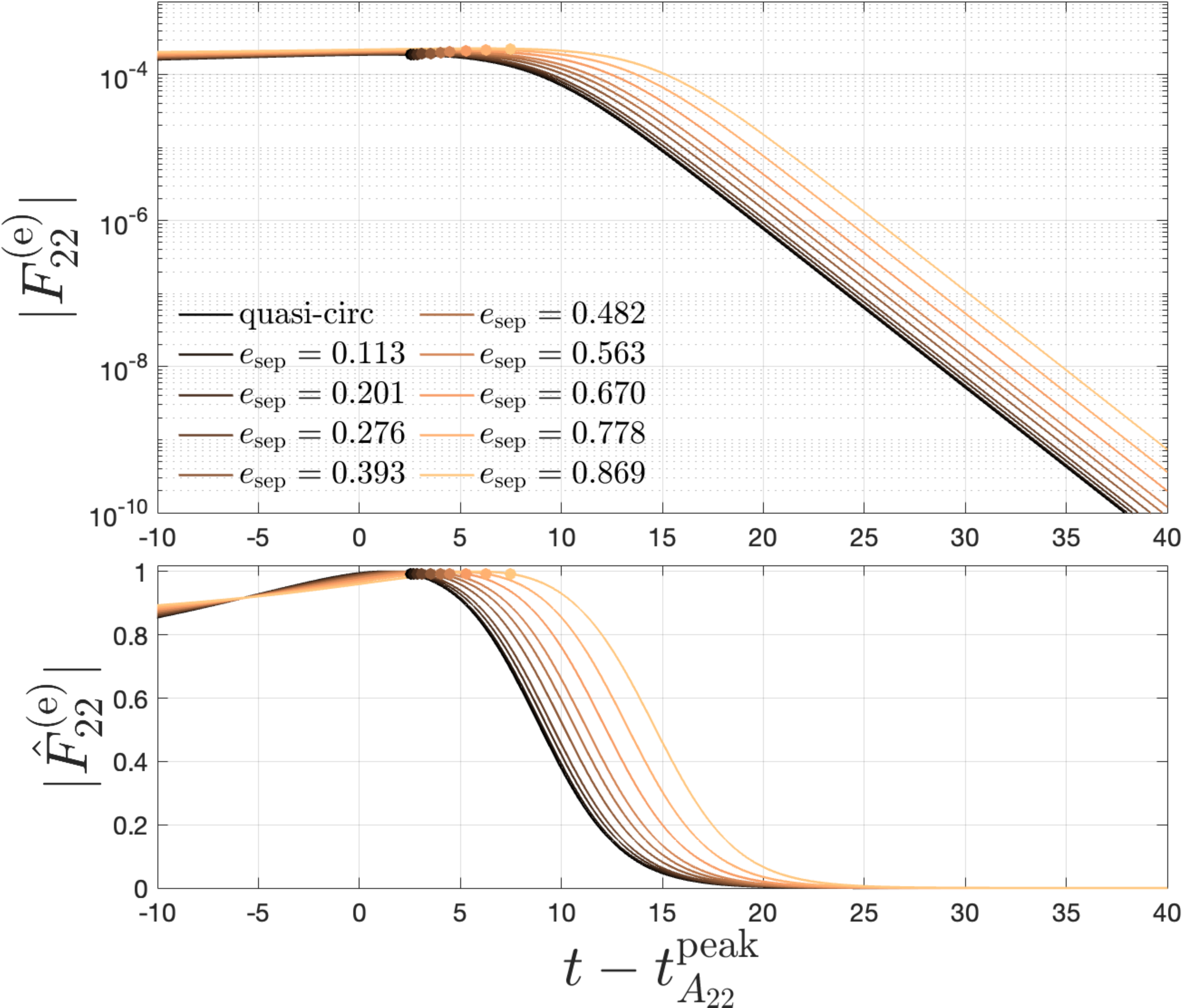}
		\caption{\label{fig:Flm} Upper panel: Zerilli source term evaluated along dynamics with different eccentricities
		for the (2,2) mode. Logarithmic vertical scale. Bottom panel: plot of
		$\hat{F}^{\rm (e)}_{22} = |F^{\rm (e)}_{22}|/\max(|F^{\rm (e)}_{22}|)$. 
		The retarded time is shifted using the peak time of the quadrupolar amplitude, $\tA22$. 
		In both panels, the dots mark the light-ring crossing, $\tLR$. }
	\end{center}
\end{figure}
%

\subsection{Iterative time-domain fit of the postpeak frequency}
\label{sec:omgfit_iter}
\begin{figure}
	\begin{center}
	\includegraphics[width=0.48\textwidth]{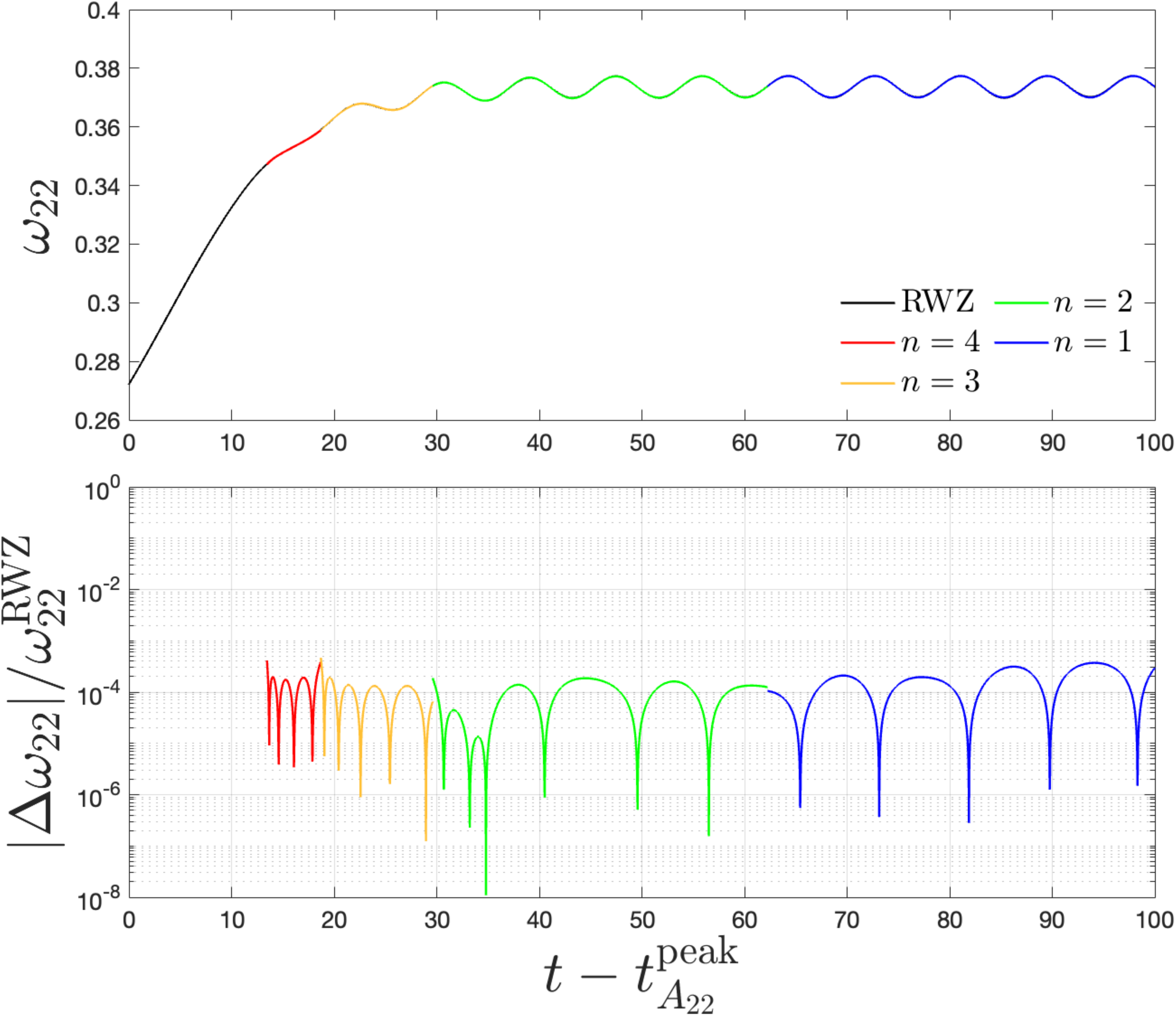}
	\caption{\label{fig:omgfit_l2m2} Iterative QNM-fit of the frequency considering the first 4 QNMs.
	In the upper panel we show the RWZ frequency (black) and the results obtained with the fits. 
	Each color represents an iteration (and thus a QNM) of the fit. In the bottom panel we show
	the relative difference between the fit and the corresponding $n$-fit.
	See Sec.~\ref{sec:omgfit_iter} for more detail on the fitting procedure.}
	\end{center}
\end{figure}
We thus proceed to fit the late ringdown waveform assuming that is given by a linear superposition
of QNMs with constant coefficients, see  Eq.~\eqref{eq:qnm_wave}, using $\tau=t-\tAlm$. 
The total QNM frequency of each multipole is obtained as
\begin{align}
\label{eq:qnm_freq}
& \omega^{\oe}_\lm = - \Im \left( \frac{\dot{\Psi}^{\oe}_\lm}{\Psi^{\oe}_\lm} \right) = \nonumber \\ 
& - \Im \left[  \frac{ \sum\limits_{n=1}^\infty 
\bar{b}_{\ell m n} \frac{\sigma^+_{\ell n}}{\sigma_{\l 1}^+}  
e^{(\sigma_{\ell 1}^+ - \sigma_{\ell n}^+) \t } 
\left(  1 + \hat{a}_{\ell m n} 
\frac{\sigma_{\l n}^-}{\sigma_{\l n}^+} 
e^{2 i \omega_{\ell n} \t} \right) }{ \sum\limits_{n=1}^\infty  \bar{b}_{\ell m n} 
e^{(\sigma_{\ell 1}^+ - \sigma_{\ell n}^+) \t } 
\left(  1 + \hat{a}_{\ell m n} e^{2 i \omega_{\ell n} \t} \right)  } \right],
\end{align}
where 
$\hat{a}_{\ell m n} = C_{\ell  m n}^-/C_{\ell m n}^+$ and 
$\bar{b}_{\ell m n} = C_{\ell m n}^+/C_{\ell m 1}^+$.
Since these are complex quantities, we define 
$\hat{a}_{\ell m n} \equiv \hat{A}_{\ell m n} e^{i \hat{\theta}_{\ell m n} }$ and 
$\bar{b}_{\ell m n} \equiv \bar{B}_{\ell m n} e^{i \bar{\phi}_{\ell m n} }$.
Recalling the hierarchy of the inverse damping times $\alpha_{\l n}$, the contributions in the frequency of the $n\geq 1$ overtones are exponentially damped with exponents
$\alpha_{\ell 1}^+ - \alpha_{\ell n}^+$. However, the contribution of the isolated fundamental frequencies
is never damped and reads 
\begin{equation}
\omega^\oe_{\lm 1} = \frac{ (1 - \hat{A}_{\ell m 1}^2) \omega_{\ell 1} }{ 1 +   
\hat{A}_{\ell m 1}^2 + 2 \hat{A}_{\ell m 1}^2 \cos(2 \omega_{\ell 1} \t + \hat{\theta}_{\ell m 1}) }.
\label{eq:beating_n1}
\end{equation}
The coefficients $\hat{A}_{\l m 1}$ and $\theta_{\l m 1}$ were already extracted from the 
late ringdown frequency in previous works~\cite{Bernuzzi:2010ty,Taracchini:2014zpa}. 
Here we extend this procedure to earlier times using Eq.~\eqref{eq:qnm_freq}.
%
\begin{table}
  \caption{\label{tab:Alm1}Coefficients $\hat{A}_{\l m 1}=|\hat{a}_{\lm 1}|$ describing the 
  beating between positive and negative frequencies fundamental QNMs for all 
  multipoles up to $\l=6$. See Table~\ref{tab:gfits_th1} for the eccentric fits of 
  the $\hat{a}_{\lm 1}$ phase.}
  \begin{center}
\begin{ruledtabular}
\begin{tabular}{c | c | c | c | c | c | c | c }
              &  $m=1$      &  $m=2$      &  $m=3$      &  $m=4$      &  $m=5$      &  $m=6$      \\
$\hat{A}_{\l m 1}$ & $[\times10^{-2}]$ & $[\times10^{-3}]$ & $[\times10^{-4}]$ & $[\times10^{-5}]$ & $[\times10^{-6}]$ & $[\times10^{-7}]$ \\
\hline
$\l=2$        &   7.30  &   4.89  &     \   &     \   &      \  &      \  \\ 
$\l=3$        &   9.34  &   7.96  &   5.53  &     \   &      \  &      \  \\ 
$\l=4$        &   9.14  &   9.11  &   8.90  &   6.28  &      \  &      \  \\ 
$\l=5$        &   9.41  &   8.97  &   9.27  &   9.96  &   7.02  &      \  \\ 
$\l=6$        &   9.46  &   9.12  &   9.11  &   9.67  &   11.15 &   8.33  \\
\end{tabular}
\end{ruledtabular}
\end{center}
\end{table}
\begin{table}
  \caption{\label{tab:gfits_th1}Global fits for the beating coefficients
  $\hat{a}_{\lm 1}=\hat{A}_{\lm 1}e^{i \theta_{\lm 1}}$. The template used for the phase is
  $\theta_{\lm 1}=\left( C^{\theta}_{\rm QC}+C^{\theta}_1\bmrg + C^{\theta}_2 \bmrg^2\right)/\left( 1 + D^{\theta}_1 \bmrg \right)$, while the modulus does not depend on the nature of the perturbation. }
  \begin{center}
\begin{ruledtabular}
\begin{tabular}{c | c | c | c | c | c }
 $(\l,m)$ & $\hat{A}_{\lm 1}$  &  $C^{\theta}_{\rm QC}$ & $C^{\theta}_1$ & $C^{\theta}_2$ & $D^{\theta}_1$ \\
 \hline
(2,2) & $4.89\cdot 10^{-3}$ & $ 5.369$ & $-9.444$  &  $-37.992$  &    $\dots$     \\ 
(2,1) & $7.30\cdot 10^{-2}$ & $ 2.893$ & $-6.074$  &  $-15.134$  &  $-2.105$ \\ 
(3,3) & $5.53\cdot 10^{-4}$ & $ 2.636$ & $-11.635$ &  $-9.555$   &     $\dots$      \\ 
(3,2) & $7.96\cdot 10^{-3}$ & $ 4.649$ & $-3.890$  &  $-0.1176$  &     $\dots$      \\ 
(3,1) & $9.34\cdot 10^{-2}$ & $ 3.810$ & $ 0.3296$ &  $-0.03943$ &    $\dots$      \\ 
(4,4) & $6.28\cdot 10^{-5}$ & $ 6.503$ & $-13.096$ &  $-8.815$   &     $\dots$      \\ 
(4,3) & $8.90\cdot 10^{-4}$ & $ 2.453$ & $-6.041$  &  $-1.309$   &     $\dots$      \\ 
(4,2) & $9.11\cdot 10^{-3}$ & $ 1.186$ & $-2.772$  &  $-0.4334$  &     $\dots$      \\ 
(4,1) & $9.14\cdot 10^{-2}$ & $ 3.714$ & $ 0.3582$ &  $ 0.1166$  &     $\dots$      \\ 
(5,5) & $7.02\cdot 10^{-6}$ & $ 4.509$ & $-15.344$ &  $-4.330$   &     $\dots$      \\ 
\end{tabular}
\end{ruledtabular}
\end{center}
\end{table}
Since the overtones have higher damping coefficients, we proceed to iteratively fit the late 
ringdown frequency on different time intervals considering only the relevant QNMs.
To establish where a certain $n$ mode becomes negligible, we set a small threshold, 
typically $\epsilon = 10^{-5}$, and we say that the $n$th-mode can be neglected 
if the condition $e^{(\alpha_{\ell 1}^+ - \alpha_{\ell n}^+) \t}<\epsilon$ is satisfied.
Applying this method to the (2,2) mode we can find
$\left \lbrace \hat{A}_{22 n}, \hat{\theta}_{22 n}, \bar{B}_{22 n}, \bar{\phi}_{22 n} \right \rbrace$ 
up to $n=4$. The results are shown in Fig.~\ref{fig:omgfit_l2m2}. 
It is interesting to note that for the (2,2) mode we are not able to go beyond $n=4$ and thus at earlier times. This 
can be justified by the discussion on the source term above. Indeed, with $n=4$ and $\epsilon=10^{-5}$ we
are able to fit from $t=13.43$, but at that time we still have
$|F^{\rm (e)}_{22}|/\max(|F^{\rm (e)}_{22}|)\simeq 0.097$ and thus the source term is not completely negligible. 
Also for this reason, the values of $\ha_{22n}$ and $\bar{b}_{22n}$ found for the overtones are not robust
and are in disagreement with the results found using different fitting procedures (see Appendix~\ref{app:td_QNM_fits}).
The procedure can be applied also to the higher modes, where the number of overtones that we are able to fit  
depends on the specific multipole considered. Generally, we have that for the modes with higher $\Delta t_\lm$ (i.e.~the
for the ones that are more delayed), we are able to fit more overtones, consistently with the fact
that at later times the source terms become negligible. For example, for the (4,1) multipole
we can fit up to $n=7$ overtones (and thus from $t\simeq t^{\rm peak}_{41}+5.94$) keeping the 
relative error of the frequency fit around $10^{-4}$.

From a practical point of view, we are particularly interested in the values related to the fundamental frequencies, 
$\ha_{\lm 1}$, since we can use them to improve the phenomenological
ringdown description as discussed in Sec.~\ref{sec:modeling}, see in particular Eq.~\eqref{eq:modemix_in_hrng}.
The values found in this work are in agreement with 
previous work~\cite{Bernuzzi:2010ty} and their modulus is reported in Table~\ref{tab:Alm1}.
Note that the order of magnitude of $\hat{A}_{\lm 1}$ is strictly linked to the number $m$
and does not strongly depended on the eccentricity 
(see e.g. the right panel of Fig.~\ref{fig:ecc_together}). Therefore, we need to perform the global fits 
only on the phases $\theta_{\lm 1}$, that we report in Table~\ref{tab:gfits_th1} 
for all the multipoles considered in this work. The relevance of these fits can be particularly appreciated
looking at the analytical/numerical comparisons of the late-ringdown waveform frequencies shown in Figs.~\ref{fig:inspl_l2m2},~\ref{fig:inspl_e095} and~\ref{fig:new_rng_schw}.

Finally, we mention that since the iterative fit of the frequency did not provide satisfactory 
results for the whole quadrupolar postpeak waveform, we also attempted the same fit {\it without} the 
iterative procedure, focusing only on the (2,2) multipole. However, also this procedure did not lead 
to robust results for the whole postpeak waveform, as detailed in Appendix~\ref{app:td_QNM_fits}.

\subsection{EOB ringdown as superposition of QNMs}
\label{sec:comb}
\begin{figure}
	\begin{center}
	\includegraphics[width=0.48\textwidth]{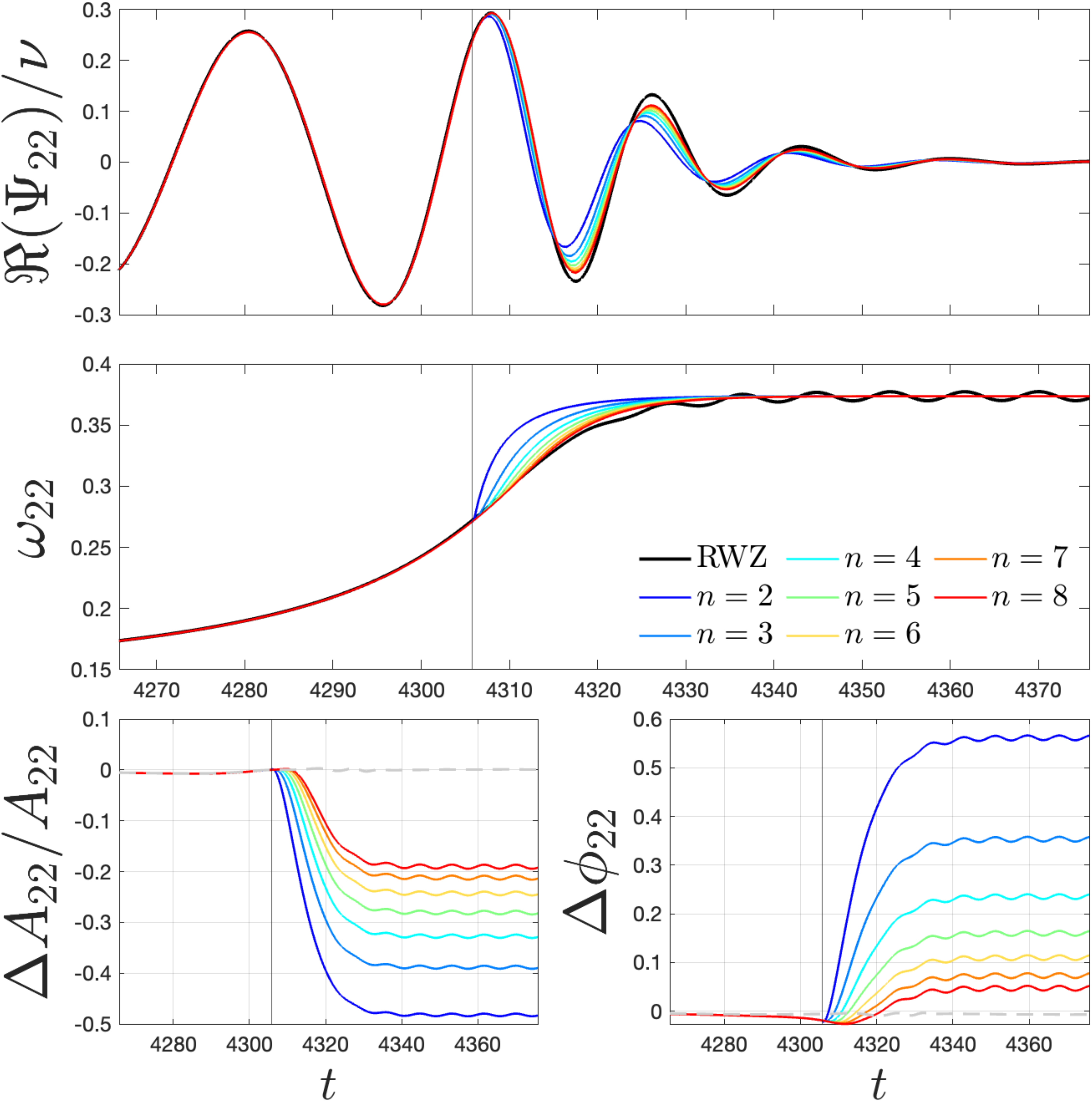}
	\caption{\label{fig:comb} EOB waveform where the ringdown has been modeled with the matching comb procedure. 
	In the bigger panels above we show the real part of the waveform
	and its frequency, black for the RWZ results and online colors for the analytical results. 
	In the two small bottom panels we show the relative amplitude difference and the phase difference
	with the same color scheme of the upper panels. 
	We also show the relative amplitude difference and the phase difference obtained with the waveform
	model discussed in Sec.~\ref{sec:model_test} (dashed gray).}
	\end{center}
\end{figure}
\begin{figure*}	
	\begin{center}
		\includegraphics[width=0.32\textwidth,height=4.85cm]{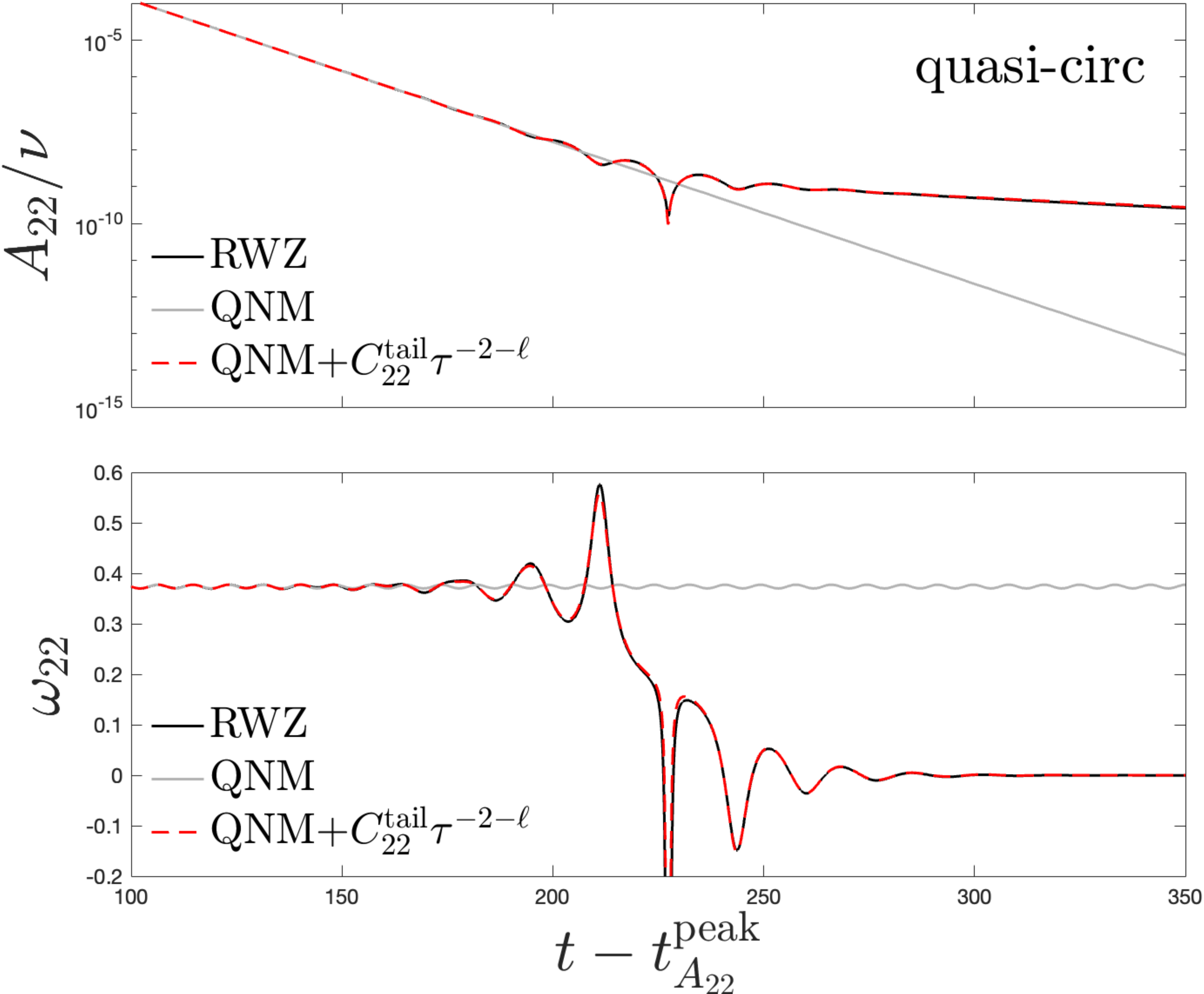}
		\includegraphics[width=0.33\textwidth]{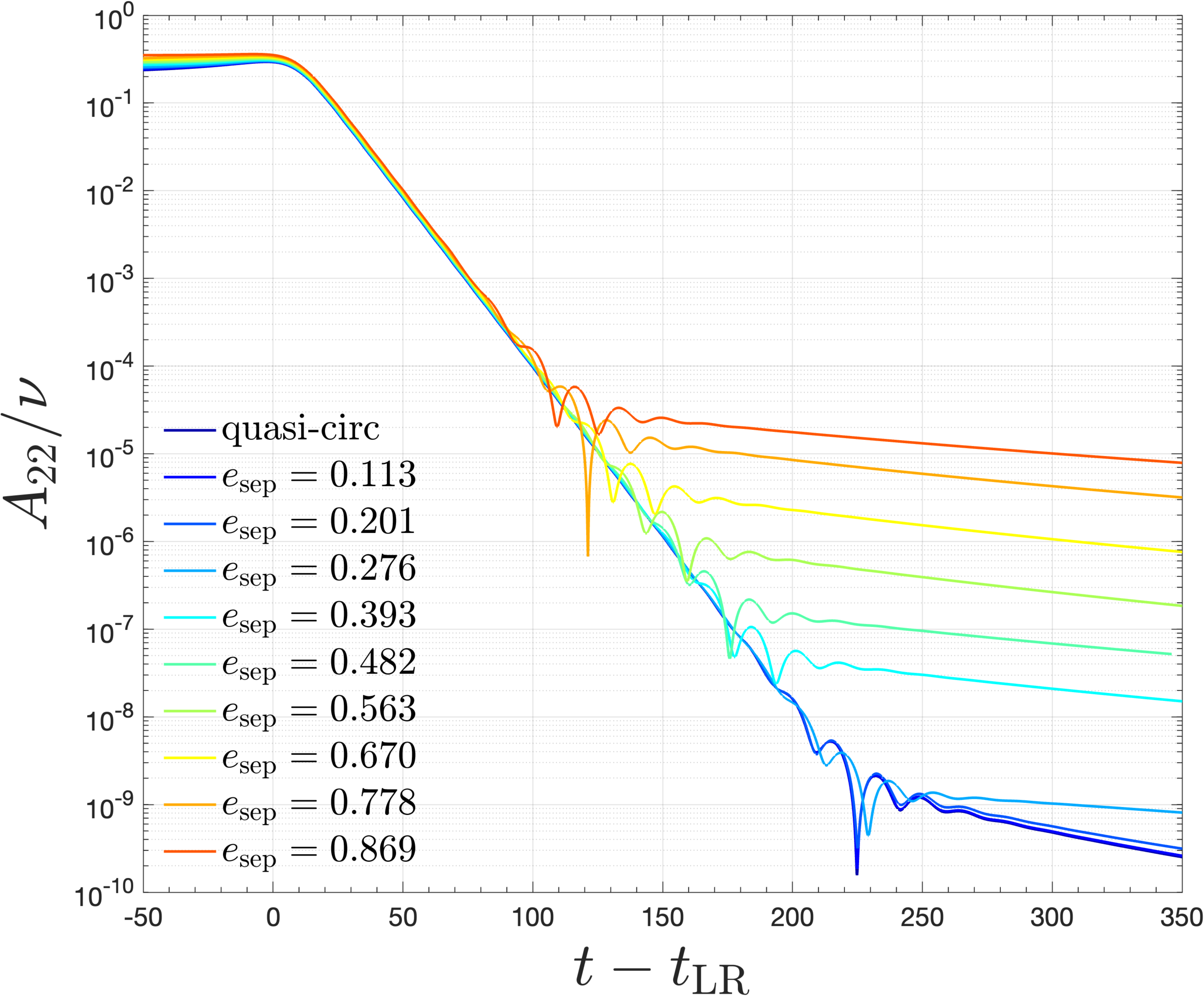}
		\includegraphics[width=0.32\textwidth,height=4.88cm]{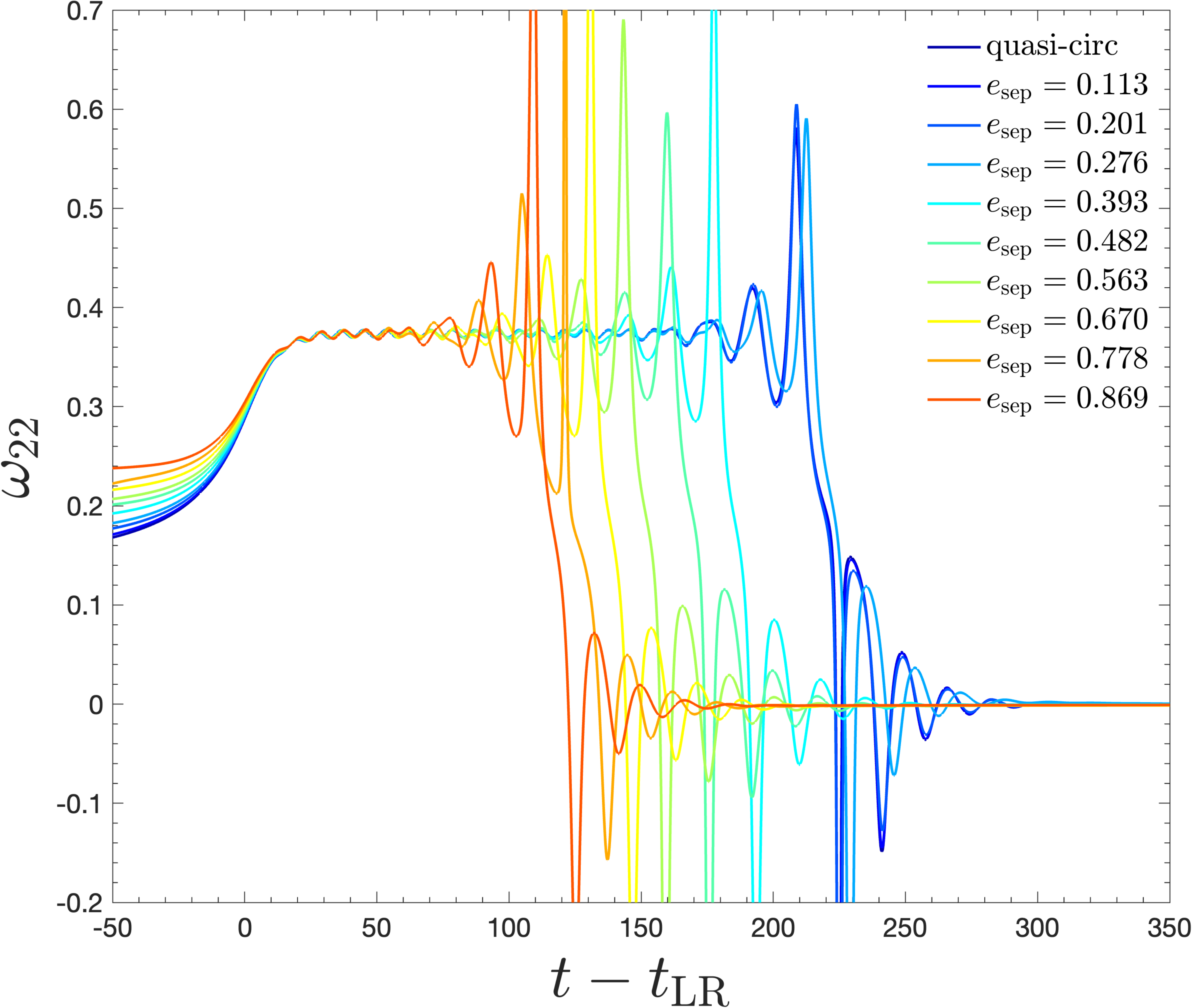}
		\caption{\label{fig:tail_l2m2} Leftmost panel: Amplitude and frequency of the Zerilli (2,2) of 
		the quasi-circular waveform (black) compared with the results obtained considering only the QNMs 
		contribution (gray) and the QNMs plus tail (dashed red). Middle and rightmost: 
		RWZ amplitudes and frequencies for different eccentricities, aligned with respect to
		the light-ring crossing.}
	\end{center}
\end{figure*}

In the previous sections we have accumulated results which indicate that the ringdown description as
linear superposition of QNMs with constants coefficients, Eq.~\eqref{eq:qnm_wave} cannot be consistently used for 
the whole postpeak waveform. However, in seminal EOB works~\cite{Damour:2007xr,Damour:2009kr} 
the ringdown was modeled precisely in this way, although it was matched with the inspiral part of the waveform on
an extended interval centered around the peak of the orbital frequency. We will now revisit 
this procedure, often referred to as ``matching comb"~\cite{Damour:2007xr}.
The basic idea of this procedure is that the match is performed on a set of points 
rather than at only one point. To determine the complex coefficients $C_{\lm n}^\pm$, 
we solve the linear system $h_\lm(t_i) = \sum C_{\lm n}^\pm e^{-\sigma_{\l n}^\pm t_i}$, 
where $t_i$ are the points of the time array used for the match, 
and $h_\lm=h^{\rm inspl}\hat{h}_\lm^{\rm NQC}$ is the NQC-corrected
inspiral waveform. If we consider $N$ QNMs (distinguished 
between positive and negative frequency), we then need $N$ points.
Note that this ringdown model does not require any tuned numerical parameters, it only needs the
QNMs frequencies since the coefficients of the ringdown are determined using the analytical waveform.
Similarly to what discussed in Sec.~\ref{sec:model_test} for the (2,2) waveform multipole,
we use $\left\lbrace A^{\rm NQC}_{22}, \dot{A}^{\rm NQC}_{22}, \ddot{A}^{\rm NQC}_{22},
\omega^{\rm NQC}_{22}, \dot{\omega}^{\rm NQC}_{22}, \ddot{\omega}^{\rm NQC}_{22}
\right\rbrace$, but here we directly extract them from the Zerilli waveform at $t_{22}^{\rm NQC}=\tA22 + 2$.
We chose as matching points the adjacent points to $\tA22$ in the time grid used 
to solve numerically the Hamilton's equations~\eqref{eq:Hamilton}. 
The results of this procedure are shown from $n=2$ up to $n=8$ in Fig.~\ref{fig:comb}. 
The configuration that reproduces the numerical waveform with the highest accuracy is the 
one with $n=8$ positive-frequency modes. For this configuration, in the late ringdown we 
have a $18\%$ relative amplitude difference and
a phase difference that oscillates around the $0.04$ radians. 
We also show the relative amplitude difference and the phase difference obtained with
the state-of-the-art waveform discussed in Sec.~\ref{sec:model_test} (dashed gray).
While the waveform obtained with the latter model is clearly more accurate
(see also Fig.~\ref{fig:inspl_l2m2} for comparison), the 
results obtained in this section are still qualitatively good for the amplitude and quite accurate for the phase.

\subsection{Tail contribution}
\label{sec:tail}
Having considered only the first part of the waveform, we have so far neglected 
the power-tail effects in the waveform~\cite{Leaver:1986gd,Andersson:1996cm}. 
However, since the QNMs are exponentially damped, there is a time where the tail effects
become dominant.
As can be seen from the frequency of the Zerilli (2,2) quasi-circular waveform shown in black in 
the left panel of Fig.~\ref{fig:tail_l2m2}, 
the effect of the tail starts to be visible at $t\sim \tA22 + 170$, and becomes dominant shortly afterward.
In order to reproduce the numerical waveform, we have to include a term of the form 
$C^{\rm tail}_{22} \t^{-2-\l}$, where the complex coefficient $C^{\rm tail}_{22}$ is determined with a fit
and $-2-\l$ is the asymptotic behavior of the tail term at future null infinity~\cite{Zenginoglu:2009ey}.
Using the (fundamental) QNMs and the power-law tail we can fully catch the behavior of 
the numerical amplitude and frequency, as shown in the left panel of Fig.~\ref{fig:tail_l2m2}.
When eccentricity is taken into account, the tail contribution becomes more significant
and starts to dominate over the QNM decay earlier, as shown in the middle and right panels
of Fig.~\ref{fig:tail_l2m2}. Moreover, the tail cannot be described as before using the ansatz 
$C^{\rm tail}_{22} \t^{-2-\l}$ since the tail  has not reached yet the asymptotic behavior. 
For example, for the configuration
with $e_0=0.5$ the decay-rate is roughly $-1.3$ instead of $-2-\ell$. Similar numbers are obtained
for the other eccentric configurations with $e \gtrsim 0.3$.

\section{Heuristic modeling of QNMs excitation}
\label{sec:toymodel}
Let us now introduce a toy model to grasp some insights on how the excitation of QNMs 
is driven by the dynamics and the related presence of a source term in the right hand side of
the RWZ equations. Following Ref.~\cite{Damour:2007xr}, and in particular Sec.~III~B, 
we base our analysis on the understanding that a Schwarzschild black hole can be seen
as a resonating object. We start by generalizing Eq.~\eqref{eq:qnm_wave}, assuming that
the constant coefficients $C^\pm_{\lm n}$ are instead time-dependent functions, $C^\pm_{\lm n}(t)$.
The ringdown waveform reads then
\begin{align}
\label{eq:general_QNM_superposition}
\Psi^\oe_\lm &=  \sum_{n} \Psi^\oe_{\lm n}, \\
\label{eq:general_QNM}
\Psi^\oe_{\lm n}(t) &\equiv C_{\lm n}^+(t) e^{-\sigma^+_{\ell n} t} + C_{\lm n}^-(t) e^{-\sigma^-_{\ell n} t}.
\end{align}
Here the origin of time, $t=0$, is assumed to be the light-ring crossing, $\tLR$.
The final goal of our investigation is to determine an approximate semi-analytical expression 
for the coefficients $C_{\lm n}^\pm(t)$ so to understand when the associated QNMs are excited. 
Since we know that the solution of the homogeneous RWZ equations is a superposition of 
QNMs with constant coefficients $C_{\lm n}^\pm$, each spherical mode is thought as 
as the solution of the homogeneous second order differential equation 
\be
\label{eq:damped_oscillator}
\ddot{\Psi}^\oe_{\lm n}+2 \alpha_{\ell n} \, \dot{\Psi}^\oe_{\lm n}+(\alpha_{\ell n}^2+\omega_{\ell n}^2 ) \, \Psi^\oe_{\lm n} = 0,
\ee 
under the ansatz $\Psi_{\lm n}(t)\propto e^{- \sigma^\pm_{\l n} t}$, with 
$\sigma^\pm_{\l n} = \alpha_{\l n} \pm  i \omega_{\l n}$. 
This is indeed the equation of an underdamped oscillator with damping coefficient 
$\alpha_{\ell n}$ and undamped angular frequency $\pm \omega_{\ell n}$. 
An external force $F(t)$ in the right-hand side of Eq.~\eqref{eq:damped_oscillator} 
yields an inhomogeneous differential equation corresponding to a driven harmonic oscillator. 
Our model is thus defined by {\it assuming} that the forcing term is given by Eq.~\eqref{eq:Flm}, 
that is $F(t)\equiv F_\lm^{\rm (e/o)}(t)$.
The solution of this inhomogeneous equation can be obtained starting from the one of 
the homogeneous problem by promoting the numerical coefficients therein to 
time-dependent functions. This is referred to as the 
\emph{method of variation of parameters} 
(or \emph{method of osculating elements})~\cite{Lagrange:1808,Lagrange:1809,Lagrange:1810}.
With this approach we know  {\it a priori} that the differential equation
\be
\label{eq:driven_oscillator}
\ddot{\Psi}^\oe_{\lm n}+2 \alpha_{\ell n} \, \dot{\Psi}^\oe_{\lm n}+(\alpha_{\ell n}^2+\omega_{\ell n}^2 ) \, \Psi^\oe_{\lm n} = F_\lm^{\rm (e/o)}(t)
\ee 
admits a solution with the precise QNM structure of Eq.~\eqref{eq:general_QNM}. 
We start by writing the time derivative of our particular solution \emph{as if} 
the coefficients $C_{\lm n}^\pm(t)$ were not time-dependent. This is equivalent to impose
\be
\label{eq:dotPsi}
\dot{\Psi}^\oe_{\lm n}(t) = -\sigma_{\ell n}^+ C_{\lm n}^+(t)e^{-\sigma^+_{\ell n} t} - \sigma_{\ell n}^- C_{\lm n}^-(t) e^{-\sigma^-_{\ell n} t},
\ee
which is true only if the time dependence of $C_{\lm n}^\pm(t)$ gives no contribution to the time derivative, namely if
the condition
\be
\label{eq:varofconst_1}
\dot{C}_{\lm n}^+(t) e^{-\sigma_{\ell n}^+ t} + \dot{C}_{\lm n}^-(t) e^{-\sigma_{\ell n}^- t} =0
\ee
is satisfied.
We can then take another time derivative on Eq.~\eqref{eq:dotPsi} in order to obtain $\ddot{\Psi}_\lm^\oe$. 
We then insert $\dot{\Psi}_\lm^\oe$ and $\ddot{\Psi}_\lm^\oe$ in Eq.~\eqref{eq:driven_oscillator}
and, considering that the 
sum of all the terms without the time derivatives $\dot{C}_{\lm n}^\pm(t)$ separately solves the associated 
homogeneous equation~\eqref{eq:damped_oscillator}, we get the second condition
\be
\label{eq:varofconst_2}
-\sigma_{\ell n}^+ \dot{C}_{\lm n}^+(t)e^{-\sigma^+_{\ell n} t} - \sigma_{\ell n}^- \dot{C}_{\lm n}^-(t) e^{-\sigma^-_{\ell n} t} = F_\lm^{\rm(e/o)}(t),
\ee
which together with Eq.~\eqref{eq:varofconst_1} builds up a system of 
two equations that can be solved for $\dot{C}_{\lm n}^\pm(t)$. 
A straightforward computation yields
\be
\label{eq:dotC+-}
\dot{C}_{\lm n}^\pm(t) = \pm i \frac{e^{\sigma^\pm_{\ell n} t} F_\lm^{\rm(e/o)}(t)}{2 \omega_{\ell n}},
\ee
and an ensuing time integral gives us the final expressions
\be
\label{eq:Clmn_integral}
C_{\lm n}^\pm(t) = C_{\lm n}^\pm(t_0) \pm \frac{i}{2 \omega_{\ell n}} \int_{t_0}^{t} dt' \, F_\lm^{\rm(e/o)}(t') e^{\sigma^\pm_{\ell n} t'},
\ee
where $t_0$ is an arbitrary initial time.
Considering that $F_\lm^{\rm(e/o)}(t')$ does not diverge during the inspiral 
and that in the integrand we have $e^{\sigma^\pm_{\ell n} t'}$, we have that 
the computation of $C_{\lm n}^\pm(t)$ is not influenced by the choice of $t_0$ as long as 
$t_0$ is not too close to $t_{\rm LR}$. In practice, we start to integrate from the beginning of our simulation.

\subsection{Results}
%
\begin{figure*}	
	\begin{center}
		\includegraphics[width=0.34\textwidth]{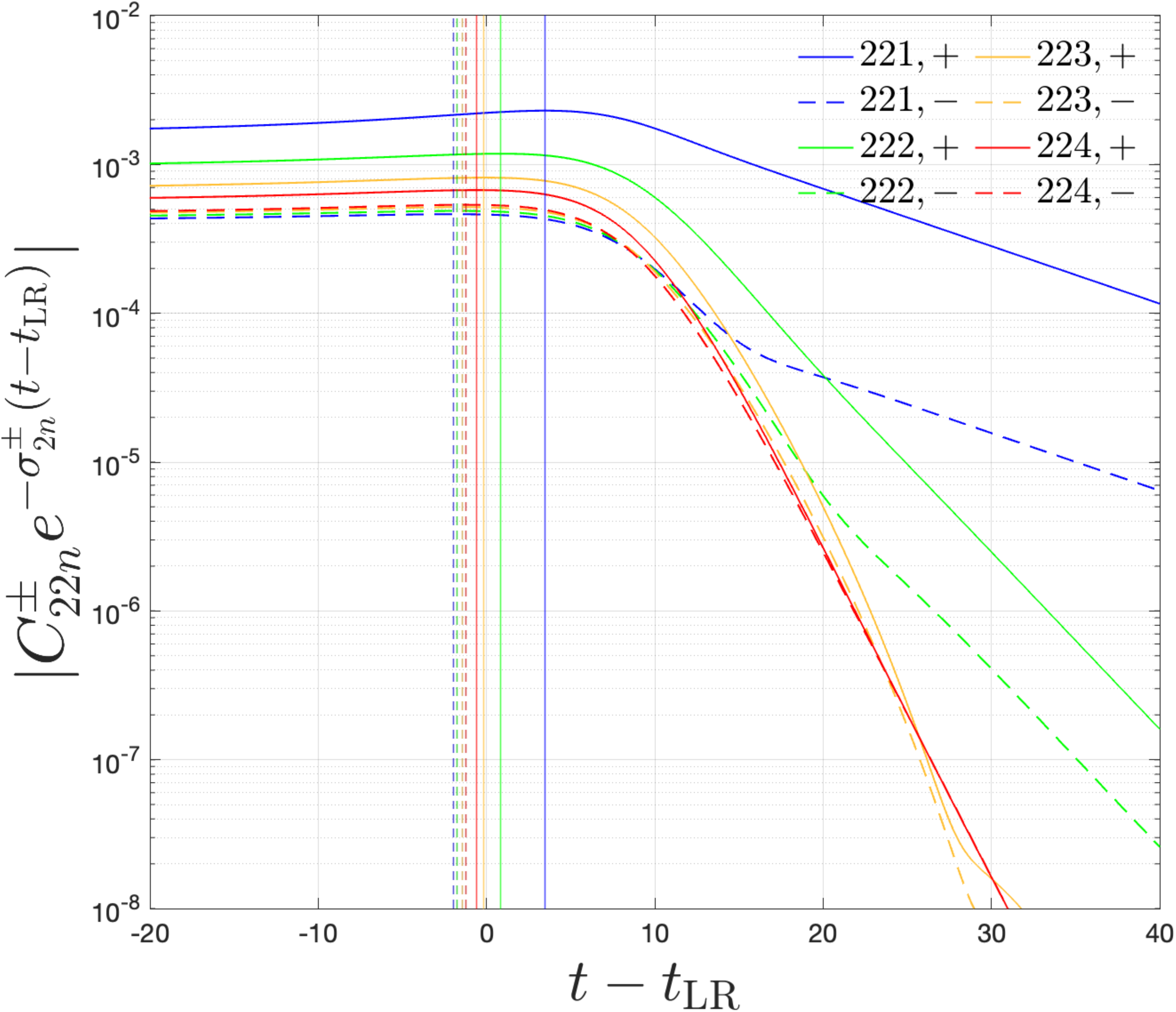}
		\includegraphics[width=0.31\textwidth]{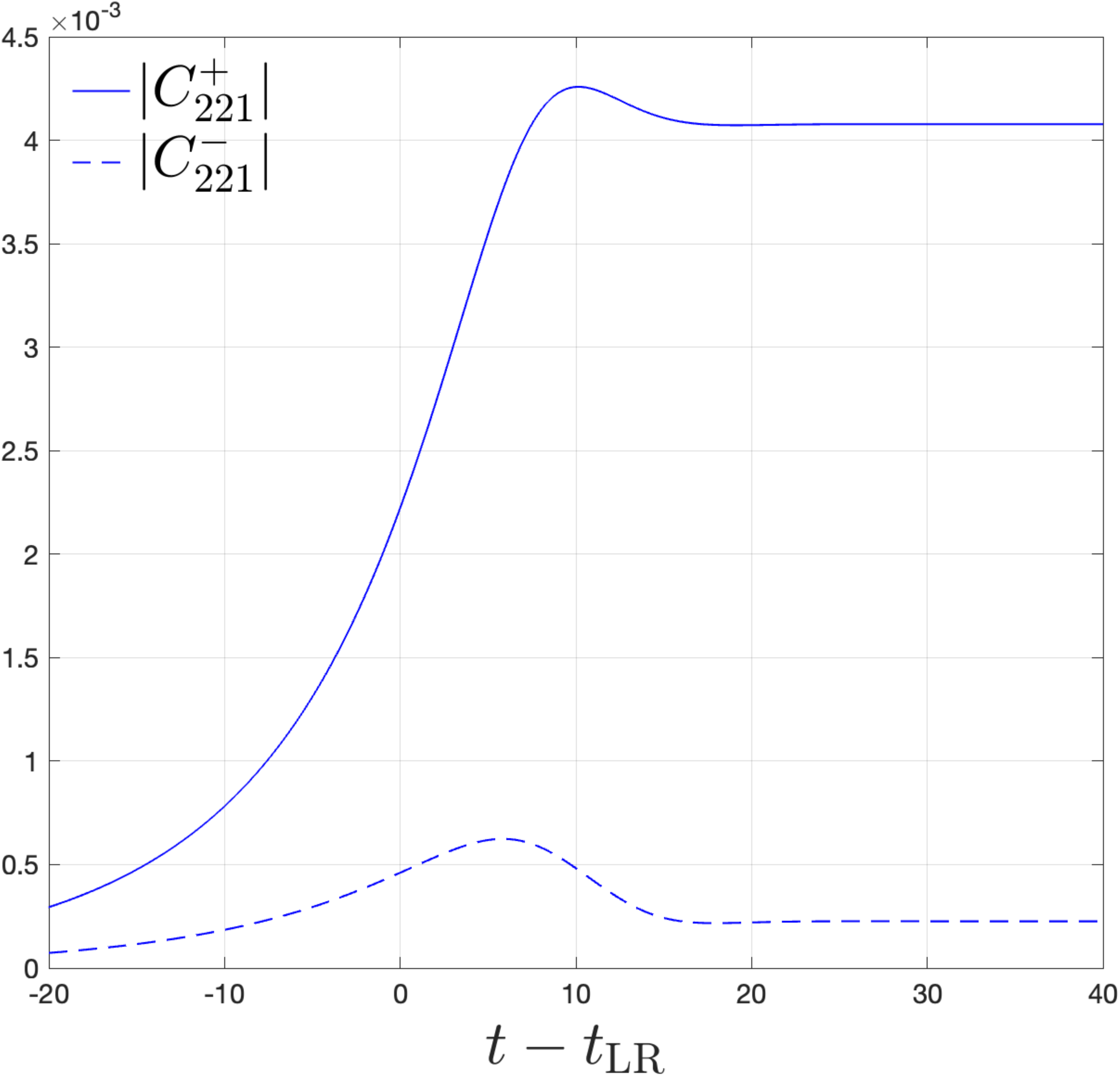}
		\includegraphics[width=0.31\textwidth]{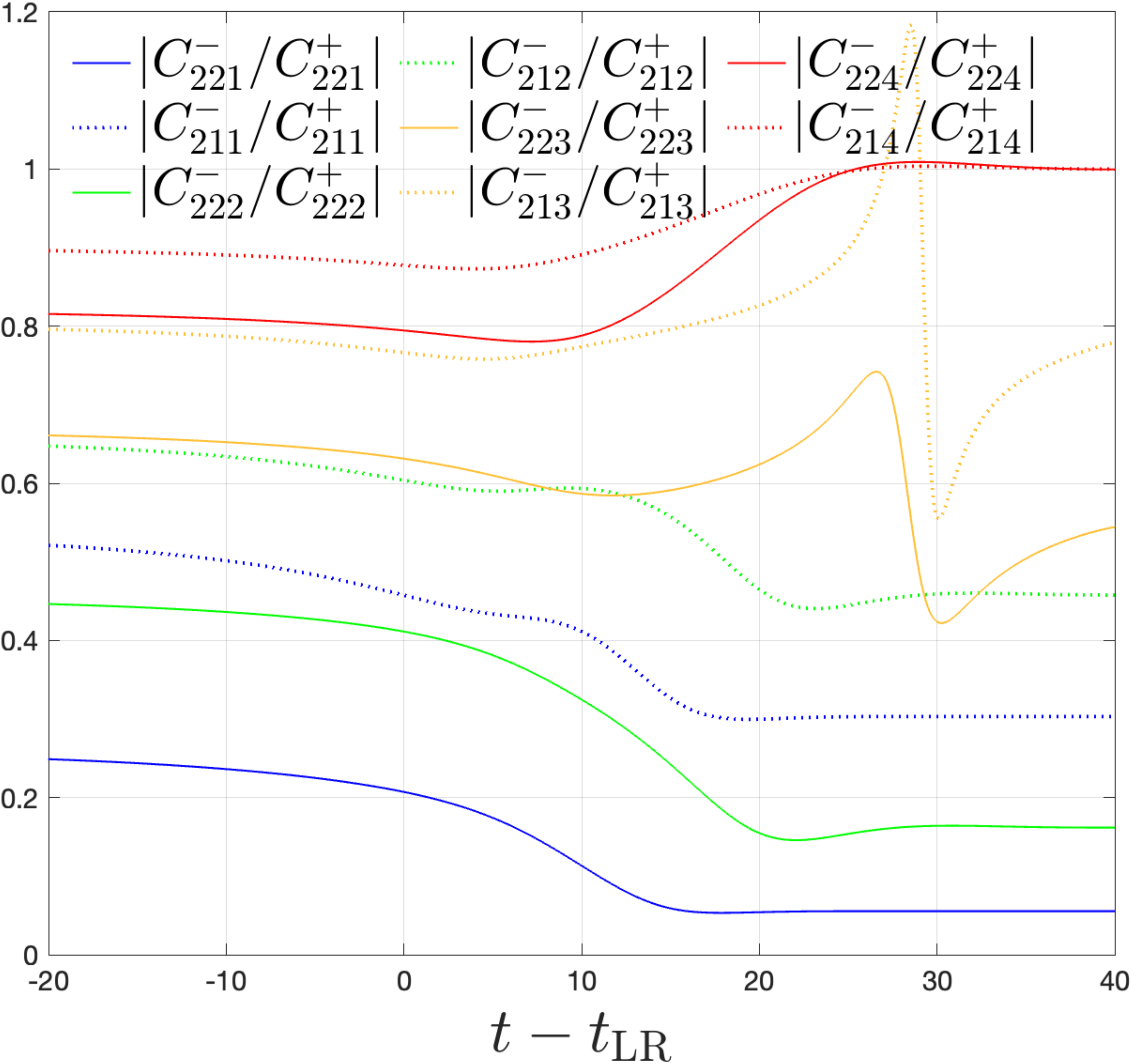}
		\caption{\label{fig:HO_l2}
		Time-dependent QNM-excitation coefficients $C_{\lm n}^\pm(t)$
		for the quasi-circular case, see Eq.~\eqref{eq:Clmn_integral}. 
		The positive-frequency modes are more relevant and
		are excited later than the negative ones, as shown by the 
		location of the amplitude maxima (vertical lines in the leftmost plot).
		In the middle panel we plot the activation-like excitation coefficients for the fundamental frequency.
	  	The rightmost panel highlights the hierarchy of the ratios $|C^-_{\lm n}/C^+_{\lm n}|$ for the overtones 
		of the (2,2) and (2,1) multipoles. Note that the beating is more relevant for the (2,1) mode (dotted line) than
		for the quadrupole (solid lines).
		}
	\end{center}
\end{figure*}

Some results  for the (2,2) and (2,1) multipoles are shown in Fig.~\ref{fig:HO_l2}.
In the leftmost panel we show the absolute value of the positive and negative parts of the solution,
$|C_{\lm n}^\pm(t) e^{- \sigma_{\l n}^\pm (t-\tLR)}|$, for $n\leq 4$. 
We see that the negative-frequency contributions are smaller than the corresponding
positive ones. The model thus predicts the negative-frequency modes  to be 
less excited than the positive ones. The same feature can be seen in the third plot
of the same figure, where we show the absolute value of the ratios of negative and positive solutions.
In addition (see the rightmost panel of the figure) the negative-frequency modes become 
more and more relevant as  $n$ grows. This behavior is consistent with the qualitative
discussion of  Ref.~\cite{Damour:2007xr}, see in particular Fig.~4. In that case, the authors argued 
that the positive QNMs are excited during the plunge since the Newtonian frequency of the 
waveform $m \Omega$ gets ``closer" to the positive QNM frequencies. Note in addition that
what is found here is consistent with the structure of the actual solution of the RWZ equation.
The second interesting finding, illustrated by the  middle panel of Fig.~\ref{fig:HO_l2}, is that 
$|C_{\lm 1}^\pm(t)|$ is reminiscent of an activation function and becomes constant after $t\sim \tLR+15$, 
showing that from that time onward the $n=1$ solution can be written as a superposition of 
QNMs with constant coefficients.
For the first overtone, we have that $C_{\lm 2}^\pm(t)$ has
a similar behavior. However, for the $n\geq 3$ overtones, the behavior of $C_{\lm n}^\pm(t)$
is less reminiscent of an activation function. 
In particular, for $n=3$ we see some oscillations in the
solution, as can be seen e.g.~from the third panel of Fig.~\ref{fig:HO_l2}. We are prone to exclude that these are
physical features that can be found also in the full-RWZ case, and we rather interpret them as expressions of the limitation of our toy model.
The third consideration regards the values of the $|C^-_{\lm n}(t)/C^+_{\lm n}(t)|$ ratios for different
multipoles. We show these ratios for the 
(2,2) and (2,1) multipoles in the third panel of Fig.~\ref{fig:HO_l2} with solid and dotted lines, respectively. 
As can be seen, the ratios of the
(2,1) multipole are higher than the ones of the (2,2) multipole for the same $n$. 
This means that, according to our toy model, the mode mixing in the (2,1) multipole should be more
evident than in the (2,2). This is precisely what happens
in the numerical solutions of the full RWZ equations, as shown in Table~\ref{tab:Alm1}.
The driven harmonic oscillator correctly predict also the qualitative relevance of the mode-mixing in the 
$(3,3)$ multipole, since the predicted ratio $|C^-_{331}(t)/C^+_{331}(t)|$ is smaller than the one of 
the (2,2) mode and, as can be seen from the numerical data, the mode-mixing is more relevant in the (2,2) multipole
rather than in the (3,3). 
The final interesting feature that we discuss is that the positive and negative solutions reach their peaks at different
times. This is shown for the (2,2) multipole in the first plot, where we mark the peak-times with vertical lines
(solid for positive-frequency modes and dashed for negative ones).
As can be seen, the negative modes with same $n$ are excited before the corresponding positive modes, and
the overtones are excited before the fundamental QNMs. 

To conclude, with this toy model we have reproduced some features that are observed when solving
the actual RWZ equations and we have also argued that the overtones are excited before the
fundamental frequency. However, since these results are only qualitative, we cannot
exploit them to improve the description of the postpeak waveform.

\section{Conclusions}
\label{sec:conclusions}

The results of this paper are twofold: (i) on the one hand, it is studied the transition from eccentric inspiral to plunge, merger and ringdown
of a binary black hole coalescence in the large mass ratio limit (or particle limit) and its gravitational wave emission; (ii) on the other hand, 
it is introduced and tested an EOB waveform valid in this limit. This improves previous results~\cite{Damour:2007xr} and generalizes 
them to the eccentric case, that was not studied systematically so far. Since the large mass ratio limit can be seen as a well controlled
theoretical laboratory to learn and explore new ideas, this work should be seen as part of the currently  ongoing effort of building accurate 
waveform templates for eccentric (comparable-mass) binaries within the EOB 
formalism~\cite{Chiaramello:2020ehz,Nagar:2020xsk,Nagar:2021gss,Joshi:2022ocr,Bonino:2022hkj,Romero-Shaw:2022fbf,Shaikh:2023ypz}.
More in detail, our results can be summarized as follows:
\begin{itemize}
\item[(i)]
We performed a systematic survey of eccentric nonspinning binaries in the test-particle limit,
focusing in particular on the transition from inspiral to plunge and on the phenomenology of the
corresponding waveform, computed solving numerically the RWZ equations, Eqs.~\eqref{eq:RWZ}, 
with  the \RWZ{} code~\cite{Bernuzzi:2010ty,Bernuzzi:2010ty,Bernuzzi:2011aj,Bernuzzi:2012ku}.
\item[(ii)]  
We discussed a new ringdown model for eccentric binaries in the test-particle limit, modifying the primary fitting templates
used in previous works~\cite{Damour:2014yha,Albanesi:2021rby}. Notably, the ringdown model includes
the mode-mixing between positive and negative frequency fundamental QNMs. The global fits on the parameter 
space are performed using as fitting parameter the gauge-invariant quantity $\bmrg$ (Eq.~\eqref{eq:bmrg}), 
that is defined as the ratio between the angular momentum and energy at the peak of the orbital 
frequency (i.e., at the light-ring crossing). This  moment can be  considered a good definition of the merger 
time within our context.
\item[(iii)]
We then built a complete test-mass EOB model for eccentric nonspinning binaries, including also higher modes
and 2PN non-circular corrections in the $\ell=m=2$ waveform mode. We analyzed in details 
the impact of several analytical building blocks, in particular we tested: (i) the relevance of the generic Newtonian prefactor, 
also in the quasi-circular case, (ii) the contribution of the 2PN non-circular corrections introduced in 
Refs.~\cite{Khalil:2021txt,Placidi:2021rkh,Albanesi:2022xge},
(iii) different prescriptions for the NQC corrections, and in particular the point at which they are computed; 
(iv) the ringdown attachment point.
All these ingredients together provide an accurate complete EOB waveform,  yielding $\Delta \phi_{22} \lesssim 0.01$ 
rad in the quasi-circular case, and $\Delta \phi_{22} \lesssim 0.05$ rad during the merger-ringdown for 
all the eccentric configurations considered. 
Building upon methods introduced in the \texttt{SEOBNR}-family~\cite{Cotesta:2018fcv,Pompili:2023tna,Ramos-Buades:2023ehm}, 
we also explored the performance of a different ringdown model for the higher modes 
that starts from $\tA22$ rather than from $\tAlm$.
\item[(iv)]We analyzed the build-up of QNMs excitation, revisiting also the matching comb procedure 
that was used in former EOB models~\cite{Damour:2007xr,Damour:2007vq,Damour:2008te,Damour:2009kr,Damour:2012ky}. 
Within this context, we also introduced a heuristic model  to compute the excitation coefficients of the QNMs 
in the presence of a driving source. The results of this model are in qualitatively and semi-quantitative agreement with the actual 
structure of the waveform. We also discussed how the power-law tail contribution presents at the end of the ringdown 
changes in the presence of eccentricity. 
\end{itemize}

\section{Acknowledgment}
S.~A. and A.~P. acknowledge the stimulating environment of IHES, were part of this research was conducted.
In particular, it was partly supported by the {\it “2021 Balzan Prize for Gravitation: Physical and Astrophysical Aspects”}, 
awarded to Thibault Damour. We are also grateful to D.~Chiaramello for producing most of the nonspinning simulations
discussed in this work. S.~B. acknowledges support by the EU H2020 under ERC Starting Grant, no. BinGraSp-714626 
and by the EU Horizon under ERC Consolidator Grant, no. InspiReM-101043372.
S.~A. is grateful to Tilly Gatto and Stefano Belisari for continuous support during the development of this work.
We are also grateful to P.~Micca for, never trivial, inspiring suggestions.

\appendix

\section{Regridding used for NQC corrections determination and matching}
\label{app:regridding}
\begin{figure}[t]
	\begin{center}
	\includegraphics[width=0.48\textwidth]{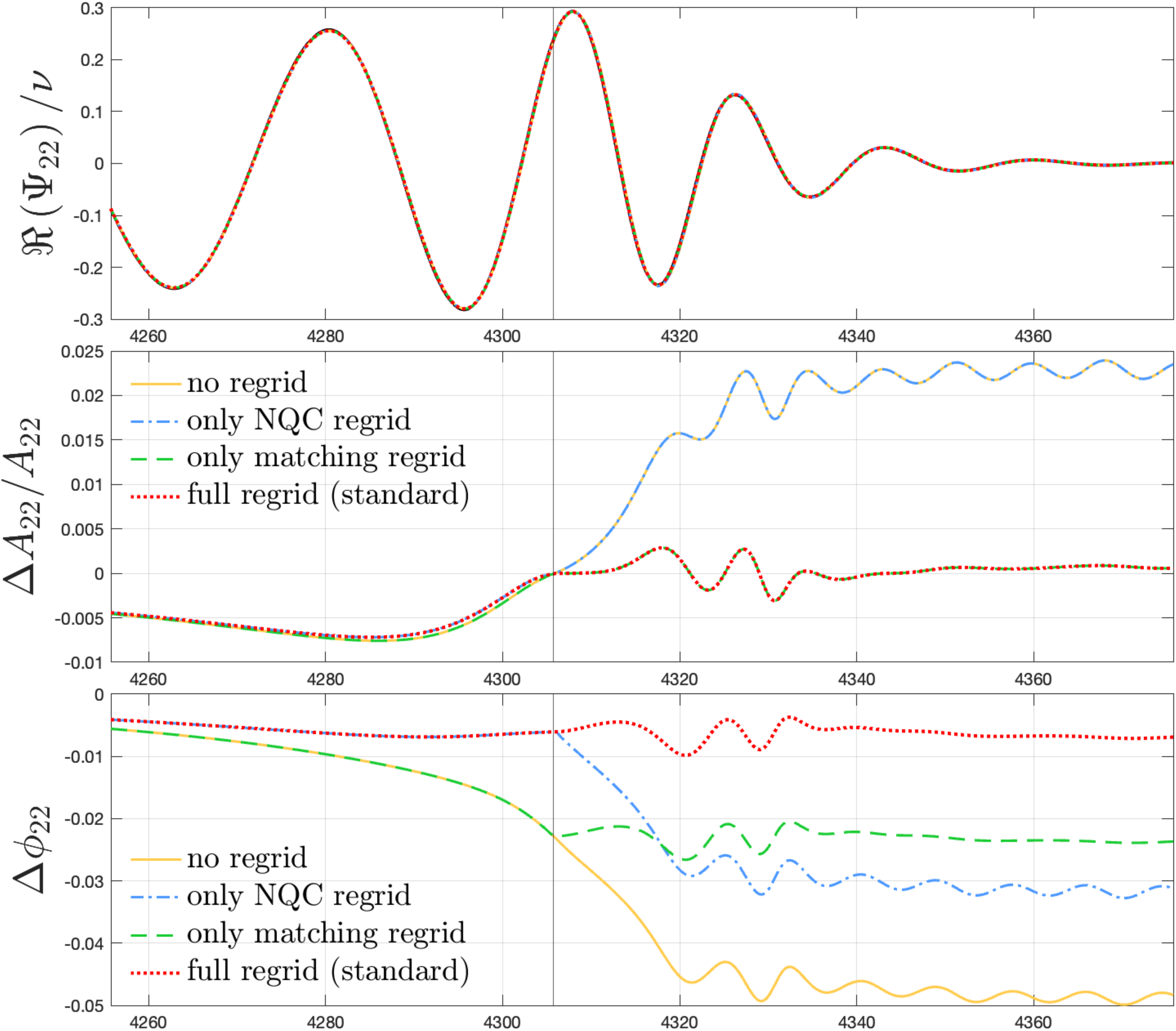}
	\caption{\label{fig:regrid} Analytical/numerical comparison for the (2,2) waveform multipole
	considering different regridding configurations. In the top panel we show the RWZ numerical waveform 
	(black, almost indistinguishable) and the analytical ones using different colors. 
	See legend and text for more details. Note that the red analytical (2,2) waveform 
	is the standard one, i.e.~the one shown in Fig.~\ref{fig:inspl_l2m2}.
	In the middle and bottom panels we show the relative amplitude difference and the phase difference in radians, 
	with same color scheme. The vertical black line marks the peak of $A_{22}$.}
	\end{center}
\end{figure}
As anticipated in Sec.~\ref{sec:nqc} and Sec.~\ref{sec:matchingpoint}, 
the fact that the dynamics is obtained solving the Hamilton's equations~\eqref{eq:Hamilton} numerically
with an ODE solver implies that we have a discrete time array. All the dynamical
quantities and the waveform are thus computed on this array.
This discretization can lead to issues when matching the inspiral waveform
to the ringdown. 

The first issue of this kind is related to the NQC corrections determination. 
Indeed, in order to determine the coefficients $a_i$ and $b_i$ 
of Eq.~\eqref{eq:hnqc}, we have to solve the linear system~\eqref{eq:nqc_syst}. 
However, the rhs is computed exactly at $\tNQC$ (since if $\tNQC>\tAlm$ we compute it
from the analytical ringdown waveform discussed in Sec.~\ref{sec:ringdown_primary},
while if $\tNQC<\tAlm$ we extract it at $\tNQC$ using a cubic spline procedure on the numerical data),
but the lhs. is not. The latter is indeed evaluated using the inspiral waveform $h_\lm^{\rm inspl}$, 
that is in turn computed from the EOB dynamics, which is found solving the Hamilton's equation with a discrete-step 
ODE-solver. Therefore, $\tNQC$ is not guaranteed to be a point of the time array on which the inspiral
waveform is computed. 
This implies that the lhs and the rhs might be computed at
slightly different times and this can introduce a systematic error in the computation
of the NQC coefficients $a_i$ and $b_i$, and thus in the complete EOB waveform.
The second issue is related to the matching procedure. Indeed, in order to match the NQC-corrected
inspiral-plunge waveform to the ringdown model we need to compute $\phi_\lm^{\rm peak}$, i.e.
the phase of the inspiral-plunge waveform at the matching-point, see Eq.~\eqref{eq:barh}.
However, it is not guaranteed that the matching time $\tAlm$ is an element
of the time array used to compute the dynamics. If we operate without performing a regrid,
the phase $\phi_\lm^{\rm peak}$ found can be flawed.

In order to solve the aforementioned issues, we apply for each multipole a regridding procedure in the 
vicinity of $\tNQC$ and $\tAlm$, so that these two points are enforced to be elements of the time grid. 
We then interpolate the chunks of the waveform that we need on the refined time arrays using a cubic spline algorithm. 
We then compute the NQC coefficients, $a_i$ and $b_i$, and we perform the match between the inspiral-plunge waveform 
and the ringdown model on the refined time grid where $\tAlm$ is a grid-point. 
We prefer to reinterpolate the obtained waveform
back on the original time array used to solve the Hamilton's equations in order to have a uniform time-step,
but note that this step is not necessary.
The results for different regridding configurations are shown in Fig.~\ref{fig:regrid}
We consider the analytical waveform without regridding (solid yellow), the one with regridding only for the NQC corrections
(dash-dotted blue), the one with regridding only for the matching procedure (dashed green)
and the one with regridding in both the NQC corrections and matching (dotted red). The latter is the default option
and it is also the one shown in Fig.~\ref{fig:inspl_l2m2}. 
As can be seen from the middle panel, for the amplitude the most relevant regridding is 
the one associated to the matching, while the one associated to the NQC does not seem important. 
The regridding in the matching zone drives the relative amplitude difference in the late 
ringdown from the $2.3\%$ to be below the $0.1\%$. 
For the phase, which analytical/numerical difference is shown in the third panel, we have that both regriddings
are important. Indeed, if we do not consider any regridding, the analytical/numerical
phase difference during the ringdown is about $-0.05$ radians, while with the full
regridding procedure the phase difference drops of one order of magnitude, reaching $-0.006$ radians. 
\begin{figure}[t]
	\begin{center}
	\includegraphics[width=0.48\textwidth]{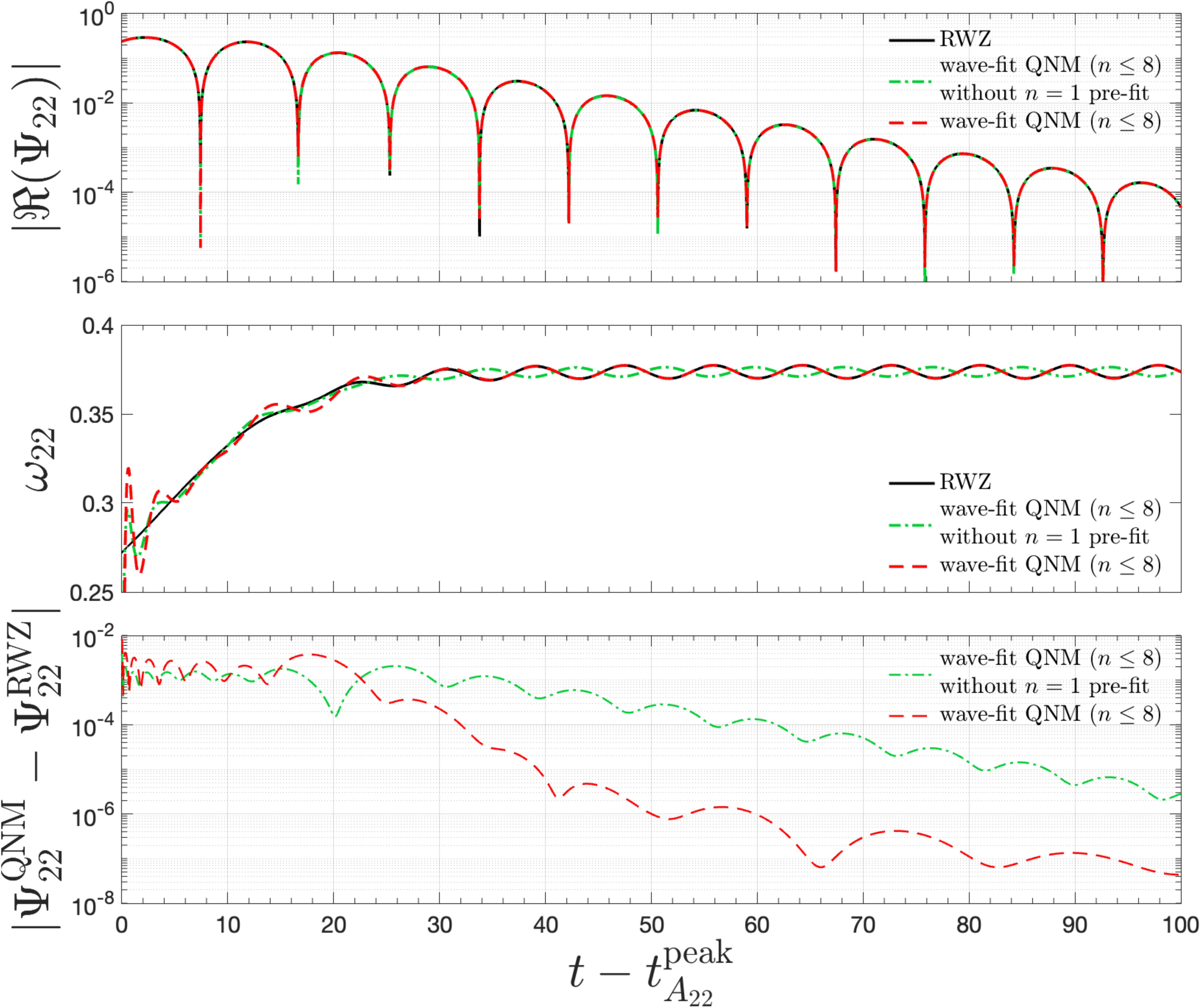}
	\caption{\label{fig:psi_fit_residual} Upper panels: RWZ waveform and frequency (black) compared with the results from
	the waveform fits performed (1) considering all the QNMs together (dash-dotted green), and  (2) prefitting
	the fundamental QNMs (dashed red) and then fitting all the overtones. In the bottom panel we
	show the residuals with same color scheme. See discussion in Appendix~\ref{app:td_QNM_fits}.}
	\end{center}
\end{figure}
\begin{figure*}[t]
	\begin{center}
	\includegraphics[width=0.48\textwidth]{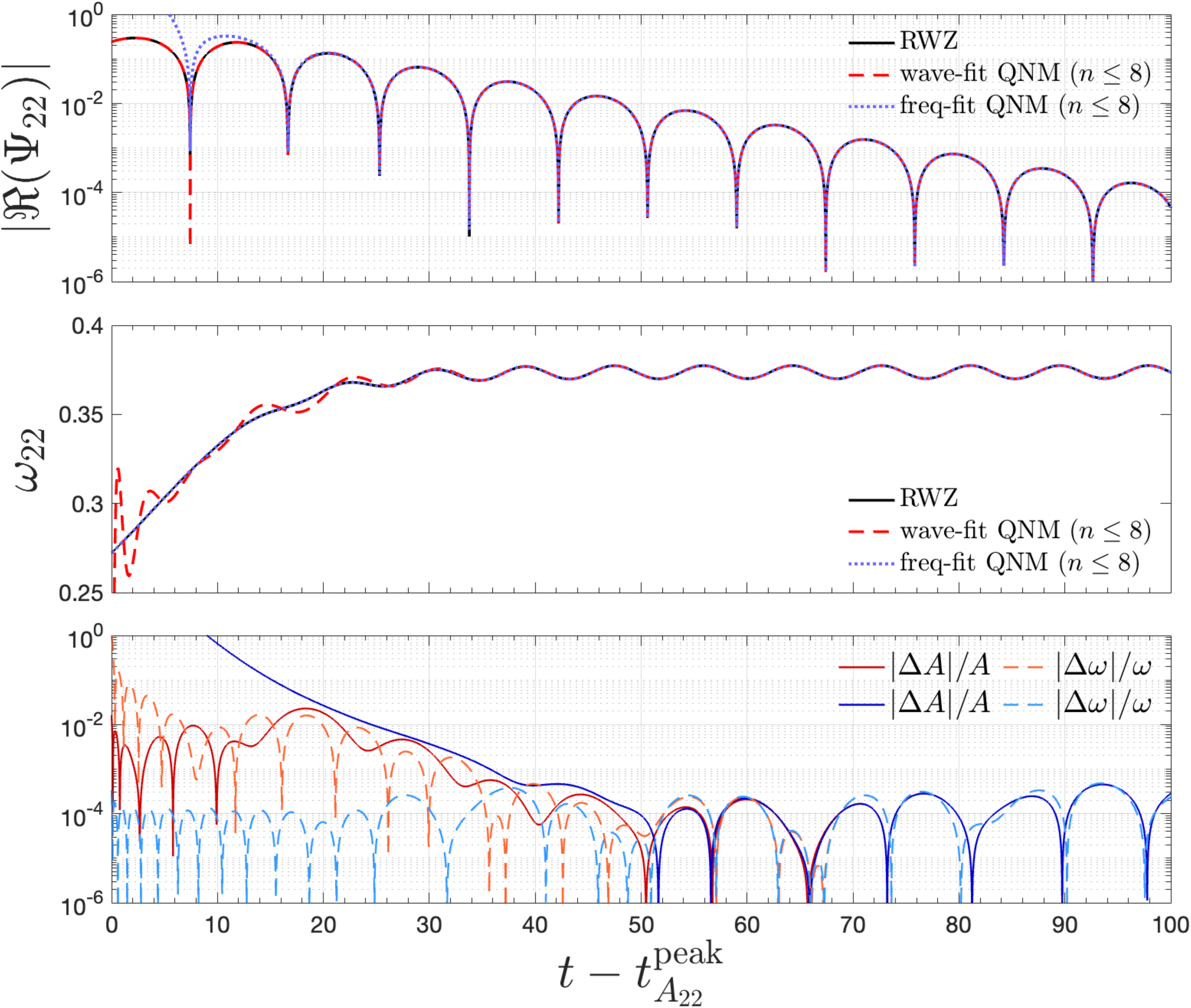}
	\includegraphics[width=0.48\textwidth]{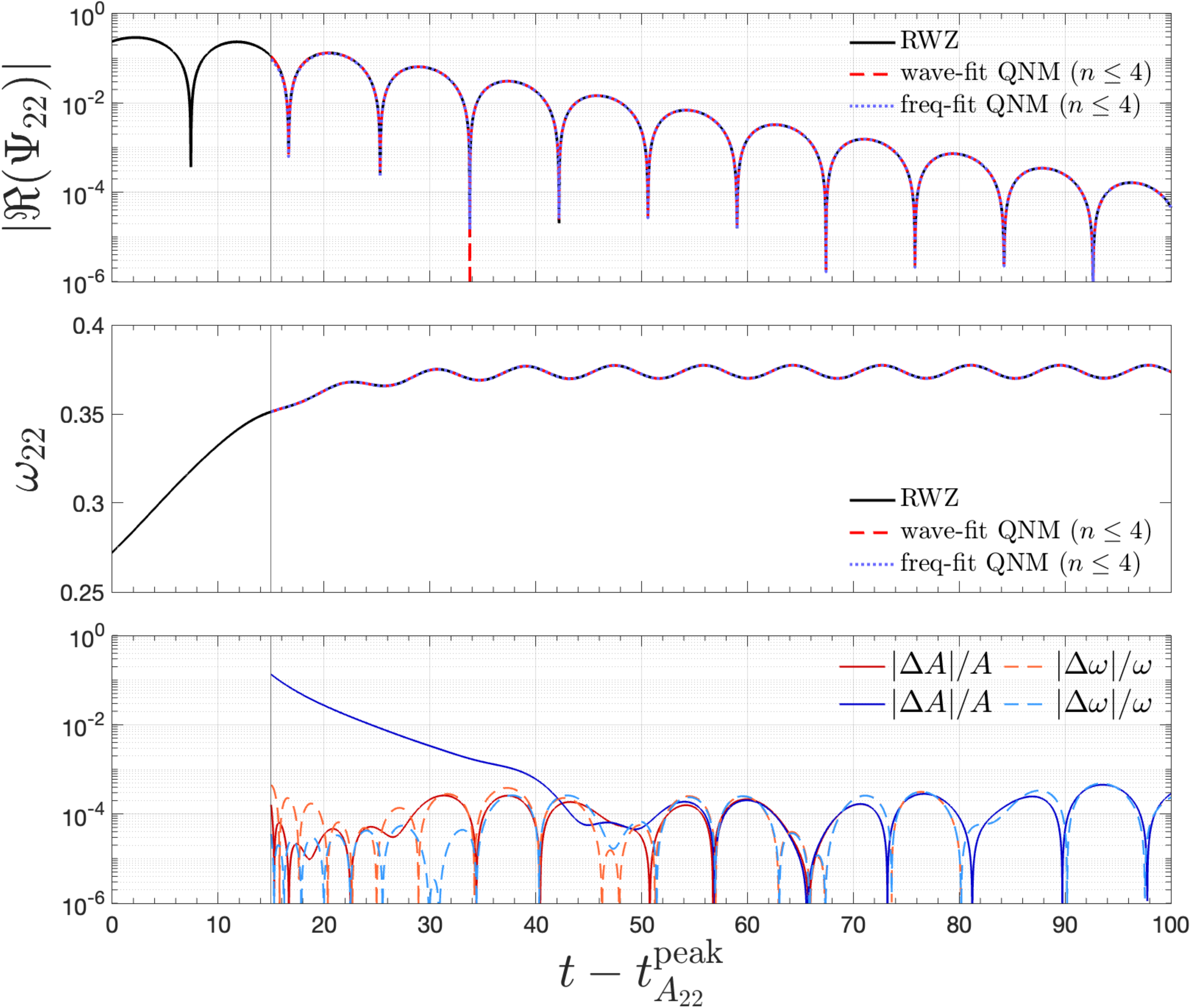}
	\caption{\label{fig:psifreq_fit} Upper panels: RWZ waveform and frequency (black) compared with the results from
	the wave-fit (red) and frequency-fit (blue). In the bottom panels we
	show the relative differences of the amplitude and frequency (hot colors for the waveform-fit, cold colors
	for the frequency-fit). On the left, we fit the whole postpeak waveform using 8 QNMs, while on the right we
	fit from $t\geq15+\tA22$ using 4 QNMs. See discussion in Appendix~\ref{app:td_QNM_fits}.}
	\end{center}
\end{figure*}
%

\section{Time-domain fits of the postpeak waveform using QNMs}
\label{app:td_QNM_fits}
In this appendix we attempt to fit the complete postpeak waveform using the pure QNMs 
ansatz of Eq.~\eqref{eq:qnm_wave} on a time interval $\tau \equiv t - \tA22 \in [\tQNM, 100]$.
We start by performing the fit of the real and imaginary parts of the waveform using the fundamental
QNMs and 7 overtones starting from the peak of the amplitude, i.e. considering $n=8$ and $\tQNM=0$; 
the results of this procedure are shown in Fig.~\ref{fig:psi_fit_residual} (dash-dotted green lines). 
The QNMs waveform overlaps quite well with the numerical RWZ waveform (black), and the corresponding residual is
around $10^{-3}$ shortly after the amplitude peak. However, it is
easy to see that i) the beating between positive and negative frequency QNMs in the late ringdown 
is not well reproduced, as shown by the waveform frequency in the second panel,  
ii) the frequency of the fitted wave shows spurious oscillations in the early ringdown. 
In order to ensure the correct late-ringdown behavior, we refine this procedure 
by prefitting only the fundamental QNMs on $\tau\in[62,100]$,
so that we can easily find $C^\pm_{221}$ since the overtones are negligible in the late ringdown, and
then performing the fit of Eq.~\eqref{eq:qnm_wave} on the whole time interval $\tau \in [0, 100]$, 
so that we can find the remaining $C^\pm_{\lm n}$ coefficients. 
The result of this procedure is shown with dashed red lines in Fig.~\ref{fig:psi_fit_residual}. 
While the beating in the late ringdown frequency is now well reproduced, the
spurious oscillations in the early ringdown frequency are still present. Note that
in this case the residual is a little bit higher in the early ringdown, but way lower in the late evolution.
We thus concluded that, while the residual of the waveforms in quite low in both fits, the waveform 
frequency shows that there are some inaccuracies in the early ringdown.

In an attempt to solve the issue of the spurious frequency oscillations, 
we also explore the possibility to find the $C^\pm_{\lm n}$ 
coefficients fitting directly the frequency  using Eq.~\eqref{eq:qnm_freq}.
Note that if we proceed this way, we cannot find $C_{22 1}^+$
since the frequency is invariant under global normalizations and phase shifts of the waveform. We thus
find $C_{22 1}^+$ as done before, i.e. fitting Eq~\eqref{eq:qnm_wave} with $n=1$ in the late time 
interval $\tau\in[62,100]$.
In the left panel of Fig.~\ref{fig:psifreq_fit} we show this frequency-fit compared with the 
waveform-fit discussed above. To evaluate the goodness of the fits, we show in this case
the relative differences of amplitude and frequency, that we consider more informative than the residual. 
As can be seen, while the frequency-fit solves the issue of the early spurious oscillations, 
produce a waveform that is way less accurate than the one found with the waveform-fit. 

Finally, in the right panels of Fig.~\ref{fig:psifreq_fit} we show 
the comparison between the frequency and waveform fits performed considering $n=4$ and $\tQNM=15$.
In this case the waveform fit is accurate and the recovered frequency is consistent with the numerical result, 
leading to relative amplitude and frequency errors around $10^{-4}$. 
However, the amplitude of the waveform recovered from the frequency-fit reaches a $10\%$ error at $\tau\sim 15$, 
and thus the frequency-fit seems less robust than the waveform-fit, even if we do not fit the early ringdown. 

We thus conclude that i) it is a good practice to check the waveform frequency in order to evaluate 
the goodness of the QNMs fits of the ringdown waveform, ii) within our methods the waveform can be 
consistently fitted only starting from later times. This is again consistent
with the discussion of Sec.~\ref{sec:source} on the RWZ source terms and with the iterative frequency-fit
of Sec.~\ref{sec:omgfit_iter}.

\section{Global fits}
\label{sec:gfit_tables}
In this appendix we report all the global fits that we use in our model, see Sec.~\ref{sec:gfits} for a general discussion
on how we perform the global fits.
We start by reporting the global
fits of the quantities needed to reconstruct the postpeak waveform for all the modes in Table~\ref{tab:gfits_primary1}
and~\ref{tab:gfits_primary2}, see definitions in Sec.~\ref{sec:ringdown_primary}. 
We then proceed to show the global fits for 
$\left\lbrace A^{\rm NQC}_{22}, \dot{A}^{\rm NQC}_{22}, \ddot{A}^{\rm NQC}_{22},
\omega^{\rm NQC}_{22}, \dot{\omega}^{\rm NQC}_{22}, \ddot{\omega}^{\rm NQC}_{22}
\right\rbrace$ evaluated at $\tNQC = \tAlm - 2$ in Table~\ref{tab:gfits_nqc1} and~\ref{tab:gfits_nqc2}.
Note that we report the fits also for $\l=m=2$, 
but for the quadrupolar waveform we do not use them since we use $t_{22}^{\rm NQC}=\tA22+2$, see 
Sec.~\ref{sec:nqc} for discussion.
Note that in all the cases we use a parabolic global template.

\begin{table}[ht!]
    \centering
    \caption{\label{tab:gfits_primary1}
    Global fits of the coefficients entering the ringdown template. 
    For each quantity $y$, the fitting function is: $y = C_{\rm QC} + C_1 \bmrg + C_2 \bmrg^2$,
    where QC indicates the quasi-circular value. 
    The amplitude considered in this table is the amplitude of the strain, $A_\lm=|h_\lm|$, 
    rather than the amplitude of the RWZ-normalized waveform.}
    \begin{ruledtabular}
    \begin{tabular}{ c | c | c | c | c }
 $(\l,m)$ & coeff & $C_{\rm QC}$ & $C_1$ & $C_2$ \\
 \hline
(2,2) &   $c_2^\phi$  &  $ 1.561\cdot 10^{-1}$ &  $-5.067\cdot 10^{-2}$ &  $-5.493\cdot 10^{-2}$ \\ 
      &   $c_3^\phi$  &                  3.272 &                  5.021 &                  6.554 \\ 
      &    $c_4^\phi$ &                  2.592 &                  8.305 &  $ 1.108\cdot 10^{ 2}$ \\ 
      &      $c_2^A$  &  $ 2.161\cdot 10^{-1}$ &  $-1.999\cdot 10^{-2}$ &  $ 7.199\cdot 10^{-3}$ \\ 
      &      $c_3^A$  &                  2.334 &                  3.030 &                  5.583 \\ 
      &      $\Amx$   &                  1.444 &  $ 9.914\cdot 10^{-1}$ &  $ 9.177\cdot 10^{-1}$ \\ 
      &    $\d2Amx$   &  $-2.359\cdot 10^{-3}$ &  $ 6.830\cdot 10^{-3}$ &  $-3.238\cdot 10^{-3}$ \\ 
      & $\Deltaomgmx$ &  $ 1.015\cdot 10^{-1}$ &  $-1.061\cdot 10^{-2}$ &  $ 4.131\cdot 10^{-3}$ \\
\hline 
(2,1) &   $c_2^\phi$  &  $ 1.539\cdot 10^{-1}$ &  $ 4.563\cdot 10^{-2}$ &  $ 1.250\cdot 10^{-1}$ \\
      &   $c_3^\phi$  &                  1.185 &                  1.721 &                 10.167 \\ 
      &    $c_4^\phi$ &                  3.866 &                  7.989 &                 80.460 \\ 
      &      $c_2^A$  &  $ 3.656\cdot 10^{-1}$ &  $ 2.191\cdot 10^{-1}$ &                 -2.017 \\ 
      &      $c_3^A$  &  $-1.535\cdot 10^{-1}$ &                 -3.204 &                 31.896 \\ 
      &      $\Amx$   &  $ 5.238\cdot 10^{-1}$ &  $ 2.332\cdot 10^{-1}$ &  $ 2.119\cdot 10^{-1}$ \\ 
      &    $\d2Amx$   &  $-2.624\cdot 10^{-3}$ &  $ 5.624\cdot 10^{-3}$ &  $ 2.423\cdot 10^{-2}$ \\ 
      & $\Deltaomgmx$ &  $ 8.302\cdot 10^{-2}$ &  $-8.936\cdot 10^{-2}$ &  $ 5.023\cdot 10^{-1}$ \\
\hline 
(3,3) & $c_2^\phi$    &  $ 1.783\cdot 10^{-1}$ &  $-1.361\cdot 10^{-2}$ &  $-1.330\cdot 10^{-3}$ \\ 
      &    $c_3^\phi$ &                  2.818 &                  4.107 &                  9.523 \\ 
      &    $c_4^\phi$ &                  1.536 &                  4.592 &                 28.099 \\ 
      &      $c_2^A$  &  $ 2.192\cdot 10^{-1}$ &  $-1.750\cdot 10^{-2}$ &  $ 3.036\cdot 10^{-2}$ \\ 
      &      $c_3^A$  &                  1.585 &                  1.715 &                  2.634 \\ 
      &      $\Amx$   &  $ 5.635\cdot 10^{-1}$ &  $ 4.299\cdot 10^{-1}$ &  $ 3.883\cdot 10^{-1}$ \\ 
      &    $\d2Amx$   &  $-1.823\cdot 10^{-3}$ &  $ 2.287\cdot 10^{-3}$ &  $ 1.026\cdot 10^{-3}$ \\ 
      & $\Deltaomgmx$ &  $ 1.462\cdot 10^{-1}$ &  $-1.916\cdot 10^{-2}$ &  $ 1.800\cdot 10^{-2}$ \\ 
\hline
(3,2) &    $c_2^\phi$ &  $ 1.851\cdot 10^{-1}$ &  $-8.128\cdot 10^{-3}$ &  $ 9.774\cdot 10^{-3}$ \\ 
      &    $c_3^\phi$ &                  1.308 &  $ 3.395\cdot 10^{-1}$ &  $ 7.985\cdot 10^{-1}$ \\ 
      &    $c_4^\phi$ &  $ 2.172\cdot 10^{-1}$ &  $ 1.135\cdot 10^{-1}$ &  $ 2.334\cdot 10^{-1}$ \\ 
      &      $c_2^A$  &  $ 2.425\cdot 10^{-1}$ &  $ 6.597\cdot 10^{-4}$ &  $-7.605\cdot 10^{-3}$ \\ 
      &      $c_3^A$  &  $ 9.188\cdot 10^{-1}$ &  $ 8.761\cdot 10^{-1}$ &  $-4.018\cdot 10^{-1}$ \\ 
      &      $\Amx$   &  $ 1.991\cdot 10^{-1}$ &  $ 1.289\cdot 10^{-1}$ &  $ 1.017\cdot 10^{-1}$ \\ 
      &    $\d2Amx$   &  $-1.561\cdot 10^{-3}$ &  $-5.955\cdot 10^{-4}$ &  $ 6.250\cdot 10^{-4}$ \\ 
      & $\Deltaomgmx$ &  $ 1.476\cdot 10^{-1}$ &  $ 1.704\cdot 10^{-4}$ &  $ 9.627\cdot 10^{-3}$ \\
\hline
(3,1) &    $c_2^\phi$ &  $ 1.483\cdot 10^{-1}$ &  $-7.316\cdot 10^{-3}$ &  $ 2.289\cdot 10^{-2}$ \\ 
      &    $c_3^\phi$ &  $ 1.406\cdot 10^{-4}$ &  $-2.424\cdot 10^{-4}$ &  $ 6.491\cdot 10^{-4}$ \\ 
      &    $c_4^\phi$ &  $ 6.005\cdot 10^{-5}$ &  $-4.839\cdot 10^{-5}$ &  $ 1.522\cdot 10^{-4}$ \\ 
      &      $c_2^A$  &  $ 2.917\cdot 10^{-1}$ &  $-1.330\cdot 10^{-1}$ &  $ 2.315\cdot 10^{-1}$ \\ 
      &      $c_3^A$  &                  2.068 &                 -1.589 &                  2.079 \\ 
      &      $\Amx$   &  $ 6.230\cdot 10^{-2}$ &  $ 2.460\cdot 10^{-2}$ &  $ 1.575\cdot 10^{-2}$ \\ 
      &   $\d2Amx$    &  $-2.678\cdot 10^{-3}$ &  $-1.660\cdot 10^{-3}$ &  $ 5.739\cdot 10^{-4}$ \\ 
      & $\Deltaomgmx$ &  $ 1.881\cdot 10^{-1}$ &  $-3.904\cdot 10^{-2}$ &  $ 1.385\cdot 10^{-2}$ \\ 
    \end{tabular}
    \end{ruledtabular}
\end{table}
\begin{table}[ht!]
    \caption{\label{tab:gfits_primary2}
     Global fits of the coefficients entering the ringdown template. 
    For each quantity $y$, the fitting function is: $y = C_{\rm QC} + C_1 \bmrg + C_2 \bmrg^2$,
    where QC indicates the quasi-circular value. 
    The amplitude considered in this table is the amplitude of the strain, $A_\lm=|h_\lm|$, 
    rather than the amplitude of the RWZ-normalized waveform.}
    \begin{ruledtabular}
    \begin{tabular}{ c | c | c | c | c }
 $(\l,m)$ & coeff & $C_{\rm QC}$ & $C_1$ & $C_2$ \\
 \hline
(4,4) &    $c_2^\phi$ &  $ 1.845\cdot 10^{-1}$ &  $-6.271\cdot 10^{-3}$ &  $-2.849\cdot 10^{-3}$ \\ 
      &    $c_3^\phi$ &                  2.249 &                  2.991 &                  5.909 \\ 
      &    $c_4^\phi$ &                  1.025 &                  2.617 &                 12.471 \\ 
      &      $c_2^A$  &  $ 2.178\cdot 10^{-1}$ &  $-1.276\cdot 10^{-2}$ &  $ 2.823\cdot 10^{-2}$ \\ 
      &      $c_3^A$  &                  1.145 &                  1.476 &                  2.040 \\ 
      &      $\Amx$   &  $ 2.754\cdot 10^{-1}$ &  $ 2.374\cdot 10^{-1}$ &  $ 2.296\cdot 10^{-1}$ \\ 
      &    $\d2Amx$   &  $-1.258\cdot 10^{-3}$ &  $ 8.999\cdot 10^{-4}$ &  $ 8.471\cdot 10^{-4}$ \\ 
      & $\Deltaomgmx$ &  $ 1.751\cdot 10^{-1}$ &  $-1.720\cdot 10^{-2}$ &  $ 1.743\cdot 10^{-2}$ \\ 
\hline
(4,3) &    $c_2^\phi$ &  $ 1.876\cdot 10^{-1}$ &  $-2.091\cdot 10^{-3}$ &  $ 2.781\cdot 10^{-4}$ \\ 
      &    $c_3^\phi$ &                  1.320 &                  1.129 &  $-7.914\cdot 10^{-4}$ \\ 
      &    $c_4^\phi$ &  $ 3.513\cdot 10^{-1}$ &  $ 6.880\cdot 10^{-1}$ &  $ 3.418\cdot 10^{-1}$ \\ 
      &      $c_2^A$  &  $ 2.252\cdot 10^{-1}$ &  $ 4.376\cdot 10^{-3}$ &  $-6.836\cdot 10^{-2}$ \\ 
      &      $c_3^A$  &  $ 3.600\cdot 10^{-1}$ &                  1.192 &                 -3.611 \\ 
      &      $\Amx$   &  $ 9.418\cdot 10^{-2}$ &  $ 7.365\cdot 10^{-2}$ &  $ 6.538\cdot 10^{-2}$ \\ 
      &    $\d2Amx$   &  $-7.851\cdot 10^{-4}$ &  $-4.334\cdot 10^{-5}$ &  $-1.645\cdot 10^{-4}$ \\ 
      & $\Deltaomgmx$ &  $ 1.722\cdot 10^{-1}$ &  $ 8.015\cdot 10^{-3}$ &  $-2.205\cdot 10^{-2}$ \\ 
\hline
(4,2) &    $c_2^\phi$ &  $ 1.848\cdot 10^{-1}$ &  $-1.045\cdot 10^{-2}$ &  $ 2.014\cdot 10^{-2}$ \\ 
      &    $c_3^\phi$ &                  1.316 &  $ 4.910\cdot 10^{-1}$ &  $-3.244\cdot 10^{-1}$ \\ 
      &    $c_4^\phi$ &  $ 8.103\cdot 10^{-1}$ &                  1.244 &                 -1.645 \\ 
      &      $c_2^A$  &  $ 2.589\cdot 10^{-1}$ &  $ 4.823\cdot 10^{-2}$ &  $-1.436\cdot 10^{-1}$ \\ 
      &      $c_3^A$  &  $-7.454\cdot 10^{-1}$ &                  1.157 &                 -1.954 \\ 
      &      $\Amx$   &  $ 3.138\cdot 10^{-2}$ &  $ 1.544\cdot 10^{-2}$ &  $ 1.181\cdot 10^{-2}$ \\ 
      &    $\d2Amx$   &  $-2.743\cdot 10^{-4}$ &  $ 1.623\cdot 10^{-4}$ &  $-3.799\cdot 10^{-4}$ \\ 
      & $\Deltaomgmx$ &  $ 1.836\cdot 10^{-1}$ &  $-2.595\cdot 10^{-2}$ &  $ 2.780\cdot 10^{-2}$ \\ 
\hline
(4,1) &    $c_2^\phi$ &  $ 1.428\cdot 10^{-1}$ &  $-1.797\cdot 10^{-2}$ &  $ 3.577\cdot 10^{-2}$ \\ 
      &    $c_3^\phi$ &  $-7.583\cdot 10^{-1}$ &  $-3.777\cdot 10^{-1}$ &  $ 6.408\cdot 10^{-1}$ \\ 
      &    $c_4^\phi$ &  $ 1.155\cdot 10^{-1}$ &  $ 1.294\cdot 10^{-1}$ &  $-2.141\cdot 10^{-1}$ \\ 
      &      $c_2^A$  &                  1.067 &  $-4.888\cdot 10^{-1}$ &  $ 8.161\cdot 10^{-1}$ \\ 
      &      $c_3^A$  &                 10.890 &                 -4.684 &                  8.425 \\ 
      &      $\Amx$   &  $ 9.263\cdot 10^{-3}$ &  $ 3.514\cdot 10^{-3}$ &  $ 2.333\cdot 10^{-3}$ \\ 
      &    $\d2Amx$   &  $-7.568\cdot 10^{-4}$ &  $-4.506\cdot 10^{-4}$ &  $-3.366\cdot 10^{-5}$ \\ 
      & $\Deltaomgmx$ &  $ 2.572\cdot 10^{-1}$ &  $-5.064\cdot 10^{-2}$ &  $ 6.728\cdot 10^{-2}$ \\ 
\hline 
(5,5) &    $c_2^\phi$ &  $ 1.872\cdot 10^{-1}$ &  $-3.773\cdot 10^{-3}$ &  $-1.675\cdot 10^{-3}$ \\ 
      &    $c_3^\phi$ &                  1.844 &                  2.360 &                  3.290 \\ 
      &    $c_4^\phi$ &  $ 7.272\cdot 10^{-1}$ &                  1.786 &                  5.737 \\ 
      &     $c_2^A$   &  $ 2.157\cdot 10^{-1}$ &  $-6.055\cdot 10^{-3}$ &  $-7.261\cdot 10^{-3}$ \\ 
      &     $c_3^A$   &  $ 8.165\cdot 10^{-1}$ &                  1.536 &  $ 3.433\cdot 10^{-1}$ \\ 
      &      $\Amx$   &  $ 1.509\cdot 10^{-1}$ &  $ 1.458\cdot 10^{-1}$ &  $ 1.537\cdot 10^{-1}$ \\ 
      &    $\d2Amx$   &  $-8.648\cdot 10^{-4}$ &  $ 3.099\cdot 10^{-4}$ &  $ 4.012\cdot 10^{-4}$ \\ 
      & $\Deltaomgmx$ &  $ 1.954\cdot 10^{-1}$ &  $-1.188\cdot 10^{-2}$ &  $-5.822\cdot 10^{-3}$ \\ 
    \end{tabular}
 \end{ruledtabular}
 \end{table}
 
 \begin{table}[ht!]
    \centering
    \caption{ \label{tab:gfits_nqc1} 
    Global fits of the coefficients for the quantities used to determine NQC corrections 
    for $\tNQC=\tAlm-2$. The fitting template is a quadratic function: $y = C_{\rm QC} + C_1 \bmrg + C_2 \bmrg^2$.
    The amplitude considered in this table is RWZ-normalized as usual, 
    i.e. $A_\lm=|\Psi_\lm|=|h_\lm|/\sqrt{(\l+2)(\l+1)\l(\l-1)}$.}
    \begin{ruledtabular}
    \begin{tabular}{ c | c | c | c | c }
 $(\l,m)$ & $X^{\rm NQC}_\lm$ & $C_{\rm QC}$ & $C_1$ & $C_2$ \\
 \hline
(2,2) &  $            A$ &  $ 2.938\cdot 10^{-1}$ &  $ 2.049\cdot 10^{-1}$ &  $ 1.860\cdot 10^{-1}$ \\ 
      &  $      \dot{A}$ &  $ 7.898\cdot 10^{-4}$ &  $-2.420\cdot 10^{-3}$ &  $ 1.338\cdot 10^{-3}$ \\ 
      &  $     \ddot{A}$ &  $-3.160\cdot 10^{-4}$ &  $ 1.042\cdot 10^{-3}$ &  $-7.056\cdot 10^{-4}$ \\ 
      &  $       \omega$ &  $ 2.609\cdot 10^{-1}$ &  $ 3.121\cdot 10^{-2}$ &  $-2.361\cdot 10^{-3}$ \\ 
      &  $ \dot{\omega}$ &  $ 5.367\cdot 10^{-3}$ &  $-1.031\cdot 10^{-2}$ &  $-5.196\cdot 10^{-4}$ \\ 
      &  $\ddot{\omega}$ &  $ 2.649\cdot 10^{-4}$ &  $ 1.256\cdot 10^{-4}$ &  $-1.332\cdot 10^{-3}$ \\ 
\hline
(2,1) &  $            A$ &  $ 1.063\cdot 10^{-1}$ &  $ 5.236\cdot 10^{-2}$ &  $ 2.594\cdot 10^{-2}$ \\ 
      &  $      \dot{A}$ &  $ 4.784\cdot 10^{-4}$ &  $-1.940\cdot 10^{-3}$ &  $ 1.196\cdot 10^{-3}$ \\ 
      &  $     \ddot{A}$ &  $-1.140\cdot 10^{-4}$ &  $ 4.152\cdot 10^{-4}$ &  $-3.622\cdot 10^{-3}$ \\ 
      &  $       \omega$ &  $ 2.809\cdot 10^{-1}$ &  $ 2.037\cdot 10^{-2}$ &  $-1.893\cdot 10^{-1}$ \\ 
      &  $ \dot{\omega}$ &  $ 9.270\cdot 10^{-3}$ &  $ 7.587\cdot 10^{-3}$ &  $ 3.664\cdot 10^{-2}$ \\ 
      &  $\ddot{\omega}$ &  $-3.868\cdot 10^{-3}$ &  $ 9.083\cdot 10^{-4}$ &  $ 3.541\cdot 10^{-2}$ \\ 
\hline
(3,3) &  $            A$ &  $ 5.115\cdot 10^{-2}$ &  $ 3.967\cdot 10^{-2}$ &  $ 3.547\cdot 10^{-2}$ \\ 
      &  $      \dot{A}$ &  $ 2.767\cdot 10^{-4}$ &  $-4.255\cdot 10^{-4}$ &  $ 1.457\cdot 10^{-5}$ \\ 
      &  $     \ddot{A}$ &  $-1.109\cdot 10^{-4}$ &  $ 1.999\cdot 10^{-4}$ &  $-5.411\cdot 10^{-5}$ \\ 
      &  $       \omega$ &  $ 4.322\cdot 10^{-1}$ &  $ 4.023\cdot 10^{-2}$ &  $-2.155\cdot 10^{-3}$ \\ 
      &  $ \dot{\omega}$ &  $ 1.026\cdot 10^{-2}$ &  $-1.279\cdot 10^{-2}$ &  $ 3.809\cdot 10^{-4}$ \\ 
      &  $\ddot{\omega}$ &  $ 3.777\cdot 10^{-4}$ &  $ 5.998\cdot 10^{-4}$ &  $-1.296\cdot 10^{-3}$ \\
\hline
(3,2) &  $            A$ &  $ 1.790\cdot 10^{-2}$ &  $ 1.174\cdot 10^{-2}$ &  $ 9.344\cdot 10^{-3}$ \\ 
      &  $      \dot{A}$ &  $ 2.582\cdot 10^{-4}$ &  $-3.442\cdot 10^{-7}$ &  $-3.865\cdot 10^{-5}$ \\ 
      &  $     \ddot{A}$ &  $-1.076\cdot 10^{-4}$ &  $ 4.345\cdot 10^{-5}$ &  $-2.468\cdot 10^{-5}$ \\ 
      &  $       \omega$ &  $ 4.205\cdot 10^{-1}$ &  $ 1.124\cdot 10^{-2}$ &  $ 2.154\cdot 10^{-3}$ \\ 
      &  $ \dot{\omega}$ &  $ 1.534\cdot 10^{-2}$ &  $-6.496\cdot 10^{-3}$ &  $ 2.843\cdot 10^{-3}$ \\ 
      &  $\ddot{\omega}$ &  $ 2.792\cdot 10^{-4}$ &  $-3.190\cdot 10^{-4}$ &  $-8.332\cdot 10^{-5}$ \\ 
\hline
(3,1) &  $            A$ &  $ 5.358\cdot 10^{-3}$ &  $ 2.113\cdot 10^{-3}$ &  $ 1.367\cdot 10^{-3}$ \\ 
      &  $      \dot{A}$ &  $ 2.413\cdot 10^{-4}$ &  $ 5.772\cdot 10^{-5}$ &  $ 5.734\cdot 10^{-5}$ \\ 
      &  $     \ddot{A}$ &  $ 4.851\cdot 10^{-6}$ &  $ 7.180\cdot 10^{-5}$ &  $ 6.059\cdot 10^{-6}$ \\ 
      &  $       \omega$ &  $ 4.189\cdot 10^{-1}$ &  $ 7.247\cdot 10^{-2}$ &  $-2.821\cdot 10^{-2}$ \\ 
      &  $ \dot{\omega}$ &  $-2.106\cdot 10^{-3}$ &  $-1.196\cdot 10^{-2}$ &  $ 7.362\cdot 10^{-3}$ \\ 
      &  $\ddot{\omega}$ &  $-1.256\cdot 10^{-2}$ &  $-1.138\cdot 10^{-2}$ &  $ 1.408\cdot 10^{-3}$ \\ 
    \end{tabular}
          \end{ruledtabular}
    \end{table}
\begin{table}[ht!]
    \caption{\label{tab:gfits_nqc2} 
    Global fits of the coefficients for the quantities used to determine NQC corrections 
    for $\tNQC=\tAlm-2$. The fitting template is a quadratic function: $y = C_{\rm QC} + C_1 \bmrg + C_2 \bmrg^2$.
    The amplitude considered in this table is RWZ-normalized as usual, 
    i.e. $A_\lm=|\Psi_\lm|=|h_\lm|/\sqrt{(\l+2)(\l+1)\l(\l-1)}$.}
    \begin{ruledtabular}
    \begin{tabular}{ c | c | c | c | c }
 $(\l,m)$ & $X^{\rm NQC}_\lm$ & $C_{\rm QC}$ & $C_1$ & $C_2$ \\
 \hline
(4,4) &  $            A$ &  $ 1.440\cdot 10^{-2}$ &  $ 1.261\cdot 10^{-2}$ &  $ 1.215\cdot 10^{-2}$ \\ 
      &  $      \dot{A}$ &  $ 1.124\cdot 10^{-4}$ &  $-1.072\cdot 10^{-4}$ &  $-3.334\cdot 10^{-5}$ \\ 
      &  $     \ddot{A}$ &  $-4.584\cdot 10^{-5}$ &  $ 5.504\cdot 10^{-5}$ &  $ 1.442\cdot 10^{-6}$ \\ 
      &  $       \omega$ &  $ 6.041\cdot 10^{-1}$ &  $ 4.204\cdot 10^{-2}$ &  $ 1.016\cdot 10^{-3}$ \\ 
      &  $ \dot{\omega}$ &  $ 1.476\cdot 10^{-2}$ &  $-1.500\cdot 10^{-2}$ &  $-1.774\cdot 10^{-4}$ \\ 
      &  $\ddot{\omega}$ &  $ 3.536\cdot 10^{-4}$ &  $ 1.146\cdot 10^{-3}$ &  $-1.417\cdot 10^{-3}$ \\ 
\hline
(4,3) &  $            A$ &  $ 4.887\cdot 10^{-3}$ &  $ 3.881\cdot 10^{-3}$ &  $ 3.445\cdot 10^{-3}$ \\ 
      &  $      \dot{A}$ &  $ 7.313\cdot 10^{-5}$ &  $-3.014\cdot 10^{-6}$ &  $ 5.836\cdot 10^{-6}$ \\ 
      &  $     \ddot{A}$ &  $-3.104\cdot 10^{-5}$ &  $ 5.333\cdot 10^{-6}$ &  $-3.956\cdot 10^{-6}$ \\ 
      &  $       \omega$ &  $ 5.980\cdot 10^{-1}$ &  $ 1.029\cdot 10^{-2}$ &  $-1.114\cdot 10^{-2}$ \\ 
      &  $ \dot{\omega}$ &  $ 1.979\cdot 10^{-2}$ &  $-7.181\cdot 10^{-3}$ &  $-3.824\cdot 10^{-4}$ \\ 
      &  $\ddot{\omega}$ &  $-1.791\cdot 10^{-4}$ &  $ 1.279\cdot 10^{-3}$ &  $ 3.702\cdot 10^{-4}$ \\ 
\hline
(4,2) &  $            A$ &  $ 1.627\cdot 10^{-3}$ &  $ 8.138\cdot 10^{-4}$ &  $ 5.752\cdot 10^{-4}$ \\ 
      &  $      \dot{A}$ &  $ 2.687\cdot 10^{-5}$ &  $ 7.489\cdot 10^{-6}$ &  $ 4.104\cdot 10^{-5}$ \\ 
      &  $     \ddot{A}$ &  $-1.470\cdot 10^{-5}$ &  $-1.090\cdot 10^{-5}$ &  $-7.254\cdot 10^{-6}$ \\ 
      &  $       \omega$ &  $ 5.784\cdot 10^{-1}$ &  $ 2.331\cdot 10^{-2}$ &  $-2.894\cdot 10^{-2}$ \\ 
      &  $ \dot{\omega}$ &  $ 2.622\cdot 10^{-2}$ &  $-4.593\cdot 10^{-3}$ &  $-3.936\cdot 10^{-3}$ \\ 
      &  $\ddot{\omega}$ &  $-4.671\cdot 10^{-4}$ &  $ 6.784\cdot 10^{-3}$ &  $-4.209\cdot 10^{-3}$ \\ 
\hline
(4,1) &  $            A$ &  $ 4.311\cdot 10^{-4}$ &  $ 1.636\cdot 10^{-4}$ &  $ 1.081\cdot 10^{-4}$ \\ 
      &  $      \dot{A}$ &  $ 4.004\cdot 10^{-5}$ &  $ 7.867\cdot 10^{-6}$ &  $ 1.042\cdot 10^{-5}$ \\ 
      &  $     \ddot{A}$ &  $ 5.997\cdot 10^{-6}$ &  $ 1.533\cdot 10^{-5}$ &  $ 9.482\cdot 10^{-8}$ \\ 
      &  $       \omega$ &  $ 5.673\cdot 10^{-1}$ &  $ 9.191\cdot 10^{-2}$ &  $-3.606\cdot 10^{-2}$ \\ 
      &  $ \dot{\omega}$ &  $-1.560\cdot 10^{-2}$ &  $-1.071\cdot 10^{-2}$ &  $ 8.617\cdot 10^{-3}$ \\ 
      &  $\ddot{\omega}$ &  $-3.018\cdot 10^{-2}$ &  $-3.355\cdot 10^{-2}$ &  $ 5.444\cdot 10^{-3}$ \\ 
\hline
(5,5) &  $            A$ &  $ 5.151\cdot 10^{-3}$ &  $ 5.057\cdot 10^{-3}$ &  $ 5.331\cdot 10^{-3}$ \\ 
      &  $      \dot{A}$ &  $ 5.170\cdot 10^{-5}$ &  $-2.949\cdot 10^{-5}$ &  $-2.260\cdot 10^{-5}$ \\ 
      &  $     \ddot{A}$ &  $-2.150\cdot 10^{-5}$ &  $ 1.723\cdot 10^{-5}$ &  $ 6.342\cdot 10^{-6}$ \\ 
      &  $       \omega$ &  $ 7.789\cdot 10^{-1}$ &  $ 4.215\cdot 10^{-2}$ &  $ 6.786\cdot 10^{-4}$ \\ 
      &  $ \dot{\omega}$ &  $ 1.894\cdot 10^{-2}$ &  $-1.663\cdot 10^{-2}$ &  $-1.112\cdot 10^{-3}$ \\ 
      &  $\ddot{\omega}$ &  $ 2.169\cdot 10^{-4}$ &  $ 1.746\cdot 10^{-3}$ &  $-1.577\cdot 10^{-3}$ \\  
    \end{tabular}
\end{ruledtabular}
\end{table}

\clearpage
\bibliography{refs20231023.bib,refs_loc20231023.bib}

\end{document}